\documentclass[traditabstract]{aa} % for the abstract without structuration 
\usepackage{amsmath}
\usepackage{graphicx}
\usepackage{url}
\usepackage{lscape}
\usepackage{longtable}
\usepackage{multirow}
\usepackage{txfonts}

\begin{document}
   \title{H$\alpha$ imaging of the \textit{Herschel} Reference Survey}

   \subtitle{The star formation properties of a volume-limited, K-band-selected 
   sample of nearby late-type galaxies\thanks{Tables 1-7  
   are also available in electronic form at the CDS via anonymous ftp to cdsarc.u-strasbg.fr(130.79.128.5) or via http://cdsweb.u-strasbg.fr/cgi-bin/qcat?J/A+A/}}

  \author{A. Boselli\inst{1}
          ,
	  M. Fossati\inst{2,3}
	  ,
	  G. Gavazzi\inst{4}
	  ,
	  L. Ciesla\inst{5}
	  ,
	  V. Buat\inst{1}
	  ,
	  S. Boissier\inst{1}
	  ,
	  T. M. Hughes\inst{6}
         }

\institute{	Aix Marseille Universit\'e, CNRS, LAM (Laboratoire d'Astrophysique de Marseille) UMR 7326, 13388, Marseille, France 
              \email{alessandro.boselli@lam.fr}
         \and 
	 	Universit{\"a}ts-Sternwarte M{\"u}nchen, Schenierstrasse 1, D-81679 M{\"u}nchen, Germany
	 \and
		Max-Planck-Institut f\"{u}r Extraterrestrische Physik, Giessenbachstrasse, 85748, Garching, Germany 
		\email{mfossati@mpe.mpg.de}
	 \and
	 	Universita' di Milano-Bicocca, piazza della Scienza 3, 20100, Milano, Italy 
              \email{Giuseppe.Gavazzi@mib.infn.it}
	 \and
	        University of Crete, Department of Physics, Heraklion 71003, Greece 
		\email{ciesla@physics.uoc.gr}   
	 \and
	        Sterrenkundig Observatorium, Universiteit Gent, Krijgslaan 281-S9, Gent 9000, Belgium
		\email{thomas.hughes@ugent.be}
}
 
\authorrunning{Boselli et al.}
\titlerunning{H$\alpha$ imaging of the \textit{Herschel} Reference Survey}

   \date{}

% \abstract{}{}{}{}{} 
% 5 {} token are mandatory
 
  \abstract
   {We present new H$\alpha$+[NII] imaging data of late-type galaxies in the \textit{Herschel} Reference Sample aimed at studying the star formation properties of 
   a K-band-selected, volume-limited sample of nearby galaxies.
   The H$\alpha$+[NII] data are corrected for [NII] contamination and dust attenuation using different recipes based on the Balmer decrement
   and the 24 $\mu$m luminosities. We show that the H$\alpha$ luminosities derived with different corrections give consistent results only whenever the 
   the uncertainty on the estimate of the Balmer decrement is $\sigma[C(H\beta)]\leq$ 0.1. 
   We use these data to derive the star formation rate of the late-type galaxies of the sample, and compare these estimates to those determined using independent 
   monochromatic tracers (FUV, radio continuum) or the output of spectral energy distribution (SED) fitting codes. This comparison suggests that the 24 $\mu$m based dust extinction correction 
   for the H$\alpha$ data might be non universal, and that it should be used with caution in all objects with a low star formation activity, where dust heating can be 
   dominated by the old stellar population. Furthermore, because of the sudden truncation of the star formation activity of cluster galaxies occurring after their interaction with the 
   surrounding environment, the stationarity conditions required to transform monochromatic fluxes into star formation rates 
   might not always be satisfied in tracers other than the H$\alpha$ luminosity. In a similar way, the parametrisation of the star
   formation history generally used in SED fitting codes might not be adequate for these recently interacting systems.
   We then use the derived star formation rates to study the $SFR$ luminosity distribution and the typical scaling relations of the late-type galaxies of the HRS. 
   We observe a systematic decrease of the specific star formation rate
   with increasing stellar mass, stellar mass surface density, and metallicity. We also observe an increase of the asymmetry and smoothness 
   parameters measured in the H$\alpha$-band with increasing $SSFR$, probably induced by an increase of the contribution of giant 
   HII regions to the H$\alpha$ luminosity function in star-forming low-luminosity galaxies.    }
   {}
   {}
   {}
   {}


   \maketitle
%
%________________________________________________________________

\section{Introduction}

Star formation is a key process in the study of galaxy evolution. Stars are formed within giant molecular clouds 
through the collapse of the gaseous component. Massive stars, once formed, produce and inject metals into the interstellar medium, that later aggregate to form dust (Valiante et al. 2009). The various ingredients of the interstellar medium, including those produced 
during stellar evolution, all contribute in regulating the matter cycle in galaxies. The formation of the molecular gas
occurs primarily on dust grains (Hollenbach \& Salpeter 1971; Wolfire et al. 2008). Dust also absorbs the interstellar radiation field, and is thus an
important parameter in the cooling process of the gas (Bakes \& Tielens 1994; Wolfire et al. 1995; Hollenbach \& Tielens 1997). Massive stars can also inject a large amount of kinetic 
energy into the interstellar medium, favoring the ionisation of the surrounding gas and the dissociation of the molecular 
component, but also cloud-cloud collisions important in the process of star formation.

The hydrogen recombination lines are due to the cascade of electrons captured by the hydrogen nucleus once photoionised by the far-UV radiation ($\lambda$ $<$ 912 \AA)
in HII regions. This highly energetic UV radiation is mainly emitted by massive ($m_{star}$ $>$ 8 M$_{\odot}$) O-B stars, whose
life on the main sequence is very short ($<$ 10$^7$ yr). Their presence thus indicates recent episodes of star formation. 
The H$\alpha$ Balmer line ($\lambda$ 6563 \AA) is the brightest of the hydrogen recombination lines. 
This line is easily accessible from ground based facilities in local galaxies since it is located in the 
visible spectral domain. Under specific conditions, its emission is proportional to 
the number of newly formed stars and can thus be used as a direct tracer of star formation
(Kennicutt et al. 1994; Kennicutt 1998; Boselli et al. 2001). Star formation rates are proportional to H$\alpha$ luminosities
if the star formation activity of the targets is constant over a timescale at least as long as the time that the ionising stars spend 
on the main sequence ($\sim$ 10$^7$ yr). Only under these conditions does the number of stars that leave the main sequence equal that of newly formed stars.
The constant of proportionality between the H$\alpha$ luminosity and the star formation rate depends on the initial mass function (IMF), on the
metallicity, and on several assumptions in the photoionisation models, and can
be estimated using population synthesis models. 

H$\alpha$ luminosities, however, can be converted into star formation rates only once corrected for dust attenuation. This is generally done using the Balmer decrement 
(Lequeux et al. 1981), determined by comparing the observed H$\alpha$/H$\beta$ flux ratio to the value expected for the typical conditions in HII regions (2.86, 
Case B; Osterbrock \& Ferland 2006). Spectroscopic observations can be used for this purpose.
An intermediate spectral resolution ($R$ $\sim$ 1000) is required to separate the emission of the H$\alpha$ line from that of the two bracketing [NII] lines ($\lambda\lambda$ 
6548, 6584 \AA), while good signal-to-noise is necessary to determine the underlying Balmer absorption produced by the stellar atmosphere of young stars. 
The accurate determination of the Balmer decrement is thus non trivial, in particular in normal nearby galaxies
characterised by a relatively low or moderate activity of star formation. Indeed, in these galaxies the intensity of the emission lines, and in particular that of 
H$\beta$, is relatively low and often comparable to the intensity of the underlying absorption. 
It is thus fundamental to understand up to which limit in signal-to-noise
the Balmer decrement can be accurately determined without introducing systematic errors in the estimate of the star formation rate (Groves et al. 2012). 
This is also crucial to quantify the uncertainties on the determination of the dust attenuation of galaxies up to $z$ $\sim$ 1, 
which is often estimated using higher order Balmer lines (H$\beta$,$\lambda$4861\AA; H$\gamma$,$\lambda$4340\AA; H$\delta$, $\lambda$4101\AA). These lines
are characterised by a lower intensity and a higher underlying absorption with respect to H$\alpha$ (e.g. Momcheva et al. 2013). 
It is also critical to estimate 
whether the exclusion of objects with low H$\beta$ emission from the analysis of the star formation properties of complete samples of galaxies does not bias 
the results. Galaxies with low H$\beta$ emission, indeed, are objects with a low star formation activity and/or high dust attenuation.

To overcome these technical difficulties, different tracers have been proposed in the literature either for correcting the observed H$\alpha$ luminosities
or for measuring star formation rates (e.g. Kennicutt 1998; Kennicutt et al. 2009; Hao et al. 2011; Kennicutt \& Evans 2012). Using the same arguments as for the H$\alpha$ line, any tracer of the young stellar population can be converted, under some assumptions,
into star formation rates. The most widely used tracers are the dust-corrected far-ultraviolet (FUV) and the radio continuum luminosities. 
At 20 cm, the radio continuum emission of galaxies is mainly due to the synchrotron emission of relativistic electrons spinning in weak magnetic fields (e.g. Lequeux 1971; Condon 1992). 
These electrons are accelerated in supernovae remnants, and are thus tightly related
to the youngest stellar populations of galaxies. The FUV and radio continuum luminosities can be converted into star formation rates whenever the star formation activity 
of galaxies is constant over $\sim$ 10$^8$ yrs, a timescale ten times longer than necessary when using the H$\alpha$ luminosity, making these tracers more uncertain in objects suddenly 
changing their star formation activity with time.
When multifrequency data are available, the star formation activity of galaxies can also be determined through the fitting of their spectral energy
distribution (SED) with specific codes. The accuracy of this technique, which has the advantage of providing a consistent estimate of the contribution of dust attenuation to the stellar
emission when UV, optical and infrared data are available, depends on the sampling of the different photometric bands. 
It also depends on the choosen parametrisation of the star formation history of the galaxies, which is generally done with simple empirical relations.
Compared to monochromatic tracers, this method has the advantage to account for possible variations of the star formation history of galaxies, even though these variations are not 
easily constrained (e.g. Buat et al. 2014).\\

The direct comparison of these different tracers is therefore crucial for identifying and quantifying their limits and uncertainties,
as well as for understanding whether the use of a specific correction or calibration can introduce important systematic biases in the derived star formation rates
(Kennicutt 1998; Kennicutt \& Evans 2012; Kennicutt et al. 2009; Calzetti et al. 2007, 2010; Salim et al. 2007; Lee et al. 2009; Boselli et al. 2009).
The comparison of these tracers must be carried out on well defined samples of galaxies spanning the widest possible range in the parameter space and having the largest
possible data coverage over all wavelengths (e.g. Buat et al. 2014).

The \textit{Herschel} Reference Sample (Boselli et al. 2010) is ideal for this purpose. Composed of 323 nearby galaxies, this sample is volume-limited (15 $\leq$ $dist$ $\leq$ 25
Mpc) and K-band-selected, 
which roughly corresponds to a stellar mass selection (Gavazzi et al. 1996). It also includes galaxies of all morphological types in the stellar mass range 5 $\times$ 10$^8$ $\leq$ $M_{star}$
$\leq$ 10$^{11}$ M$_{\odot}$. The sample has been defined to study the physical properties of the interstellar medium, the star formation process, and the effects of the environment on galaxy evolution
in normal galaxies. It thus includes galaxies in different density regions, from the sparse field to the rich core of the Virgo cluster. We have been collecting  
multifrequency data covering the entire electromagnetic spectrum, including UV \textit{GALEX} and visible SDSS data (Boselli et al. 2011; Cortese et al. 2012),
near- (2MASS), mid- and far-IR \textit{WISE} (Ciesla et al. 2014), \textit{Spitzer} (Bendo et al. 2012; Ciesla et al. 2014), and \textit{Herschel} (Ciesla et al. 2012; 
Cortese et al. 2014) data, while radio continuum data at 20 cm are available from the NVSS survey (Condon et al. 1998). Medium resolution ($R$ $\sim$ 1000) integrated spectroscopic data 
are also available (Boselli et al. 2013), as well as HI and CO data (Boselli et al. 2014a). The purpose of this article is to use this unique sample and set of data to 
determine and compare different tracers of star formation, derived from both monochromatic luminosities and SED fitting techniques, 
in order to determine their range of validity, their strengths and limits. We then use these data to trace the statistical properties of the star formation activity of the 
late-type galaxies of the sample, including their star formation rate distribution, scaling relations, and structural and morphological CAS parameters (Conselice 2003), 
for both normal and cluster galaxies. There are indeed strong indications that the 
star formation properties of cluster galaxies are strongly affected by the hostile environment in which they reside (e.g. Boselli \& Gavazzi 2006, 2014).

The paper is structured as follows: in section 2 we describe the sample, in section 3 the observations and in section 4 the data reduction. The data are analysed in section 5 and 6, 
and the conclusions summarised in section 7. In three different appendices, we present the new spectroscopic data determined using the {\sc GANDALF} code, the radio continuum data at
20 cm taken from the literature and used in the analysis, and we list the recipes used to convert observed luminosities into star formation rates.
We limit our analysis to the late-type galaxies of the sample (Sa-Im-BCD). It is indeed known that the H$\alpha$ emission in early-types (E-S0) 
does not necessary come from the photoionisation of the gas by the young stellar populations (several
of these early-type objects are also strong X-ray emitters). Furthermore, their FUV emission is generally due to very evolved stars 
(O'Connell 1999; Boselli et al. 2005) not associated to any event of star formation, while their radio continuum emission might be dominated by the
contribution of the central AGN (M87 and M84 are well know powerful radio galaxies).

\begin{figure}
\centering
\includegraphics[width=8cm]{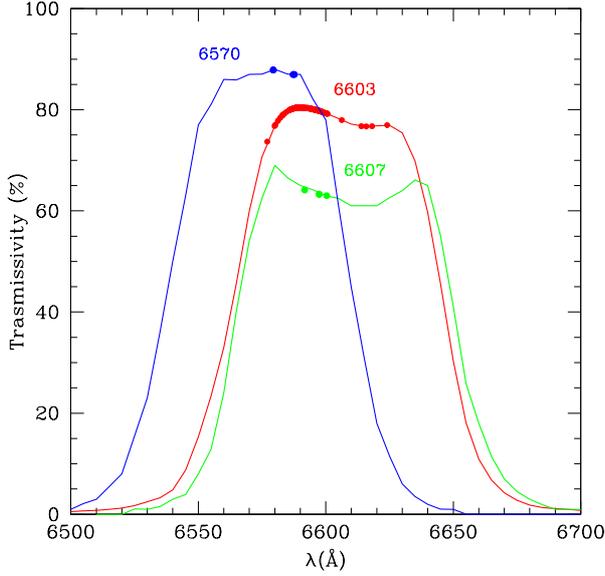}
\caption{Transmissivity of the ON-band (6603, 6607, 6570 \AA) filters. Points mark the
throughput at the wavelength corresponding to the H$\alpha$ line at the redshift of the target galaxies.}
\label{filter}
\end{figure}

\section{The sample}

The observed sample is composed of those HRS late-type galaxies without H$\alpha$ imaging
data available in the literature (see Table \ref{TabHRS}). It also includes a few objects
with H$\alpha$+[NII] data from aperture photometry (Kennicutt \& Kent 1983; Romanishin 1990). 
Combined with H$\alpha$+[NII] imaging data available in the literature, 
mostly gathered during our previous survey of the Virgo cluster (Boselli \& Gavazzi 2002; Boselli et al.
2002a; Gavazzi et al. 2002, 2006), the HRS sample is now complete at the 87\% level, and to 98\% if limited to late-type galaxies. 
Because of the presence of bright stars close to the target, whose reflection causes unwanted extended low 
surface brightness structures on the images, six objects could not be observed.
The HRS sample is listed in Table \ref{TabHRS}, arranged as follows:

\begin{itemize}
\item {Column 1: \textit{Herschel} Reference Sample (HRS) name, from Boselli et al. (2010).}
\item {Column 2: Zwicky name, from the Catalogue of Galaxies and of Cluster of Galaxies (CGCG; Zwicky et al. 1961-1968).}
\item {Column 3: Virgo Cluster Catalogue (VCC) name, from Binggeli et al. (1985).}
\item {Column 4: Uppsala General Catalog (UGC) name (Nilson 1973).}
\item {Column 5: New General Catalogue (NGC) name (Dreyer 1888).}
\item {Column 6: Index Catalogue (IC) name (Dreyer 1895).}
\item {Columns 7 and 8: J2000 right ascension and declination, from NED.}
\item {Column 9: Morphological type, from NED, or from our own classification if not available.}
\item {Column 10: Distance, in Mpc. Distances have been determined from the recessional velocity assuming a Hubble constant $H_0$ = 70 km s$^{-1}$ Mpc$^{-1}$
for galaxies outside the Virgo cluster, and assumed to be 17 Mpc for galaxies belonging to Virgo, with exception to those located in the Virgo cluster B substructure 
(23 Mpc; Gavazzi et al. 1999).}
\item {Column 11: Stellar mass, from Cortese et al. (2012a), determined following the prescription of Zibetti et al. (2009) based on the $i$-band luminosity and $g-i$ mass-to-light ratio. 
For galaxies without SDSS $g$ and $i$-band data (11 objects, marked with $^a$ in Table 1), 
stellar masses have been computed using the prescription of Boselli et al. (2009) based on the H-band luminosity and $B-H$ mass-to-light ratio. }
\item {Column 12: $g$-band optical isophotal diameter (24.5 mag arcsec$^{-2}$), from Cortese et al. (2012a). For the HRS galaxies without SDSS images, the $g$-band isophotal diameter
was determined from the relation $r_{24.5}(g)$ = 0.871($\pm$0.017)$r_{25}(B)$ + 6.041($\pm$2.101), where $r_{25}(B)$
is the radius given in NED (Boselli et al. 2014a).}
\item {Column 13: inclination of the galaxy, determined using the prescription based on the morphological type described in Haynes \& Giovanelli (1984)
and the $i$-band ellipticity given in Cortese et al. (2012a).}
\item {Column 14: Heliocentric radial velocity (in km s$^{-1}$), from HI data when available (Boselli et al. 2014a), otherwise from NED.}
\item {Column 15: Cluster or cloud membership, from Gavazzi et al. (1999) for Virgo and Tully (1987) or Nolthenius (1993) whenever available, 
or from our own estimate (Boselli et al. 2010). }
\item {Column 16: Code to indicate whether H$\alpha$+[NII] data are available (1) or not (0). }
\end{itemize}

\section{Observations}

H$\alpha$+[NII] narrow band imaging of 138 HRS late-type galaxies has been obtained during 
different observing runs, from 2006 to 2012, with the 2.1m and the 1.5m telescopes at San Pedro
Martir (SPM; Baja California, Mexico). These galaxies have been observed as fillers during an 
H$\alpha$+[NII] imaging survey of HI selected galaxies in the nearby universe
(H$\alpha$3; Gavazzi et al. 2012; Gavazzi et al. 2015b). All galaxies observed at the 
2.1m SPM telescope (133 objects) were observed through the narrow band 
interferometric filter $\lambda$=6603
\AA, $\Delta$$\lambda$ 70 \AA ~(ON-band frame)
whose spectral coverage is optimal for HRS objects with recessional velocity 
160 $<$ $vel$ $<$ 3500 km s$^{-1}$. The 5 galaxies done at the 1.5m telescope
have been observed using the $\lambda$=6607 \AA, $\Delta$$\lambda$ 61 \AA ~(ON-band frame)
and the $\lambda$=6570 \AA, $\Delta$$\lambda$ 66 \AA ~ filters (see Fig. \ref{filter}).
Given the relatively large width of these
filters, the present H$\alpha$ images include the contribution from the [NII] lines. 
The stellar continuum (OFF-band frame) was gathered through a broad-band
$r$-Gunn filter. Typical integration times were 15-20 minutes ON-band, generally
split into shorter exposures for cosmic ray removals, and 4 minutes OFF-band.
The observations were generally taken during photometric conditions, with a 
seeing of $\sim$ 1.5-3.0 arcsec (see Gavazzi et al. 2012; Gavazzi et al. 2015b). Photometric calibrations were secured with the
observation of two standards, Feige34 and Hz44, from the catalogue of Massey et
al. (1988), observed every 2-3 hours with integrations of 1-2 minutes. The repeated
observations of the standard stars have shown that the photometric accuracy (zero
point) was stable within $<$ 5\%.

\begin{figure*}
\centering
\includegraphics[width=17cm]{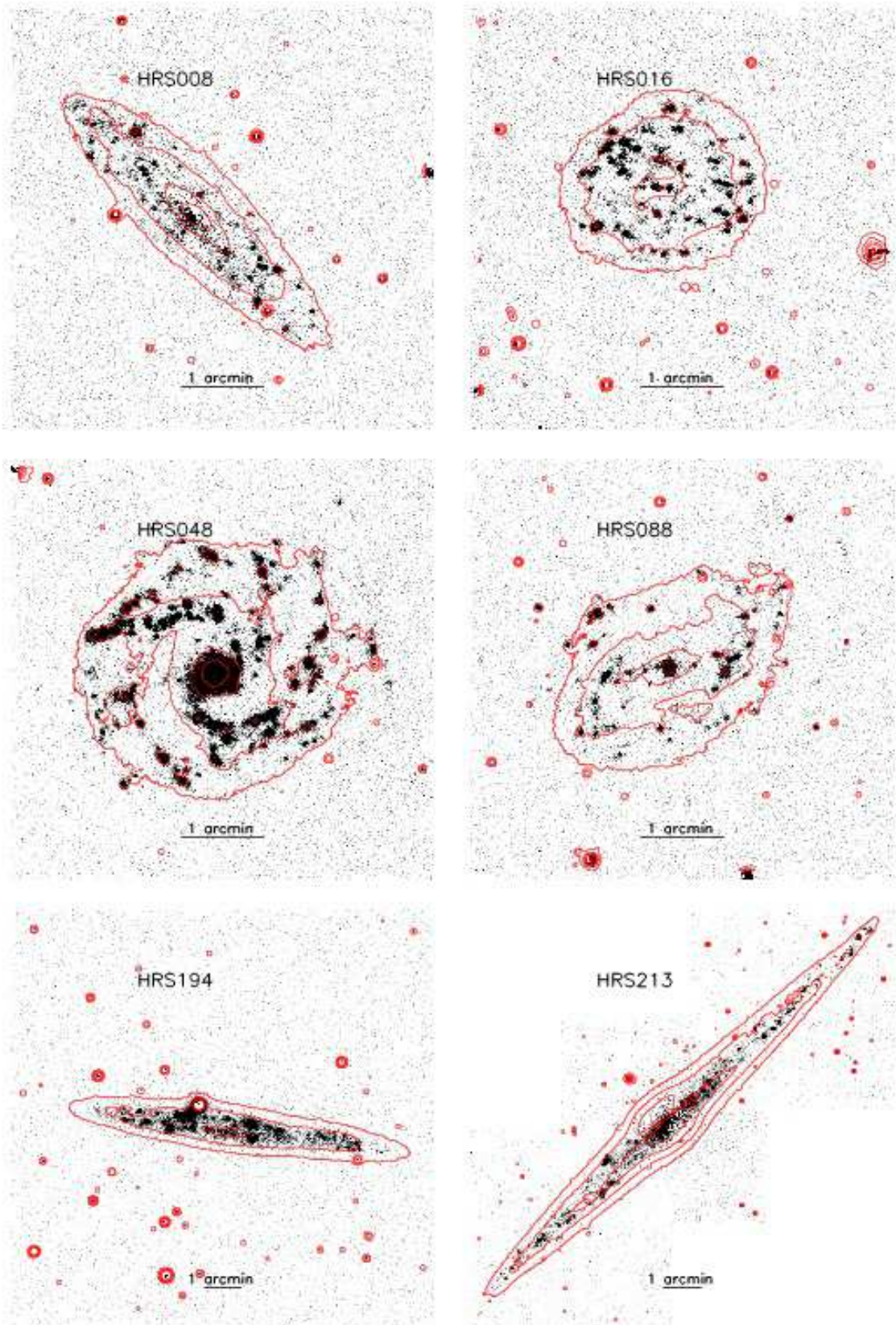}
\caption{Images of 6 galaxies observed in H$\alpha$. The OFF-band contours are logarithmically drawn at 3, 9, 27, and 81 $\times \sigma$ 
of the sky background in the OFF frame and the grey scales represent the NET flux intensity between 1 and 5 $\times \sigma$ of the sky in the NET 
frame. A 1 arcmin bar is given on all images. North is up and east is to the left.}
\label{atlas}
\end{figure*}

\section{Data reduction}

The obtained frames were reduced following the same procedures as described in 
our previous papers (e.g. Gavazzi et al. 2002, 2012). These are standard procedures generally
used in the literature (e.g. Waller 1990; Koopmann et al. 2001; James et al. 2004; Kennicutt et al. 2008). These procedures are based on {\sc IRAF}
STSDAS\footnote{{\sc IRAF} is distributed by the National Optical Astronomy Observatory, 
which is operated by the Association of Universities for Research in Astronomy (AURA) 
under cooperative agreement with the National Science Foundation.} reduction packages. Each image was bias subtracted and
divided by the median of several flat fields obtained on empty sky regions during
twilight. When three images in the same filter were available, a median combination
of the images allowed cosmic ray removal. For single images, cosmic ray removal was
secured using the COSMICRAY {\sc IRAF} task and by direct inspection of the frame.
Unwanted foreground stars were removed on each ON- and OFF-band frame. The sky
background was measured in concentric, uncontaminated annuli around the object, and
subtracted from the flat-fielded images. \\
Total counts in the two frames have been obtained
by integrating the pixel counts over the area covered by each galaxy, as derived by 
the optical major and minor diameters.
If $C_{ON}$ and $C_{OFF}$ are the integrated pixel counts in the ON and OFF-band filter respectively,
${C_{NET} = C_{ON} - n C_{OFF}}$,
then the NET flux in the observed H$\alpha$+[NII] line is given by:

\begin{equation}
{F(H\alpha+[NII])_o ~~~~[\rm{erg~ cm^{-2} sec^{-1}}]= 10^{Zp} \frac{C_{NET}}{T R_{ON}(H\alpha)}}
\end{equation}

\noindent
and the equivalent width by:

\begin{equation}
{H\alpha+[NII] E.W._o ~~~~[\rm{\AA}]= \frac{\int R_{ON}(\lambda)d\lambda}{R_{ON}(H\alpha)} \frac{C_{NET}}{n C_{OFF}}}
\end{equation}

\noindent
where $T$ is the integration time (sec), $10^{Zp}$ is the ON-band zero point ($\rm erg~ cm^{-2} sec^{-1}$) 
corrected for atmospheric extinction 
and $R_{ON}(\lambda)$ is the transmissivity of the ON-filter at the wavelength of the redshifted H$\alpha$ line (Fig. \ref{filter}).
Eq. (2) shows that the H$\alpha$ equivalent width does not depend on $Zp$, but only
on the normalization constant $n$ measured using several stars in both frames, and so it can also be estimated in marginal photometric conditions. 
The normalisation factor $n$ has been multiplied by $\sim$ 0.95 
as indicated by Spector et al. (2012) to account for the fact
that field stars are generally redder than the stellar continuum of the observed galaxies. 
%With this correction the $r$-band photometry of the target galaxies matches that determined 
%using SDSS images (Gavazzi et al. 2015b). }\\

\noindent
We corrected for the contamination of the H$\alpha$+[NII] line emission in the 
broad band filter (OFF-band) following the prescription given in Boselli et al. (2002a): 

\begin{displaymath}
{F(H\alpha+[NII])_c=}
\end{displaymath}
\begin{equation}
{=F(H\alpha+[NII])_o (1+{\frac{\int R_{ON}(\lambda)d\lambda}{\int R_{OFF}(\lambda)d\lambda}})}
\end{equation}

\noindent
and

\begin{displaymath}
{H\alpha+[NII]E.W._c=}
\end{displaymath}
\begin{displaymath}
{=H\alpha+[NII]E.W._o (1+{\frac{H\alpha+[NII]E.W._o}{\int R_{OFF}}}) \times}
\end{displaymath}
\begin{equation}
{(1+{\frac{\int R_{ON}(\lambda)d\lambda}{\int R_{OFF}(\lambda)d\lambda}})}
\end{equation}

\noindent
where $F(H\alpha+[NII])_o$ and $H\alpha+[NII]E.W._o$ are the observed values 
(from eq. (1) and (2)), $F(H\alpha+[NII])_c$ and $H\alpha+[NII]E.W._c$ the corrected ones,
and $R_{ON}$ and $R_{OFF}$ the transmissivity of the ON band and $r$-Gunn filters.\\
For extended sources, the dominant source of error is the variation 
of the background on angular scales similar to the size of the source on the plane of the sky.
The error thus depends primarily on the quality of the  
flat-fielding. We measured the background in several regions comparable with the size
of the galaxies and determined that its fluctuation (per pixel) is 10 \% of 
the purely statistical $rms$ on the individual pixels. The total uncertainty on the ON and OFF
counts is thus proportional to the area $A$ (in pixels) covered by each galaxy,
estimated from the optical major and minor axes, $a$ and $b$:\\

\noindent{\smallskip}
$\sigma_{ON} = 0.1 ~rms_{ON}~ A$\\
\noindent{\smallskip}
$\sigma_{OFF} = 0.1 ~rms_{OFF}~ A$ \\ 
\noindent{\smallskip}
which add up to:\\
$\sigma_{NET} = \sqrt{(\sigma_{ON})^2 + (\sigma_{OFF})^2 + (0.1~C_{NET})^2}$ \\
\noindent
The term $(0.1~C_{NET})^2$ accounts for 
the uncertainty on the photometric calibration.\\
The errors on the H$\alpha$ +[NII] flux $\sigma_{F}$ and equivalent width $\sigma_{E.W.}$ are finally:
\begin{equation}
{\sigma_{F} = \frac{F(H\alpha)}{C_{NET}} \sigma_{NET}}
\end{equation}

\begin{equation}
{\sigma_{E.W.} = \frac{\int R_{ON}(\lambda)d\lambda}{R_{ON}(H\alpha) (n C_{OFF})^2} \sqrt{(n C_{OFF})^2 \sigma_{ON}^2 + C_{ON}^2 \sigma_{OFF}^2}}
\end{equation}

\noindent
We recall that equations 5 and 6 do not take into account the uncertainty on the normalisation factor $n$ which might depend on the colour of each galaxy
and can be as large as 10-30\%~ (Spector et al. 2012).
The derived parameters of the observed galaxies are listed in Table 2, arranged as follows:

\begin{itemize}
\item {Column 1:\textit{Herschel} Reference Sample (HRS) name.}
\item {Column 2: ON-band filter.}
\item {Column 3: Telescope.}
\item {Column 4: used CCD.}
\item {Column 5: Pixel size, in arcseconds.}
\item {Column 6: Observing run.}
\item {Column 7: Number of exposures.}
\item {Column 8: ON band exposure time per pose, in seconds.}
\item {Column 9: Air mass.}
\item {Column 10: Photometric quality of the sky: 1 stands for photometric conditions, 0 unclear conditions (thin cirrus).}
\item {Column 11: Zero point of the observations, in erg cm$^{-2}$ s$^{-1}$.}
\item {Column 12: Normalisation factor $n$ between the ON- and the OFF-band ($r$-Gunn) filter.}
\end{itemize}

\noindent
A few galaxies have been observed during different observing runs. For these galaxies Table \ref{TabHaobs} gives a mean value.

\begin{figure*}
\centering
\includegraphics[width=17cm]{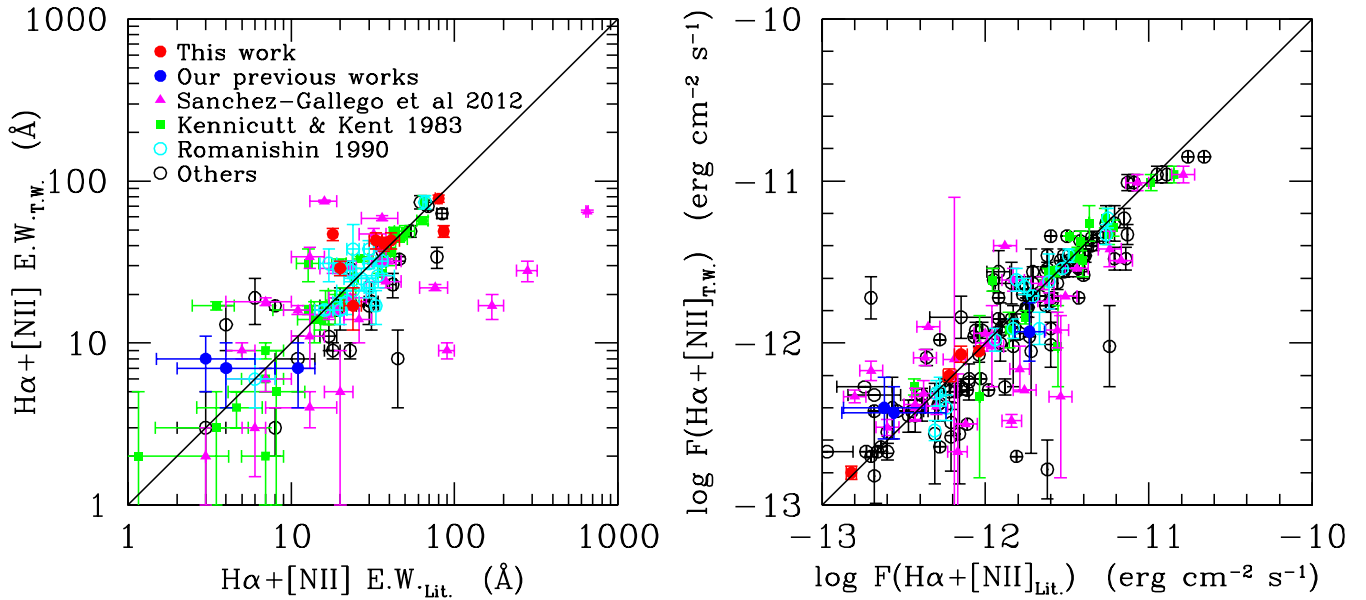}
\caption{Comparison of the H$\alpha$+[NII] equivalent widths (left) and fluxes (right) of the HRS galaxies 
with independent measurements available in the literature. Red filled dots indicate galaxies with multiple observations done in this work, 
blue filled dots galaxies with data already published by our team (Boselli \& Gavazzi 2002; Boselli et al.
2002a; Gavazzi et al. 2002, 2006), magenta, green, cyan, and black symbols H$\alpha$+[NII] 
measurements from Sanchez-Gallego et al. (2012), Kennicutt \& Kent (1983) (corrected by a factor of 16\% as suggested by Kennicutt et al. 1994), 
Romanishin (1990) or from other references in the literature, respectively. 
The solid line shows the 1:1 relation.}
\label{comp}
\end{figure*}

\subsection{H$\alpha$+[NII] data for HRS galaxies}

We combine the new set of H$\alpha$+[NII] imaging data with those collected in the literature. 
With our new observations, 281 of the 323 galaxies of the sample now have H$\alpha$+[NII] imaging data. 
The sample is almost complete if limited to late-type systems (254/260 objects, 98 \%). Table \ref{TabHa}
lists the H$\alpha$+[NII] equivalent widths and fluxes for the whole HRS sample. Table \ref{TabHa} is arranged as follows:

\begin{itemize}
\item {Column 1:\textit{Herschel} Reference Sample (HRS) name.}
\item {Column 2 and 3: H$\alpha$+[NII] equivalent width and error, in \AA.}
\item {Column 4 and 5: observed H$\alpha$+[NII] flux and error, in erg cm$^{-2}$ s$^{-1}$.}
\item {Column 6: Reference to the data. When two references are given, the first refers to the equivalent width, the second to the flux.
References are coded as follows:
TW: this work,
1: Boselli \& Gavazzi (2002),
2: Boselli et al. (2002a),
3: Gavazzi et al. (2002),
4: Gavazzi et al. (2006),
5: Koopmann et al. (2001),
6: Young et al. (1996),
7: Kennicutt et al. (1987),
8: Macchetto et al. (1996),
9: James et al. (2004),
10: Hameed et al. 2005),
11: Koopmann \& Kenney (2006),
12: Usui et al. (1998),
13: Domingue et al. (2003),
14: Trinchieri \& Di Serego Alighieri (1991),
15: Finkelman et al. (2010),
16: Kim (1989),
17: Martel et al. (2004),
18: Shields (1991),
19: Singh et al. (1995),
20: Kennicutt \& Kent (1983),
21: Romanishin (1990),
22: Sanchez-Gallego et al. (2012).
}
\item {Column 7: Alternative references, if available.}
\item {Column 8: Notes to individual objects: $c$ indicates that the flux of the galaxy has been determined by indirectly calibrating
the image using the published flux of the companion galaxy, $m$ indicates that the published value is a mean value of two independent
measurements, $v$ is for vignetted images where the total flux cannot be properly extracted.}
\end{itemize}

We also determined the H$\alpha$+[NII] CAS (concentration, asymmetry and 
clumpiness; Conselice 2003) structural parameters for all galaxies with available images. These parameters have been determined following the same procedures 
described in Fossati et al. (2013) in both the NET- and the $r$-band images. These parameters are given in Table \ref{Tabstruc},
arranged as follow:

\begin{itemize}
\item {Column 1:\textit{Herschel} Reference Sample (HRS) name.}
\item {Columns 2-4: $r$-band, H$\alpha$+[NII] and EW$_{H\alpha+[NII]}$ effective radii, in arcsec.}
\item {Columns 5-7: Concentration, asymmetry, and clumpiness (CAS) parameters from the $r$-band images.}
\item {Columns 8-10: Concentration, asymmetry, and clumpiness (CAS) parameters from the H$\alpha$+[NII] narrow band images images.}
\end{itemize}

Figure \ref{atlas} shows the H$\alpha$+[NII] image of six representative galaxies of the sample.
All the Tables presented in this work, as well as the H$\alpha$+[NII] images of the whole sample, will be made available to the community through the HRS dedicated database (http://hedam.lam.fr).

\subsection{Comparison with the literature}

Independent sets of data are available for several HRS galaxies (see Table \ref{comparisontab}). In order to check the quality of our own measurements 
and of those collected from the literature, in Fig. \ref{comp} we compare the different sets of published data. 
A comparison between the equivalent width of the H$\alpha$+[NII] line determined from this
set of imaging data with that obtained from integrated spectroscopy has been already presented in Boselli et al. (2013). 
Figure \ref{comp} and Table \ref{comparisontab} indicate that the different sets of imaging data
give results consistent within $\simeq$ 20\% ~ for the equivalent widths and $\simeq$ 10\% ~ for the fluxes. 
The agreement with the spectroscopic data of Boselli et al. (2013) is within $\simeq$ 5\% ~(see their Fig. 10). 
The agreement is good between our independent measurements, or with those obtained by our team
during previous observing runs (Boselli \& Gavazzi 2002; Boselli et al.
2002a; Gavazzi et al. 2002, 2006). Our new set of data is also consistent with the measurements of Kennicutt \& Kent (1983) 
(corrected by a factor of 16\% as suggested by Kennicutt et al. (1994) to take into account a possible contamination of a telluric line in their narrow band filters) and
Romanishin (1990) done using aperture photometry. They are also fairly consistent 
with the data recently published by Sanchez-Gallego et al. (2012) 
or with a few other data collected from the literature from a large variety of references. Figure \ref{comp} and Table \ref{comparisontab} also show 
that the uncertainty on the data is generally underestimated using standard error propagation (as recipes given in eq. 5 and 6). 
One possible reason is because these recipes do not take into 
account the contribution from correlated noise (which is realistically expected to be a factor of 2-3). Other possible reasons are the photometric uncertainties
on the zero point determination and uncertainties in the continuum subtraction.
Overall, the uncertainty on H$\alpha$+[NII]E.W. over the whole HRS dataset is of the order of $\sim$ 66\%, while that on the H$\alpha$+[NII] flux $\sim$ 60\%.

\section{Determination of the SFR}

\begin{figure}
\centering
\includegraphics[width=13cm]{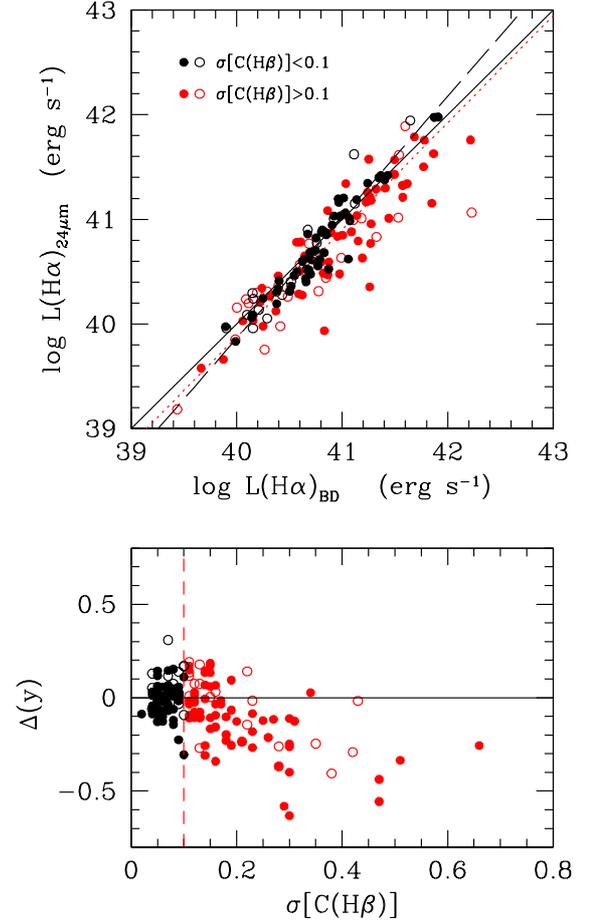}
\caption{Upper panel: relationship between the H$\alpha$ luminosity corrected for dust attenuation using the 24 $\mu$m luminosity
and the prescription of Calzetti et al. (2010) and the H$\alpha$ luminosity corrected using the Balmer decrement. 
Black symbols are for galaxies with a $\sigma[C(H\beta)]$ $\leq$ 0.1, red symbols for $\sigma[C(H\beta)]$ $>$ 0.1. 
Filled dots are for galaxies with a normal gas content ($HI-def$ $\leq$ 0.4), empty symbols for gas-poor objects ($HI-def$ $>$ 0.4).
The black solid line shows the 1:1 relation, the black long-dashed line the bisector fit (Isobe et al. 1990) determined using the best quality sample 
($\sigma[C(H\beta)]$ $\leq$ 0.1), while the red dotted line the best fit determined using the whole sample. Lower panel: relationship between the distance from the 
$L(H\alpha)_{24\mu }$ vs. $L(H\alpha)_{BD}$ relation and the uncertainty on the Balmer decrement estimate $\sigma[C(H\beta)]$. The vertical red dashed line shows 
the limit in $\sigma[C(H\beta)]$ = 0.1 above which data are asymmetrically distributed in $\Delta(y)$. }
\label{hacal}
\end{figure}

\subsection{Dust attenuation correction}

As mentioned in the introduction, the H$\alpha$ emission of late-type star forming galaxies not dominated by an AGN
is due to the gas ionised by the youngest and most massive O-B
stars (Kennicutt 1998; Boselli et al. 2001). Under some assumptions on the shape of the IMF and on the star formation history,
H$\alpha$ data can be used to measure the present day star formation activity of galaxies. To do this, the H$\alpha$+[NII] data listed in Table \ref{TabHa}
must be corrected to remove the contribution of the two [NII] lines in the narrow band filter and to account for both the Galactic and 
internal dust attenuation. We first correct the observed H$\alpha$+[NII] for the [NII] contamination using an updated version of the long slit integrated spectroscopic data
of the HRS galaxies published in Boselli et al. (2013) (see Appendix A). We then correct them for Galactic extinction using the Schlegel et al. (1998) 
extinction map combined with the Galactic extinction curve of Fitzpatrick \& Massa (2007):

\begin{equation}
{A_G(H\alpha) =  2.517 \times E(B-V)_G}
\end{equation}

\noindent
The same set of spectroscopic data is used to estimate the Balmer decrement:

\begin{equation}
{C(H\beta) = \frac{log(2.86) - log[\frac{F(H\alpha)}{F(H\beta)}]_{obs}}{f(H\alpha)}}
\end{equation}

\noindent
based on H$\alpha$-to-H$\beta$ flux ratio and the Galactic extinction law ($f(H\alpha)$=-0.297).
The attenuation in the H$\alpha$ line is then simply given by the relation:

\begin{equation}
{A(H\alpha) = 1.754 \times C(H\beta)}
\end{equation}

\noindent
To allow a direct comparison with other works in the literature, we do not apply any further correction for the 
escape fraction of ionising photons nor for the absorption of the ionising radiation by dust (see Boselli et al. 2009 for details).\\

The attenuation in the H$\alpha$ emission can also be determined using the 24 $\mu$m emission combined with one of the several prescriptions
given in the literature (Kennicutt et al. 2007, 2009; Calzetti et al. 2007, 2010; Zhu et al. 2008)\footnote{H$\alpha$ fluxes must be first 
corrected for Galactic extinction and [NII] contamination as mentioned above.}. These relations have been calibrated using nearby samples 
of galaxies with available narrow band H$\alpha$+[NII] imaging data, integrated spectroscopy, and mid-infrared data. These data are also available for the HRS sample: \textit{WISE} 22 $\mu$m data for the whole HRS have been recently published by Ciesla et al. (2014). These data can be converted into 24 $\mu$m
flux densities by multiplying them by a factor of 1.22, as prescribed in Ciesla et al. (2014) (see also Boselli et al. 2014d).

\subsubsection{Limits in the Balmer decrement determination}

Figure \ref{hacal} shows the relationship between the H$\alpha$ luminosity corrected for dust attenuation using the 24 $\mu$m emission
with the prescription of Calzetti et al. (2010) and the H$\alpha$ luminosity corrected using the Balmer decrement. Only galaxies detected 
at 22 $\mu$m and with an available estimate of the [NII]/H$\alpha$ ratio and of the Balmer decrement are included. The two dependent variables
are obviously strongly related. The determination of the Balmer decrement, however, is very uncertain in those objects with a weak Balmer 
emission, since the contamination of the underlying stellar absorption can be dominant. 
The lower panel of Fig. \ref{hacal} shows the relationship between the perpendicular distance from the $L(H\alpha)_{24\mu m}$ vs. $L(H\alpha)_{BD}$ relation
and the uncertainty on the Balmer decrement estimate $\sigma[C(H\beta)]$ given in column 12 of Table \ref{GANDALF}.
Figure \ref{hacal} shows that the points are symmetrically distributed around the mean relation for $\sigma[C(H\beta)]$ $\lesssim$ 0.1, while they systematically drop below
this relation for larger uncertainties on the estimate of the Balmer decrement.
 
\begin{figure*}
\centering
\includegraphics[width=17cm]{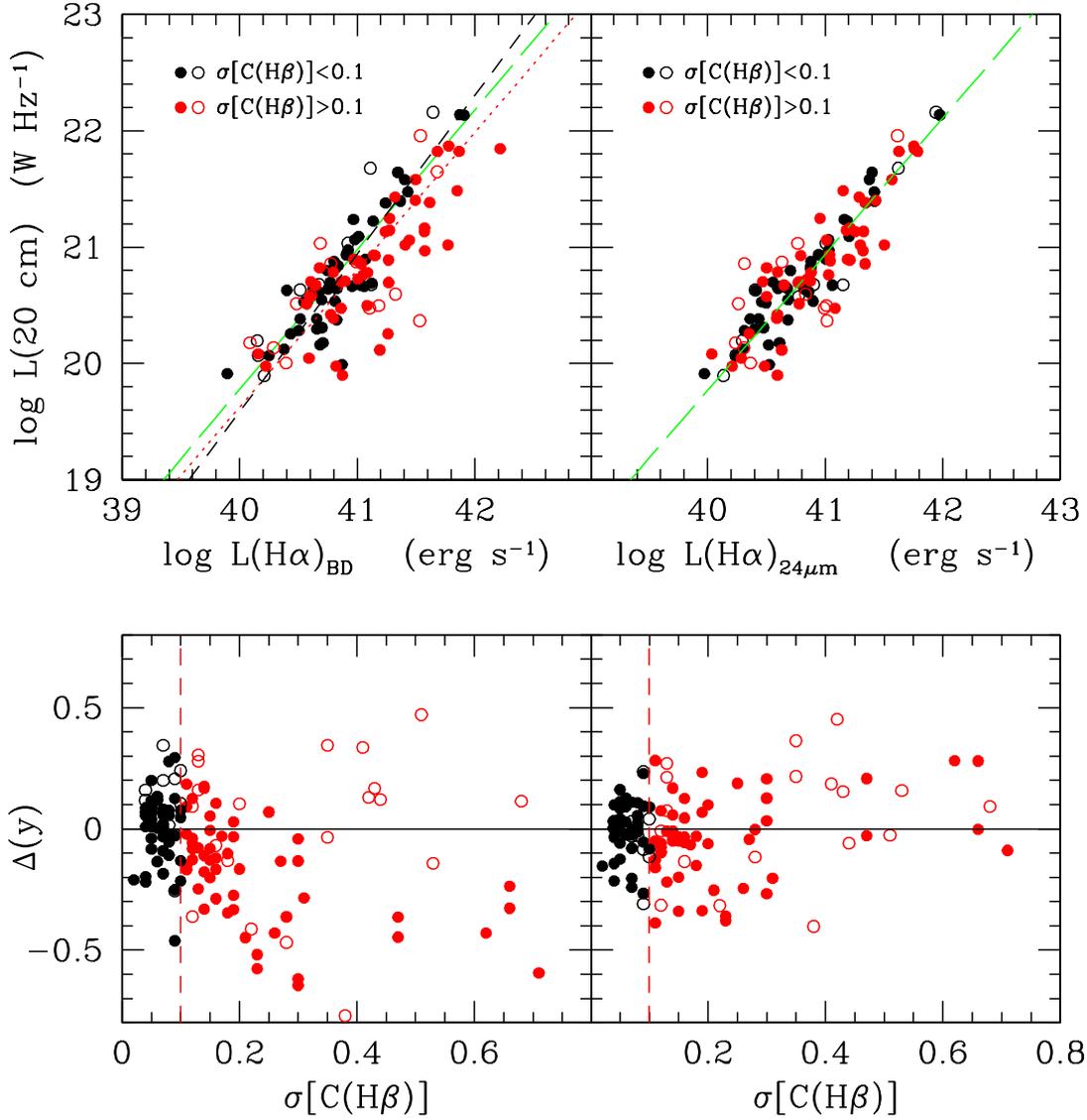}
\caption{Upper panels: relationship between the 20 cm radio luminosity and the H$\alpha$ luminosity corrected for dust attenuation using the Balmer decrement (left)
and the 24 $\mu$m luminosity following the prescription of Calzetti et al. (2010) (right). 
Black symbols are for galaxies with a $\sigma[C(H\beta)]$ $\leq$ 0.1, red symbols for $\sigma[C(H\beta)]$ $>$ 0.1. 
Filled dots are for galaxies with a normal gas content ($HI-def$ $\leq$ 0.4), empty symbols for gas-poor objects ($HI-def$ $>$ 0.4).
The black short-dashed line in the left panel shows the bisector fit determined using the best quality sample ($\sigma[C(H\beta)]$ $\leq$ 0.1), 
while the red dotted line the best fit determined using the whole sample. 
The green long-dashed line gives the bisector fit 
in the $L(20 cm)$ vs. $L(H\alpha)_{24\mu m}$ relation determined using all 
galaxies in the right panel. When plotted in the left panel, this fit is close to the one traced by the black dashed line. Lower panels: relationship
between the distance from the 
$L(20 cm)$ vs. $L(H\alpha)_{BD}$ (left) and the $L(20 cm)$ vs. $L(H\alpha)_{24\mu m}$ relations (right) and the uncertainty on the Balmer decrement estimate $\sigma[C(H\beta)]$.
}
\label{radiocal}
\end{figure*}

\begin{figure*}
\centering
\includegraphics[width=17cm]{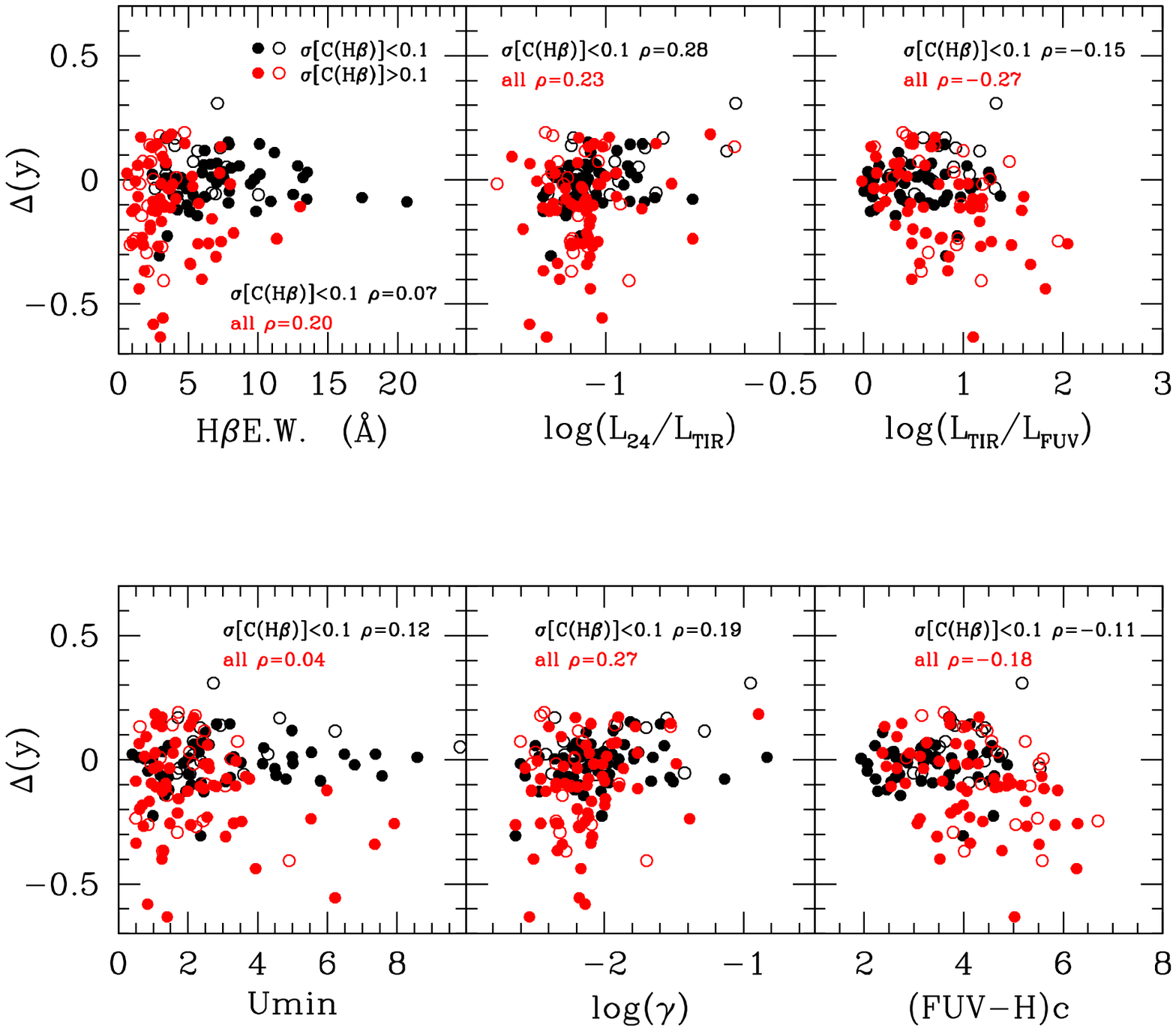}
\caption{Relationship between the distance from the $L(H\alpha)_{24\mu }$ vs. $L(H\alpha)_{BD}$ relation and 
different parameters characterising the physical properties of the target galaxies, determined as described in the text.
Filled dots are for galaxies with a normal gas content 
($HI-def$ $\leq$ 0.4), empty symbols for gas-poor objects ($HI-def$ $>$ 0.4). Black symbols are for galaxies with $\sigma[C(H\beta)]$ $\leq$ 0.1, 
red symbols for $\sigma[C(H\beta)]$ $>$ 0.1. $\rho$ gives the Spearman correlation coefficient for each panel for the whole sample (black and red; 152 objects) or for 
galaxies with high signal-to-noise in the spectroscopic data ($\sigma[C(H\beta)]$ $\leq$ 0.1; black; 69 objects).}
\label{dispersionegan}
\end{figure*}

To understand whether this trend is due to a systematic bias in the 24$\mu$m-based dust attenuation correction or 
in the Balmer decrement-based correction, we compare $L(H\alpha)_{BD}$ and $L(H\alpha)_{24\mu m}$ to the radio continuum luminosity at 20 cm,
which is an independent tracer of the star formation activity in galaxies (e.g. Kennicutt 1998; Bell 2003). 
At this frequency, the radio continuum
emission of galaxies is primarily due to the synchrotron emission of relativistic electrons spinning around weak magnetic fields. These electrons are
accelerated in supernova remnants, and are thus a direct tracer of the young stellar population (e.g. Boselli 2011).
Radio continuum data at 20 cm (1.49 GHz), collected from the literature as explained in Appendix B, are available for 169 (65\%) late-type galaxies\footnote{We consider in the
following analysis only galaxies with high-quality radio data (flag 1 in Table \ref{Tabradio}).}.
Figure \ref{radiocal} shows that the 20 cm luminosity of the HRS galaxies is tightly correlated with the H$\alpha$ luminosity. The relation with 
the H$\alpha$ luminosity corrected for dust attenuation using the 24 $\mu$m band (right panel, $\sigma$ = 0.15) is less dispersed than the one determined correcting H$\alpha$
using the Balmer decrement (left panel, $\sigma$=0.20; see Table \ref{calibrazione})\footnote{The less dispersed relation between the radio continuum emission and the H$\alpha$ luminosity corrected using the 24 $\mu$m emission 
than with the Balmer corrected H$\alpha$ luminosity might result from the  tight connection between the radio and far infrared emission of galaxies 
(far infrared-radio correlation, de Jong et al. 1985; Condon et al. 1991; Yun et al. 2001; Bell 2003).}. 
Figure \ref{radiocal} shows, however, that  as for the $L(H\alpha)_{24\mu m}$ vs. $L(H\alpha)_{BD}$ relation shown in Fig. \ref{hacal}, 
the points are symmetrically distributed around the $L(20 cm)$ vs. $L(H\alpha)_{BD}$ relation only whenever $\sigma[C(H\beta)]$ $\lesssim$ 0.1, 
while they drop below the mean relation for larger values of $\sigma[C(H\beta)]$. On the opposite,
the dispersion in the $L(20 cm)$ vs. $L(H\alpha)_{24\mu m}$ relation is symmetric. Figures \ref{hacal} and \ref{radiocal} consistently indicate that the Balmer decrement is 
systematically overestimated by $\simeq$ 0.2 dex ($A(H\alpha)$ $\simeq$ 0.5 mag) whenever $\sigma[C(H\beta)]$ $\gtrsim$ 0.1. 
\\

\begin{figure*}
\centering
\includegraphics[width=17cm]{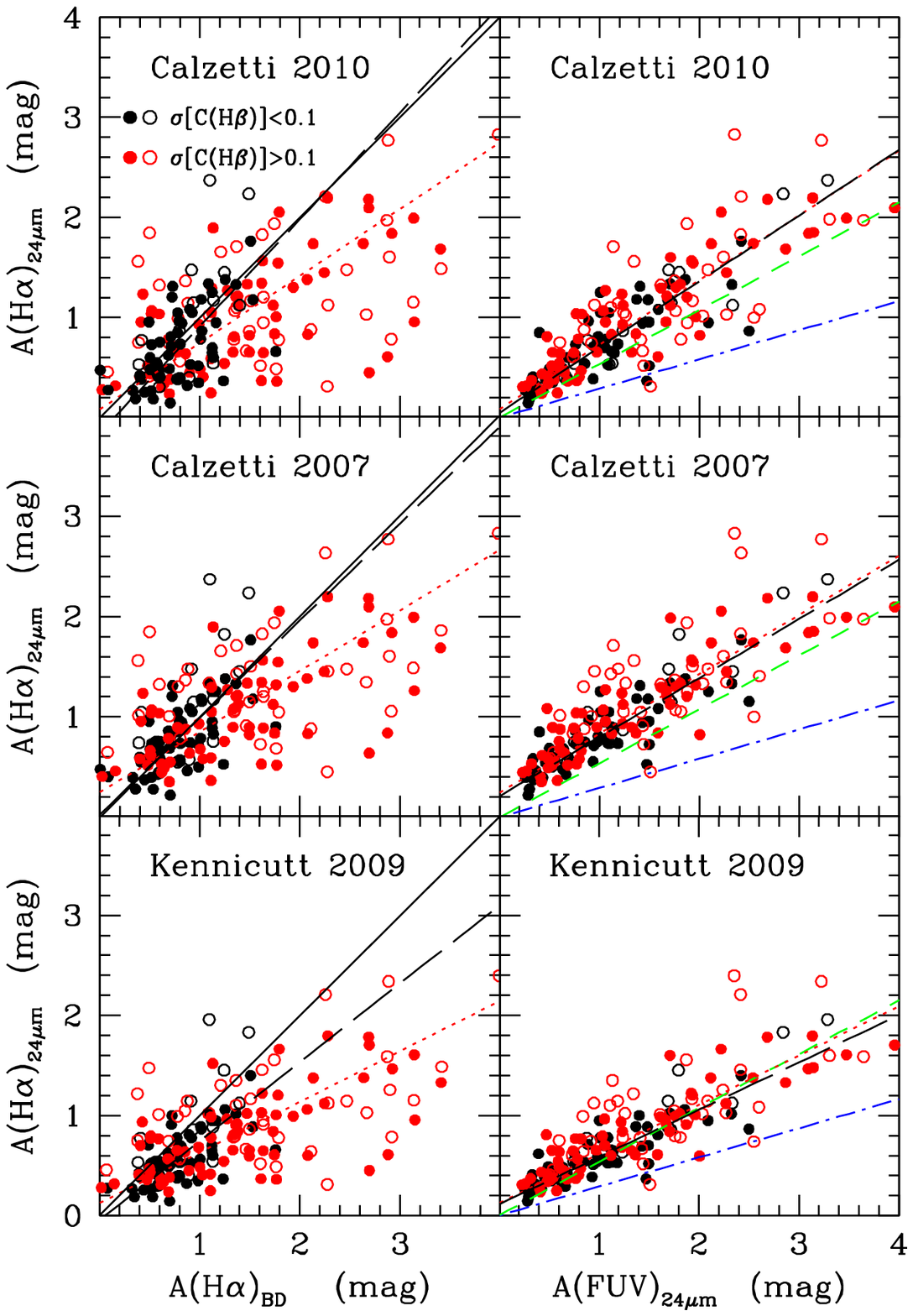}
\caption{Left column: comparison of $A(H\alpha)_{24\mu m}$, the attenuation of the H$\alpha$ measured using different recipes based on the H$\alpha$ over 24 $\mu$m flux ratio,
and that determined using the Balmer decrement. 
Black symbols are for galaxies with $\sigma[C(H\beta)]$ $\leq$ 0.1, red symbols for $\sigma[C(H\beta)]$ $>$ 0.1. Filled dots are for galaxies with a normal gas content ($HI-def$ $\leq$ 0.4), empty symbols for gas-poor objects ($HI-def$ $>$ 0.4).
The black solid lines show the 1:1 relation, the black long-dashed line the bisector fit determined using the best quality sample ($\sigma[C(H\beta)]$ $\leq$ 0.1)), while the red dotted line the best fit determined using the whole sample. 
Right: comparison of $A(H\alpha)_{24\mu m}$ and the attenuation 
$A(FUV)_{24\mu m}$ determined in the \textit{GALEX} FUV band using the prescription of Hao et al. (2011) based on the 24 $\mu$m emission band. The 
blue dotted-dashed line shows the relation expected for a screen model and the Galactic extinction law of Fitzpatrick \& Massa (2007), the green
short-dashed line the Calzetti attenuation law (Calzetti 2000), and the red dotted line the bisector fit to the data for all galaxies. }
\label{AHacor}
\end{figure*}

\subsubsection{Limits in the 24 $\mu$m dust attenuation correction}

The analysis presented in the previous section indicates that H$\alpha$ luminosities can be accurately corrected for dust attenuation using the Balmer decrement
only whenever $\sigma[C(H\beta)]$ $\lesssim$ 0.1.  
This, however, might introduce systematic biases, in particular in the comparison with the 24 $\mu$m dust attenuation corrected H$\alpha$ luminosities.
Indeed, as shown in Boselli et al. (2013), the Balmer decrement $C(H\beta)$ is tightly related with the H$\beta$E.W., 
thus omitting galaxies with low H$\beta$ emission (thus those with low signal-to-noise in H$\beta$ or equivalently with a large uncertainty on $C(H\beta)$, see Appendix A) 
might strongly bias the sample towards low attenuated objects. The H$\beta$ emission is 
also tightly connected to the specific star formation rate. The exclusion of galaxies with small values of 
H$\beta$ might thus bias the sample towards star-forming, low-mass systems. It is well known that in these systems the dust heating sources
are mainly young massive stars, while in more quiescent and massive objects the contribution to the dust heating of the older stellar component 
might be very important (e.g. Boselli et al. 2006, Cortese et al. 2008, Salim et al. 2009, Bendo et al. 2010, 2012; Boquien et al. 2011; Boselli et al. 2012). 
Thus, the calibration of dust attenuation based on the 24 $\mu$m emission, taken here as proxy for the total far-infrared luminosity, might
not be representative for the most quiescent objects of the sample. To test whether this strong assumption might introduce a systematic bias in the results, we plot in 
Fig. \ref{dispersionegan} the relationship between the perpendicular distance from the $L(H\alpha)_{24\mu m}$ vs. $L(H\alpha)_{BD}$ relation 
observed in Fig. \ref{hacal} and different variables characterising the physical properties of the interstellar radiation field of the
HRS galaxies. 
%These are the H$\beta$E.W., the ratio of the 24 $\mu$m-to-total infrared luminosity ($L_{24}/L_{TIR}$), the ratio of the 
%total infrared luminosity to the far UV luminosity ($L_{TIR}/L_{FUV}$), the $Umin$ and $\gamma$ parameters, and the FUV to H-band corrected 
%colour index ($FUV-H)_c$. 
The $L_{TIR}$, $Umin$, and $\gamma$ parameters give respectively the total infrared luminosity, the
intensity of the general interstellar radiation field responsible for the heating of the diffuse dust component
and the fraction of dust mass in PDRs heated by the energetic radiation produced by OB associations (Draine et al. 2007). 
They have been determined by Ciesla et al. (2014), by fitting the infrared SED of the HRS galaxies using the models of Draine \& Li (2007).
$L_{TIR}/L_{FUV}$ is a direct tracer of the dust attenuation in galaxies, whereas the 
$\gamma$ parameter, the H$\beta$ equivalent width, and the FUV-to-H-band colour index are tracers of the hardness of the radiation field heating the dust. 
The $L_{24}/L_{TIR}$ quantifies the contribution of the hot dust component to the total far infrared dust emission of galaxies, and is thus 
tightly connected to the shape of the SED and to the activity of star formation. \\

Figure \ref{dispersionegan} shows that, when all galaxies are considered regardless of the uncertainty on the Balmer decrement
(black and red symbols together), $\Delta(y)$ is barely anticorrelated with $L_{TIR}/L_{FUV}$ and $(FUV-H)_c$ \footnote{the probability that the two variables are correlated
is 99$<$ $P$ $<$ 99.9 \%} and shows a bimodal distribution when plotted vs. H$\beta$E.W., $L_{24}/L_{TIR}$, and $\gamma$.
In these plots, galaxies with $\Delta(y)$ $\lesssim$ -0.3/-0.4 all have low values of H$\beta$E.W. ($\lesssim$ 5 \AA), $L_{24}/L_{TIR}$ ($\lesssim$ 0.1),
and $\gamma$ ($\lesssim$ 10$^{-2}$). They also have red $(FUV-H)_c$ colours ($\gtrsim$ 4 mag) and high $L_{TIR}/L_{FUV}$ ratios ($\gtrsim$ 5).
All these properties consistently indicate that galaxies located below the standard $L(H\alpha)_{24\mu m}$ vs. $L(H\alpha)_{BD}$ relation 
are relatively quiescent objects where the dust heating is dominated by the evolved stellar population and where dust attenuation is probably important. 
If confirmed, these trends would indicate a systematic residual in the H$\alpha$ luminosity correction based on the 24 $\mu$m emission, 
making eq. B.2-B.3 non universal since they are valid only for active star forming galaxies. 
We notice, however, that if the same analysis is restricted to those objects with a low uncertainty on the Balmer decrement ($\sigma[C(H\beta)]$ $\leq$ 0.1,
black symbols in Fig. \ref{dispersionegan}), only the trend with $L_{24}/L_{TIR}$ is still statistically significant 
(the probability that the two variables are correlated is $P$ $>$ 99\%). 
As mentioned above, however, limiting the analysis to galaxies with low uncertainty on the Balmer decrement, thus with high signal-to-noise in the H$\alpha$ and H$\beta$
lines, might severely bias the sample towards active galaxies. It is thus hard to conclude 
whether there is a statistically significant indication that the proposed calibration varies with the properties of galaxies. However, it is clear
that, as first stressed by Kennicutt et al. (2009), 
the dust attenuation correction of the H$\alpha$ emission based on monochromatic far infrared tracers should be used with extreme caution in 
quiescent massive spiral galaxies. \\

\subsubsection{Comparison between different $A(H\alpha)$ and $A(FUV)$ estimators}

Different recipes have been proposed in the literature to correct H$\alpha$ luminosities using the 24 $\mu$m emission.  
Figure \ref{AHacor} shows the relationship between different H$\alpha$ attenuations determined using these recipes (Calzetti et al. 2010, 2007, and 
Kennicutt et al. 2009 from top to bottom), and the H$\alpha$ attenuations determined using the Balmer decrement 
(left column). All recipes give $A(H\alpha)_{24\mu m}$ $\lesssim$ $A(H\alpha)_{BD}$ regardless of the quality of the spectroscopic data.
Among these corrections, however, the one proposed by Calzetti et al. (2010), 
which uses two different coefficients, one for a starburst regime and one for normal star forming regions, gives values closest to the 1:1 relation. 
Whenever the Balmer decrement cannot be determined with high accuracy ($\sigma[C(H\beta)]$ $\leq$ 0.1), we adopt this correction in the following analysis.
%%QUI C'E' DA DIRE CHE VORREMMO CORREZZIONI PIU FORTI CONSISTENTE CON COSA TROVA KENNICUTT QUANDO USA BALMER DECREMENT E NON PASCHEN LINES....
Figure \ref{AHacor} also shows the relationship between $A(H\alpha)_{24\mu m}$ and $A(FUV)_{24\mu m}$ determined
using the prescription of Hao et al. (2011). $A(H\alpha)_{24\mu m}$ is generally $\lesssim$ $A(FUV)_{24\mu m}$ (see however Buat et al. 2002), which is another good reason to prefer the H$\alpha$ to the FUV luminosity as a star formation tracer, given that it is less affected by attenuation. 
The relation between $A(H\alpha)_{24\mu m}$ and $A(FUV)_{24\mu m}$ is steeper than the one
expected for a simple screen model combined with a Milky Way attenuation curve (blue dotted-dashed line). This relation is also slightly steeper than the widely used Calzetti's law
(Calzetti 2001, $A(FUV)$ = 1.86 $\times$ $A(H\alpha)$, green dashed line) when the H$\alpha$ attenuation is measured using the prescriptions of Calzetti et al. (2007, 2010),
whereas they are in agreement when the H$\alpha$ attenuation is determined using the prescription of Kennicutt et al. (2009). 

%Given that for this subsample of galaxies the Balmer decrement
%determined using the double gaussian fit adopted by Boselli et al. (2013) and the GANDALF code give consistent results, we can use eithr set of spectroscopic data. 
%We chose, however, the set of data published in Boselli et al. (2013) since it includes more galaxies than the one where the GANDALF has been run (we do not have access 
%to the spectra obtained by other teams), and thus limit the determination of $L(H\alpha)_{BD}$ to those objects with H$\beta$E.W. $>$ 5 \AA. We notice that 
%using this dataset and the Calzetti et al. (2010) calibration we get consistent $A(H\alpha)_{24\mu m}$ and $A(H\alpha)_{BD}$ attenuation (upper left panel of Fig.
%\ref{AHacor}).

\begin{figure*}
\centering
\includegraphics[width=17cm]{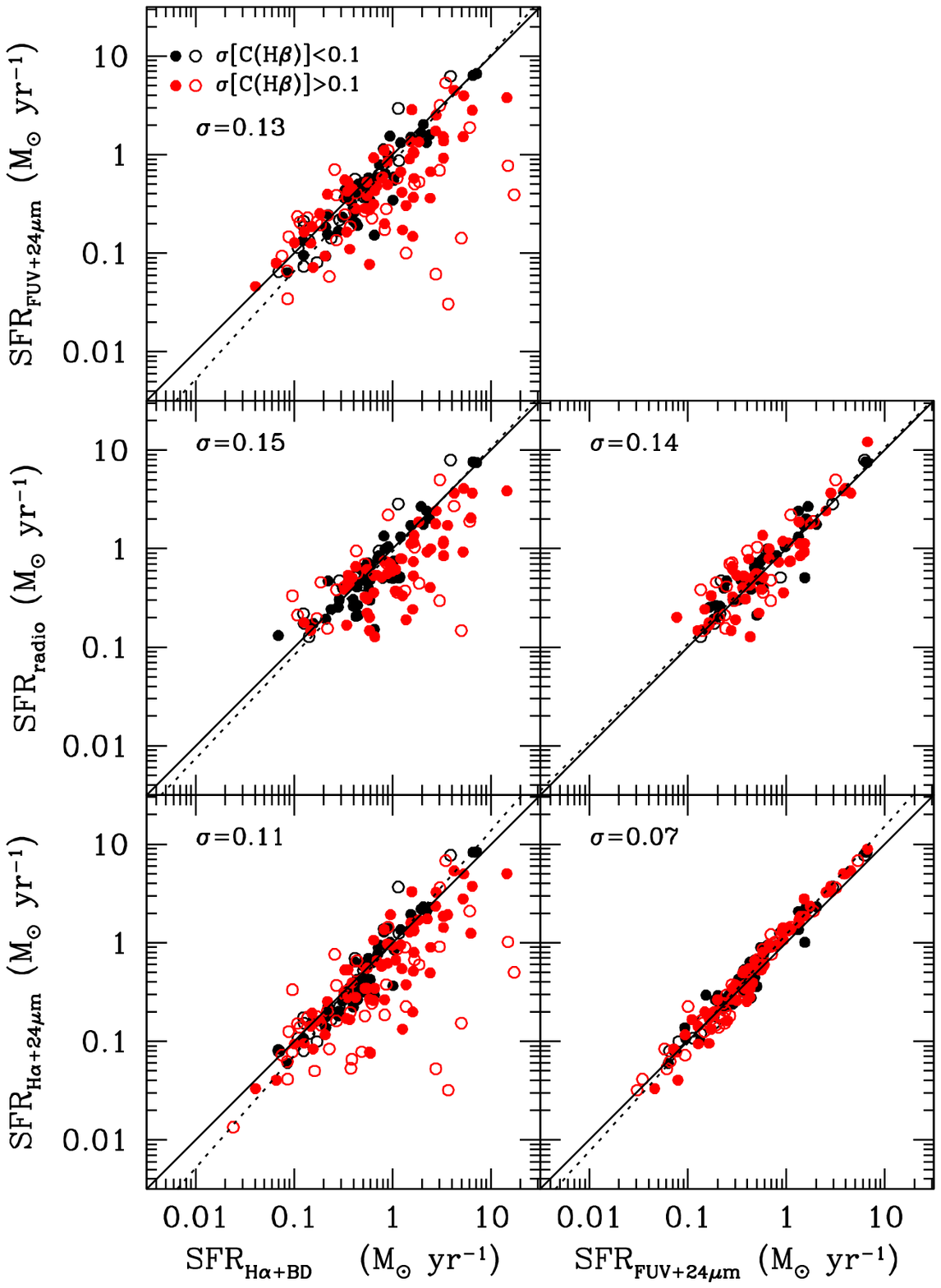}
\caption{Comparison between the star formation rate determined using different tracers. $SFR_{H\alpha +BD}$ stands for star formation rates 
determined using H$\alpha$ luminosities corrected for Balmer decrement (using the {\sc GANDALF} dataset), $SFR_{H\alpha +24\mu m}$ corrected using the prescription of Calzetti et al. (2010)
based on the 24 $\mu$m emission. $SFR_{FUV +24\mu m}$ have been determined using \textit{GALEX} FUV data corrected for dust attenuation using the 24 $\mu$m emission following Hao et al. (2011), 
and $SFR_{radio}$ using the 20 cm radio emission following Bell (2003) (see Appendix C).
Black symbols are for galaxies with a spectroscopic $\sigma[C(H\beta)]$ $\leq$ 0.1, red symbols for galaxies with $\sigma[C(H\beta)]$ $>$ 0.1. 
Filled dots are for galaxies with a normal gas content ($HI-def$ $\leq$ 0.4), empty symbols for gas-poor objects ($HI-def$ $>$ 0.4).
The black solid line shows the 1:1 relation, the black dotted line the bisector fit (and $\sigma$ its dispersion) determined using the best quality sample ($\sigma[C(H\beta)]$ $\leq$ 0.1).
}
\label{sfrha}
\end{figure*}

\subsection{SFR}

\subsubsection{Comparison between different tracers}

Once corrected for dust attenuation, H$\alpha$ luminosities can be transformed into star formation rates ($SFR$, in M$_{\odot}$ yr$^{-1}$)
using a factor which depends on the assumed IMF and stellar model\footnote{As mentioned above, we do not apply any correction for any possible escape
fraction of the ionising radiation, nor for the absorption by dust of ionising photons before the ionisation of the gas (see Boselli et al. 2009).}:

\begin{equation}
{SFR = k(H\alpha) \times L(H\alpha)_{cor}}
\end{equation}

\noindent
We recall that this relation is valid only under the assumption that the mean star formation activity of the emitting galaxies 
is constant over a timescale of a few Myr, roughly comparable to the typical time spent by the stellar population responsible for the ionisation of the gas on the main sequence
(Boselli et al. 2009; Boissier 2013; Boquien et al. 2014). The ionising stars are O and early-B stars, whose typical age is $\lesssim$ 10$^7$ years. 
The stationarity condition is generally satisfied in massive, normal, star forming galaxies undergoing a secular evolution. In these objects
the total number of OB associations is significantly larger than  the number of HII regions under formation and of
OB stars reaching the final stage of their evolution, thus their total H$\alpha$ luminosity
is fairly constant with time. This might not be the case in strongly perturbed systems or in dwarf galaxies, 
where the total star formation activity can be dominated by individual giant HII region (Boselli et al. 2009; Weisz et al. 2012)
and the IMF is only stochastically sampled (Lee et al. 2009; Fumagalli et al. 2011; da Silva et al. 2014). The HRS sample is dominated by 
relatively massive galaxies undergoing a secular evolution. For these objects, eq. (10) can thus be applied. 
The sample, however, also includes galaxies in the Virgo cluster region, where the perturbation induced by the cluster environment might 
have affected their star formation rate (e.g. Boselli \& Gavazzi 2006; 2014). Models and simulations
have shown that in these objects the suppression of star formation occurs on timescales
of the order of a few hundreds Myr (Boselli et al. 2006, 2008a,b, 2014d). These timescales are relatively long compared to the 
typical age of O-B stars. The recent work of Boquien et al. (2014) has clearly shown that the Lyman continuum emission tightly follows the 
rapid variations of the star formation activity of simulated galaxies down to timescales of a few Myrs. We can thus safely consider that 
the linear relation between the H$\alpha$ luminosity and the star formation rate given in eq. 10 is satisfied in the HRS sample.

%\begin{figure*}
%\centering
%\includegraphics[width=17cm]{sfrfuv.ps}
%\caption{Comparison between the star formation rate determined using 20 cm radio continuum data (upper left), the H$\alpha$ data corrected using the 24 $\mu$m emission
%(Calzetti's 2010 calibration; lower left), the \textit{GALEX} NUV (upper right), the H$\alpha$ data corrected using the Balmer decrement (lower right) 
%and that determined using the FUV luminosity corrceted using the 24 $\mu$m emission.
%Black symbols are for galaxies with a spectroscopic SN$>$50, red symbols for galaxies with SN$<$50. 
%Filled dots are for galaxies with a normal gas content ($HI-def$ $\leq$ 0.4), empty symbols for gas-poor objects ($HI-def$ $>$ 0.4).
%The black solid line shows the 1:1 relation, the black dotted line the bisector fit determined using the best quality sample (SN$>$50)
%and $\sigma$ its dispersion.
%}
%\label{sfrfuv}
%\end{figure*}

Figure \ref{sfrha} shows the relationship between the star formation rate determined using different tracers: the H$\alpha$ luminosity,
corrected for dust attenuation using both the Balmer decrement and the 24 $\mu$m emission, the FUV \textit{GALEX} luminosities corrected using the 24 $\mu$m emission, and the 
20 cm radio continuum luminosity. For consistency all $SFR$ have been measured using the Kennicutt (1998) prescriptions
based on a similar IMF (Salpeter in the stellar mass range 0.1 $<$ $m_{star}$ $<$ 100 M$_{\odot}$). 
For the radio continuum we use the Bell (2003) calibration, which is consistent with those
used in the other bands (see Appendix C). The different values of $SFR$ are listed in Table \ref{SFR}.

The different SFR calibrations give similar results once determined for star forming galaxies with low uncertainties in $C(H\beta)$.
This result is consistent with what was found in the previous section.
When compared to $SFR_{FUV+24 \mu m}$, the dispersion is different when different 
tracers are used: it is very small when compared to $SFR_{H\alpha+24 \mu m}$ and gradually increases with $SFR_{H\alpha+BD}$ (when
limited to $\sigma[C(H\beta)]$ $\leq$ 0.1) and $SFR_{radio}$ (see Table \ref{sfrsfr}). This increase of the dispersion in the relations can be naturally explained by considering
that some variables are not fully independent. The dispersion in the relation with the radio continuum 
tracer might also be affected by other physical factors. In fact, the radio continuum emission can be affected by the presence 
of an AGN. There is also some indication that the radio continuum emission of cluster galaxies is, on average,
stronger than that of similar objects in the field (Gavazzi et al. 1991; Gavazzi \& Boselli 1999a,b). The increase of the
radio continuum activity of cluster galaxies has been interpreted as due to the compression of the magnetic field 
during their interaction with the dense intergalactic medium (e.g. Boselli \& Gavazzi 2006).
%Figure \ref{sfrfuv} shows that the four galaxies with the highest 
%radio-determined star formation rates are all HI-deficient objects, thus galaxies which recently interacted with the hot and dense 
%intracluster medium (Boselli \& Gavazzi 2006; Boselli et al. 2014d).

\begin{figure}
\centering
\includegraphics[width=9cm]{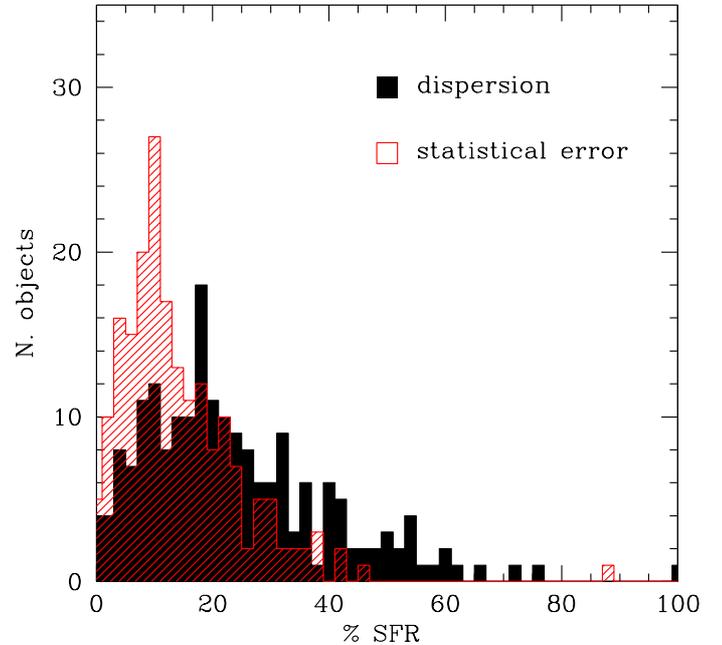}
\caption{Distribution of the dispersion $\sigma$
in the different $SFR$ tracers (black histogram) and the statistical error ($\sigma/\sqrt(N)$ when more than one estimator is available (red histogram).}
\label{error}
\end{figure}

Out of the 260 late-type galaxies in the HRS sample, 196 have more than one empirical determination of the $SFR$. 
Figure \ref{error} shows that the typical dispersion $\sigma$ in the different tracers is of the order of 24\%, while the statistical error $\sigma/\sqrt(N)$
in the final $SFR$ is of the order of 14\%\footnote{This number underestimates the statistical error since it is based on 
non fully independent values of $SFR$.}. Beside the uncertainty in the data, a part of this scatter can be due to the fact that the $SFR$ tracers based 
on the FUV and radio luminosities can be seriously affected by the fact that the stationarity conditions necessary to transform luminosities into star formation rates
are not always satisfied, as clearly indicated by Boquien et al. (2014).

\subsubsection{Comparison with SED fitting estimates}

The star formation activity of galaxies can also be estimated by fitting their observed spectral energy distribution with stellar population synthesis 
models. An energetic balance between the absorbed stellar radiation and the energy emitted in the far infrared domain 
quantifies dust attenuation, allowing the determination of several physical parameters of the studied galaxies, such as the stellar mass and the star formation rate (GRASIL, Silva et al. 1998;
MAGPHYS, da Cunha et al. 2008; CIGALE, Noll et al. 2009). 
SED fitting has several advantages with respect to the star formation rate determination based on monochromatic fluxes used in the previous 
section. First of all, it uses simultaneously several spectrophotometric bands, thus significantly reducing the observational uncertainty on the 
data. Thanks to a self consistent determination of the dust attenuation,
SED fitting also provides several physical parameters ($SFR$, $M_{star}$, $Z$, ...) suitable for any kind of statistical analysis.
The SED fitting technique, however, also has several weaknesses. First of all, the output of the SED fitting depends on the star formation history
of the galaxy, which is parametrised using simple analytic prescriptions. These are typically not 
optimised to reproduce the abrupt truncation of the star formation activity occurred in cluster galaxies (e.g. Boselli \& Gavazzi 2006; 2014). 
SED fitting is generally done assuming a constant (and fixed) metallicity. 
The star formation rate derived by SED fitting also depends on the adopted population synthesis models and IMF, as in the case of the monochromatic-based estimates. 

It is thus worth comparing the star formation rate determined using a SED fitting code to the one
derived directly from monochromatic fluxes as described in the previous section. To do this, we run the CIGALE code 
(Noll et al. 2009; Burgarella et al. in prep. Boquien et al. in prep.) on all the HRS galaxies.
The far infrared part of the spectrum is fitted using the Draine \& Li (2007) dust models, as extensively described in Ciesla et al. (2014).
The UV-visible-near infrared part of the spectrum is fitted using Bruzual \& Charlot (2003) stellar population models, 
and assuming a Salpeter IMF and solar metallicity, consistent with our approach for the monochromatic determinations. The HRS sample is ideally suited for 
SED fitting since multifrequency data (including 15 photometric bands) are available for the vast majority of the galaxies. For the present work,
we limit the comparison to those galaxies with available \textit{GALEX} FUV data. Although nebular emission lines can be added to the stellar continuum emission, 
we do not use them in the present fit, because we want to be representative of the typical data generally used in the SED fitting of galaxies extracted from cosmological surveys.

\begin{figure*}
\centering
\includegraphics[width=17cm]{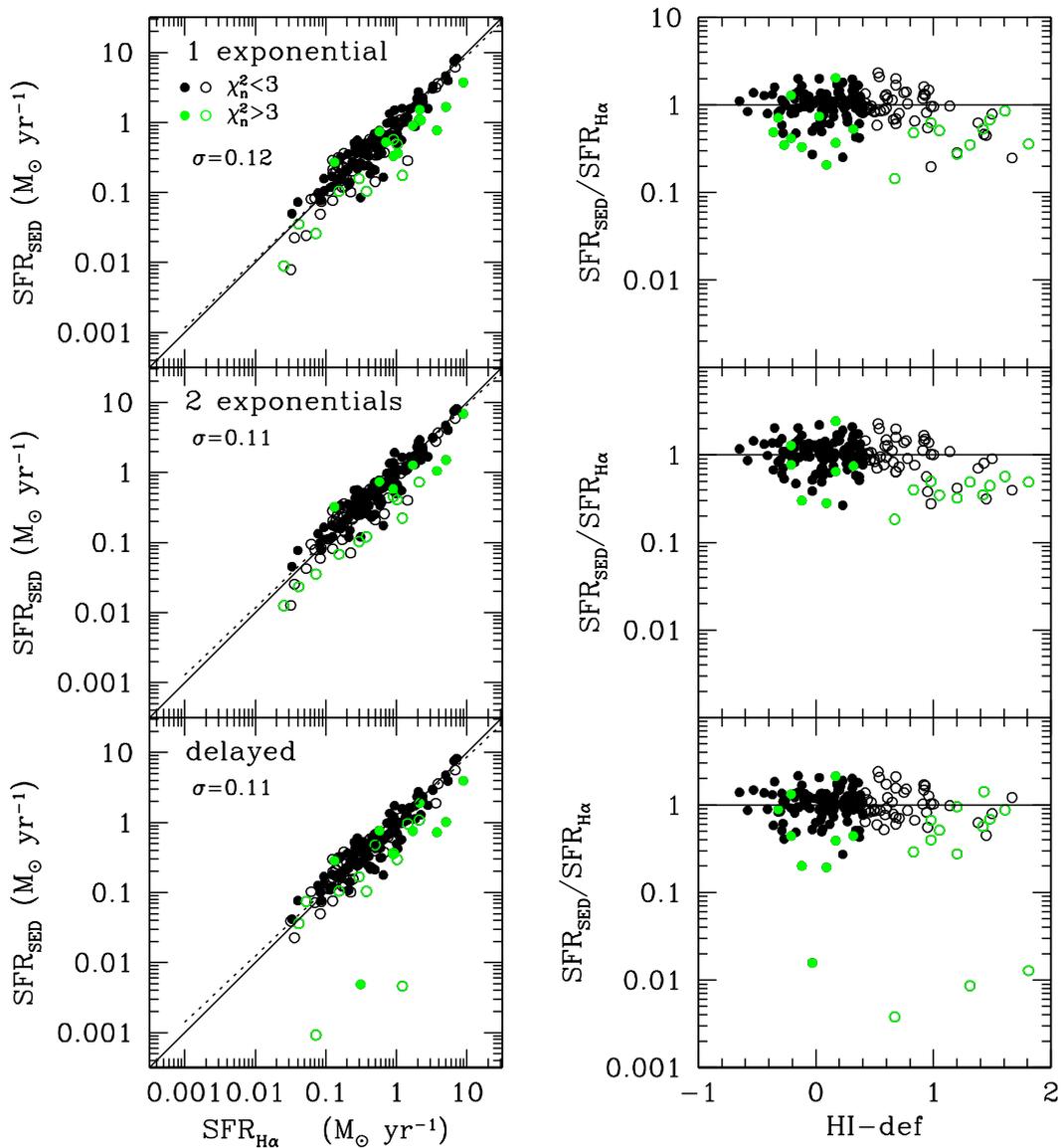}
\caption{Left panels: The comparison between the star formation rate determined using a SED fitting technique and that determined using the H$\alpha$ emission. 
For the SED fitting determination, all galaxies are assumed to be 13 Gyr old. Three different 
parametric star formation histories are assumed: exponential (upper panel), double exponential (middle), and delayed (lower). Green symbols
indicate those objects where the normalised $\chi^2_n$ $>$ 3, black symbols those with $\chi^2_n$ $<$ 3. Filled symbols are for HI-normal galaxies,
empty symbols for HI-deficient objects ($HI-def$ $\leq$ 0.4). The black solid line shows the 1:1 relation, the black dotted line the bisector fit (and $\sigma$ its dispersion)
determined using galaxies with $\chi^2_n$ $<$ 3. Right panels: relationship between the $SFR_{SED}/SFR_{H\alpha}$ ratio and the HI-deficiency parameter.}
\label{sfrcigale}
\end{figure*}

To remove any possible dependence on short timescale variations of the star formation activity of cluster galaxies, we use the 
H$\alpha$ luminosity as a monochromatic tracer (Boquien et al. 2014). Whenever possible ($\sigma[C(H\beta)]$ $\leq$ 0.1) we correct it for dust attenuation using the Balmer decrement,
otherwise using the 24 $\mu$m emission. A 24 $\mu$m based correction might indeed introduce systematic age effects since the dust emitting at this wavelength 
might be heated by stars older than those responsible for the ionisation of the gas (e.g. Bendo et al. 2012).
The variable $SFR_{H\alpha}$ is available for 196/260 of the late-type galaxies of the sample. 

Figure \ref{sfrcigale} shows the relationship between the star formation rate determined using the SED fitting procedure and that derived from the H$\alpha$ luminosity. 
The SED fitting estimates have been determined using three commonly-used parametrisations of the star formation history of
galaxies: a single exponentially declining law

\begin{equation}
{SFR(t) \propto exp(-t/\tau_1) ~~~[0.5 \leq \tau_1 \leq 100 ~\rm{Gyr}]}
\end{equation}

\noindent
a double exponentially declining law

\begin{equation}
  SFR(t) = \begin{cases}
   \exp{-t/\tau_1}  & \text{if } t < t_1 - t_2 \\
   \exp{-t/\tau_1} + k \times \exp{-t/\tau_2}     &  \text{if } t \geq t_1 - t_2\\
     [0.5 \leq \tau_1 \leq 20 ~\rm{Gyr}; ~~\tau_2 \rightarrow \infty]
  \end{cases} 
\end{equation}

\noindent
and a delayed star formation history

\begin{equation}
{SFR(t) \propto t \times exp(-t/\tau_1) ~~~[0.5 \leq \tau_1 \leq 20 ~\rm{Gyr}]}
\end{equation}

\noindent
where $\tau$ are the folding times. These star formation rates are given in Table \ref{SFR}. For the three 
fits we make the reasonable assumption that galaxies are coeval (13 Gyrs old). Overall the three star formation rate estimates determined with the SED fitting
code give results consistent with those determined using H$\alpha$, as already found in other samples of local or high-$z$ galaxies
(e.g. Wuyts et al. 2009; Pforr et al. 2012; Buat et al. 2014)\footnote{In these works composite star formation histories (exponentially declining plus burst) are often adopted.}. 
This is expected for several reasons: first of all, the single exponentially declining and delayed
star formation histories are smooth with respect of time and do not have major changes in the last hundreds of Myrs. They thus correspond to the 
stationarity condition ($SFH$ = constant) assumed for the monochromatic determination. Second, the HRS galaxies are mainly massive galaxies which underwent a 
secular evolution. The sample, indeed, does not include strong starbursts or recent mergers, where the star formation activity might have changed abruptly with time.
We notice, however, that the SED fitting derived star formation activities of the most HI-deficient galaxies
are generally smaller than those determined using H$\alpha$ luminosities. These are also objects where the quality of the fit 
is lower than the average (normalised $\chi^2_n$ $>$ 3).

As extensively discussed in Boselli \& Gavazzi (2006; 2014), Boselli et al. (2006, 2014c, 2014d), Hughes \& Cortese (2009), Cortese \& Hughes (2009)
and Gavazzi et al. (2013b), the HI-deficiency parameter is tightly connected 
to the perturbation that affected cluster galaxies. The gas removal resulting from this interaction
quenched the activity of star formation on relatively short timescales, in particular in low-mass systems ($\sim$ 100 Myr). 
On these timescales, the H$\alpha$ luminosity can still be used as an accurate tracer of the star formation rate of the perturbed galaxies
(Boquien et al. 2014). On the contrary, the parametric star formation histories adopted for the fit are not optimised to 
reproduce this rapid truncation of the star formation activity in the gas stripped cluster galaxies. It is thus conceivable that 
in these objects the SED fitting gives less accurate $SFR$ than in unperturbed systems ($HI-def$ $\leq$ 0.4). Intuitively, however, we would 
expect an opposite effect since the adopted smooth star formation history would overestimate, rather than underestimate the present day star formation rate.
Tests and simulations done so far on observed or mock catalogues of galaxies consistently indicate that the star formation rates determined using SED fitting 
codes are accurate regardless the use of different parametrisations or the bursty nature of their evolution (Wuyts et al. 2009; 
Pforr et al. 2012; Buat et al. 2014; Ciesla et al. 2015). The departure of the HI-deficient galaxies from the 1-to-1 relation might thus be due to other reasons.
Indeed, these are quiescent objects characterised by red colours and weak Balmer emission lines, where the contribution of the old stellar population to the dust heating 
is important. As discussed in sect. 5.1, the recipes used to determine the H$\alpha$ attenuation based on the 24 $\mu$m emission might introduce systematic effects in the data.
To further investigate this intriguing topic, we are planning to use parametric star formation histories ad hoc defined to reproduce the truncated star formation activity 
of cluster galaxies. The results of this analysis will be presented in a forthcoming paper (Ciesla et al. in prep.).

\section{SFR properties of the HRS}

The complete nature of the HRS, which is volume-limited and K-band-selected, makes it the ideal sample for determining the typical statistical properties of 
galaxies in the local universe. The completeness in the H$\alpha$ band has been reached in the late-type
systems (Sa-Im-BCD, 98\%), hence we can derive their statistical properties. The following analysis is thus limited to these star forming objects. Keeping in mind all the systematic biases in the different 
tracers analysed in the previous sections, we decided to use in the following analysis the star formation rate determined, whenever possible, by averaging the different 
monochromatic estimates ($SFR_{H\alpha +BD}$, $SFR_{H\alpha +24\mu m}$, $SFR_{FUV +24\mu m}$, $SFR_{radio}$ ), otherwise using the only one available 
among them. We define this variable $SFR_{MED}$. In this way we increase the number of galaxies with an available estimate of the $SFR$ from 196/260 ($SFR_{H\alpha}$)
to 236/260 ($SFR_{MED}$), making the sample complete to 91\%. We checked that the results presented in the following section are robust versus the use of different star 
formation rate tracers.

\subsection{The SFR distribution}

\begin{figure}
\centering
\includegraphics[width=9cm]{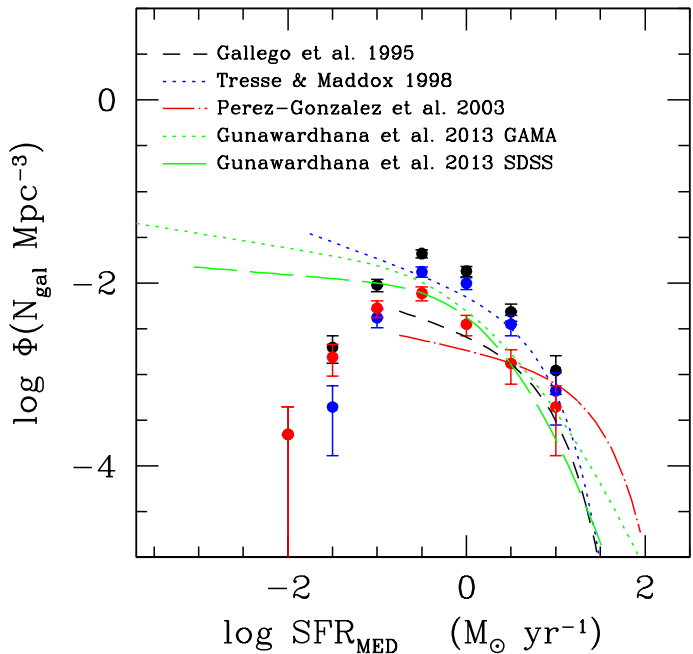}
\caption{Distribution of the SFR in number of galaxies Mpc$^{-3}$ for the whole HRS late-type galaxies (black dots) and for the subsample of HI-normal 
($HI-def$ $\leq$ 0.4; blue dots) and HI-deficient ($HI-def$ $>$ 0.4; red dots) galaxies. The distribution is compared to several SFR luminosity function derived
from H$\alpha$ measurements for the general field. SFR have been determined 
using the calibration used in this work for a Salpeter IMF. 
%The H$\alpha$ luminosity functions of Coma, A1367, and Virgo are from Iglesias-Paramo et al. (2002).
}
\label{LFSFR}
\end{figure}

The completeness of the sample allows us to estimate the star formation rate distribution of the HRS. 
This can be done by counting the number of galaxies in 0.5 dex bins of log$SFR$ (Figure \ref{LFSFR}). 
The normalisation factor is determined assuming that the volume covered by the HRS survey is of 4539 Mpc$^{3}$, 
consistently with Boselli et al. (2014b) and Andreani et al. (2014). This volume has been calculated 
considering that, according to the selection criteria described in Boselli et al. (2010)\footnote{The selection conditions are: high Galactic latitude ($b$ $>$ +55$^{o}$)
and low Galactic extinction ($A_B$ $<$ 0.2 mag) to avoid Galactic cirrus contamination.}, 
we selected galaxies in the volume between 15 and 25 Mpc over an area of 3649 sq.deg.
We recall that this is a star formation rate distribution and not a luminosity function since galaxies are first selected in the K-band
and then counted in bins of $SFR$. We determined the star formation rate 
distribution for the whole sample and separately for HI-normal ($HI-def$ $\leq$ 0.4) and HI-deficient ($HI-def$ $>$ 0.4) galaxies. The former 
can be considered as typical field sources, while the latter are cluster perturbed objects (e.g. Boselli \& Gavazzi 2006). The star formation rate distribution
is compared to several field (Gallego et al. 1995; Tresse \& Maddox 1998; Perez-Gonzalez et al. 2003; Gunawardhana et al. 2013) star formation 
rate luminosity functions determined for local galaxies. H$\alpha$ luminosity functions have been transformed into star formation rates adopting the same 
calibration for a Salpeter IMF and assuming $H_0$ = 70 km s$^{-1}$ Mpc$^{-1}$.

Figure \ref{LFSFR} shows that the shape of the HRS star formation rate distribution for the whole or for the unperturbed sample 
is very similar to most of the luminosity functions measured for local galaxies in the literature (maybe with the exception of Perez-Gonzalez et al. 2003)
for log $SFR$ $\gtrsim$ -0.5 M$_{\odot}$ yr$^{-1}$. For lower values of $SFR$, the HRS star formation rate distribution drastically drops with respect 
to the star formation rate luminosity functions published in the literature. This effect is similar to the one observed in the molecular gas mass distribution (Boselli et al. 2014b)
of the HRS, and can be explained by the incompleteness of the sample at low H$\alpha$ luminosities. Being less then unity the slope of the $SFR$ vs. $M_{star}$ relation
(Fig. \ref{main} and Table \ref{fit}), a stellar mass selection (which roughly corresponds to a K-band-selection) excludes low mass star forming objects.
Using the $SFR$ distribution of mock samples, it has been shown that the star formation rate distribution of a mass-selected sample does not 
follow the typical Schechter function, but it is better represented by a double Gaussian function with a form close to the one 
depicted in Fig. \ref{LFSFR} (Salim \& Lee 2012).
A decrease at the faint end of the H$\alpha$ luminosity function for Coma, A1367, and Virgo cluster galaxies has been found by Iglesias-Paramo et al. (2002).

%The shape of the star formation rate luminosity distribution of the HRS galaxies is much closer to the H$\alpha$ derived star formation rate 
%luminosity function of Coma, A1367, and Virgo cluster galaxies (Iglesias-Paramo et al. 2002; right panel of Fig. 
%\ref{LFSFR})\footnote{The observed shift in the Y-axis is due to the fact that the cluster luminosity functions are measured in an over dense region.}.
%For these cluster galaxies, the decrease at low luminosities has been explained by the morphology segregation effect (Iglesias-Paramo et al. 2002).

Figure \ref{LFSFR} also shows that the distribution of the HI-normal, unperturbed objects is above that drawn by the HI-deficient, cluster galaxies.
This observational evidence can be easily explained by the quenching of star formation activity in late-type galaxies entering the cluster.
Their interaction with the surrounding medium efficiently removes their atomic and molecular gas component, reducing the amount of gas available to form new stars
(e.g. Boselli \& Gavazzi 2006, 2014; Cortese \& Hughes 2009; Hughes \& Cortese 2009; Gavazzi et al. 2013a,b; Boselli et al. 2014c,d). 
This transformation is particularly efficient in dwarf systems where the shallow gravitational potential well cannot retain the gaseous component anchored 
to the disc (Boselli et al. 2008a,b; Boselli et al. 2014d; Boselli \& Gavazzi 2014). Curiously, the distribution of the HI-deficient cluster galaxies above
$SFR$ $\simeq$ 1 M$_{\odot}$ yr$^{-1}$ is similar to the one determined from the GAMA and SDSS sample of Gunawardhana et al. (2013). We recall, however, that 
the determination of the total volume sampled by the HRS is quite uncertain, thus if the shape of the luminosity distribution is robustly determined, a shift
in the Y-axis can not be totally excluded.

\subsection{SFR scaling relations}

We trace the typical star formation rate scaling relations of the HRS. These relations can be compared to those already determined on the same sample
for the optical and UV structural parameters (Cortese et al. 2012a), for the atomic (Cortese et al. 2011),
molecular, and total gas content (Boselli et al. 2014b), and for the dust component (Cortese et al. 2012b). For a fair comparison with these works,  
we use in the following section the same scaling parameters: the stellar mass $M_{star}$  (taken from Cortese et al. 2012a), the stellar mass surface density
$\mu_{star}$, the metallicity 12+log(O/H), and the specific star formation rate $SSFR$. The star formation rates given in Table \ref{SFR}
have been divided by a factor 1.58 to convert them into the Chabrier (2003) IMF. 
Stellar masses have been determined using $i$-band luminosities and $g-i$ colours combined with the prescription of Zibetti et al. (2009). Their typical uncertainty is 
0.15 dex\footnote{We choose to use these stellar masses instead of those determined from SED fitting, which are probably more accurate, because we want to be consistent with all the 
previous HRS papers presenting the stellar (Cortese et al. 2012a), gas (Cortese et al. 2011; Boselli et al. 2014b), and dust (Cortese et al. 2014) scaling relations.}. 
The stellar surface density $\mu_{star}$ is the total stellar mass divided by the circular area defined by the $i$-band effective radius (the radius enclosing 50\% ~ of 
the total light). Its typical uncertainty is 0.20 dex. Metallicities are taken from Hughes et al. (2013), and are determined using the PP04 O3N2 calibration on [NII] and [OIII] 
emission lines (Pettini \& Pagel 2004), with an uncertainty of 0.13 dex. Specific star formation rates $SSFR$ are defined as the ratio of the star formation rate 
per unit stellar mass. They correspond to the birthrate parameter $b$ once the typical age of galaxies (here assumed to be of 13 Gyr)
and a constant returned gas fraction $R$ of 0.3 (Boselli et al. 2001) is taken into account. The typical uncertainty on this variable is 0.25 dex.
%Total gas masses $M_{gas}$ are determined as indicated in Boselli et al. (2014a) and include the molecular gas component and the Helium contribution (30\%).
%The molecular gas component has been determined assuming the $H$-band luminosity dependent conversion factor of Boselli et al. (2002b). The uncertainty on this variable 
%is 0.20 dex.

\begin{figure}
\centering
\includegraphics[width=9cm]{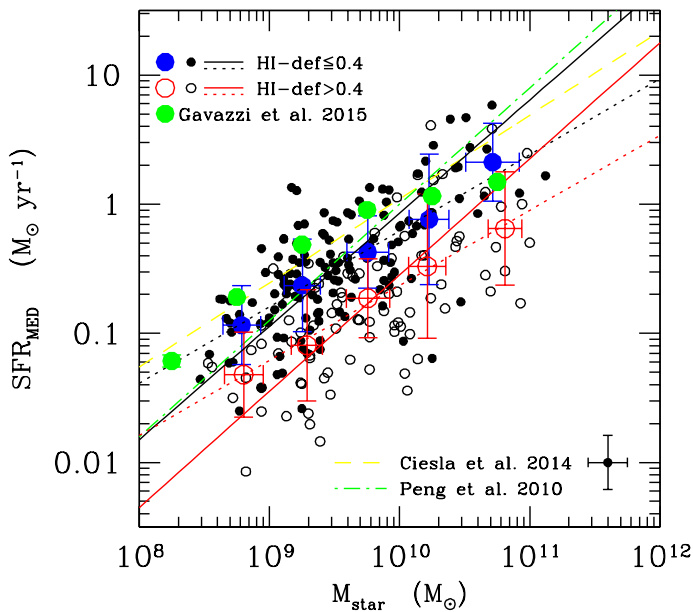}
\caption{Relationship between the star formation rate and the stellar mass for the HRS late-type galaxies. Filled dots are for HI-normal ($HI-def$ $\leq$ 0.4)
field galaxies, empty circles for HI-deficient ($HI-def$ $>$ 0.4) cluster objects. The large filled blue and empty red circles give the mean values and the standard deviations
in different bins of stellar mass determined in this work. The large green filled dots indicate
the mean values of Gavazzi et al. (2015). The bisector fit for HI-normal and HI-deficient galaxies
are given by the solid black and red lines, the linear direct fits by the dotted lines. The linear best fit of Ciesla et al. (2014) is marked with a yellow
dashed line, the best fit of Peng et al. (2010) by the green dotted-dashed line. The error bar shows the typical uncertainty on the data.}
\label{main}
\end{figure}

Figure \ref{main} shows the relationship between the star formation rate and the stellar mass for all the HRS galaxies. This scaling relation is often referred
in the literature as the main sequence (e.g., Guzman et al. 1997; Brinchmann \& Ellis 2000; Bauer
et al. 2005; Bell et al. 2005; Papovich et al. 2006; Reddy et al. 2006; Noeske et al. 2007; Salim et al. 2007; Elbaz et al. 2007; Daddi et al. 2007; 
Pannella et al. 2009; Rodighiero et al. 2010, 2011; Peng et al. 2010; Karim et al. 2011; Whitaker et al. 2012, 2014; Speagle et al. 2014). 
This relation is expected since it links two variables that scale with the size of galaxies. What is 
interesting in this relation is the determination of its slope and intercept and of its dispersion (see Tables \ref{fit} and \ref{Tabscaling}).
The bisector fit derived for the HRS (black solid line) is similar to the one determined by Peng et al. (2010) for a large sample of SDSS local galaxies (green dotted-dashed line), whereas the best linear fit (black dotted line) is similar to that determined for the HRS by Ciesla et al. (2014) using different definitions of $M_{star}$ and $SFR$ (yellow dashed line). 
Figure \ref{main} shows a systematic shift in the relation between HI-normal field (filled dots, solid black line) and HI-deficient cluster galaxies (empty circles, solid red line).
The relations determined for the two subsamples have a very similar slope, but the shift in the Y-axis is as large as 0.65 dex. The shift in the main sequence as a function of gas
content is consistent with what is observed in higher redshift samples by Tacconi et al. (2013). This trend is predictable from the Schmidt law 
and is a further confirmation that the activity of star formation is quenched in galaxies stripped of their gaseous content in dense environments (e.g. Boselli \& Gavazzi
2006, 2014). This evidence, however, contradicts the results of Peng et al. (2010), who found that there is no significant change in the 
slope and intercept of the main sequence drawn by galaxies belonging to different density environments. The sample of Peng et al. (2010), however, is
composed of star forming galaxies as defined in Brinchmann et al. (2004), which includes galaxies with high signal-to-noise (SN$>$ 3) in the H$\alpha$ and 
H$\beta$ lines. This selection criterion obviously favors active star forming objects as those located in the field but excludes the typical HI-deficient galaxies 
in clusters, where the activity of star formation is significantly quenched by the lack of gas. 
It is indeed known that these objects mainly populate the green valley and are thus, in terms of star formation, intermediate 
between normal star forming discs and passive early-type galaxies (Boselli et al. 2008a; Hughes \& Cortese 2009; Cortese \& Hughes 2009; Gavazzi et al. 2013a,b; 
Boselli et al. 2014c,d).

\begin{figure}
\centering
\includegraphics[width=9cm]{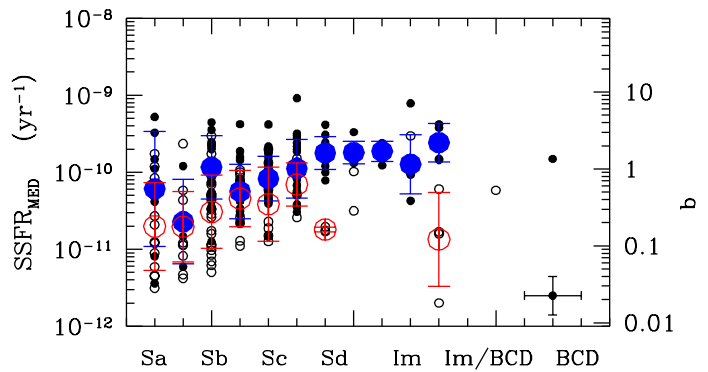}
\caption{Relationship between the specific star formation rate and the morphological type for HI-normal ($HI-def$ $\leq$ 0.4; filled dots) and HI-deficient
($HI-def$ $>$ 0.4; empty circles) galaxies. The large filled blue dots indicate the mean values per each morphological class for normal gas-rich
systems, the empty red ones for cluster HI-deficient galaxies. For the large symbols the error bar shows the standard deviation of the distribution.
The small error bar shows the typical uncertainty on the data. }
\label{sfrtype}
\end{figure}

Figure \ref{sfrtype} shows the relationship between the specific star formation rate  
and the morphological class in late-type systems. Figure \ref{sfrtype} shows that the specific star formation rate is fairly constant with morphological type
(e.g. Kennicutt et al. 1994).
It also shows that gas-deficient galaxies have, on average, lower specific star formation rates than similar objects in the field.  
The mean value of the specific star formation rate of HI-normal galaxies is $SSFR$ = -10.01$\pm$0.41 yr$^{-1}$, while that of HI-deficient objects is 
$SSFR$ = -10.52$\pm$0.49 yr$^{-1}$. Mean values for each morphological class and standard deviations are given in Table \ref{type}. 
Again, these results are fully consistent with what previously found from the H$\alpha$ imaging survey of nearby cluster galaxies (Gavazzi et al. 2002, 2006).

\begin{figure*}
\centering
\includegraphics[width=17cm]{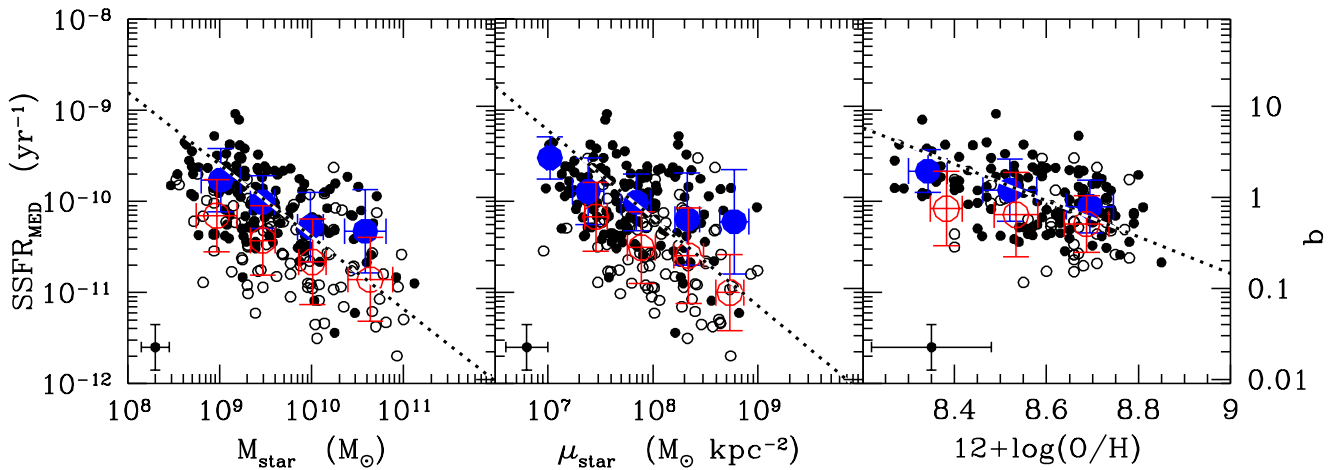}
\caption{Relationship between the specific star formation rate and the total stellar mass (left), the stellar mass surface density (centre), 
and the metallicity (right) for HI-normal ($HI-def$ $\leq$ 0.4; filled dots) and HI-deficient
($HI-def$ $>$ 0.4; empty circles) galaxies. The large filled blue dots indicate the mean values per each morphological class for normal gas-rich
systems, the empty red ones for cluster HI-deficient galaxies. For the large symbols the error bar shows the standard deviation of the distribution.
The dotted line shows the bisector fit determined for HI-normal galaxies. 
The small error bar shows the typical uncertainty on the data. }
\label{scaling}
\end{figure*}

Figure \ref{scaling} and  Tables \ref{fit} and \ref{Tabscaling} show that the specific star formation rate decreases with increasing stellar mass, 
stellar surface density, and metallicity (the probability that these variable are correlated is in all cases $P$ $>$ 99.9\%). 
The decrease of the specific star formation rate as a function of the stellar mass has already been noticed in Gavazzi et al. (1998) and explained in
the framework of a secular evolution of galaxies in Boselli et al. (2001). It also matches the trend observed in the SDSS for local galaxies 
(e.g. Brinchmann et al. 2004). The observed trends with the stellar mass surface density and the metallicity are due to the fact that $M_{star}$, $\mu_{star}$,
and 12+log(O/H) are tightly connected variables (e.g. Brinchmann et al. 2004; Boselli et al. 2008b; Tremonti et al. 2004).
As for Fig. \ref{sfrtype},
HI-deficient cluster galaxies have, on average, specific star formation rates smaller than similar field objects with a normal HI gas content (Gavazzi et al. 2002, 2006).

\subsection{Morphological properties}

\begin{figure*}
\centering
\includegraphics[width=17cm]{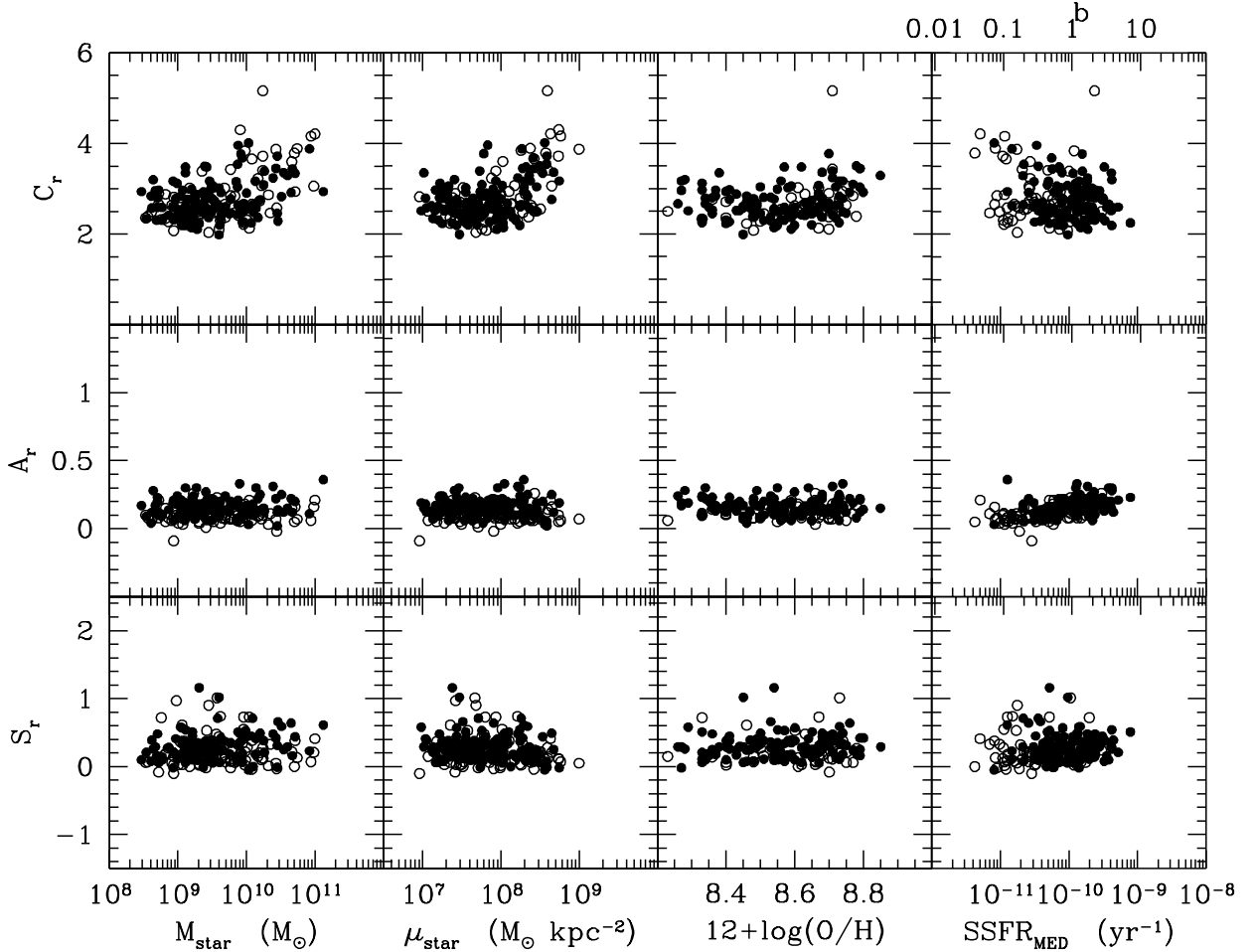}
\caption{Relationship between the CAS parameters determined in the $r$-band and the total stellar mass (left), the stellar mass surface density (centre left), 
the metallicity (centre right), and the specific star formation rate (right) for HI-normal ($HI-def$ $\leq$ 0.4; filled dots) and HI-deficient
($HI-def$ $>$ 0.4; empty circles) galaxies. }
\label{casr}
\end{figure*}

\begin{figure*}
\centering
\includegraphics[width=17cm]{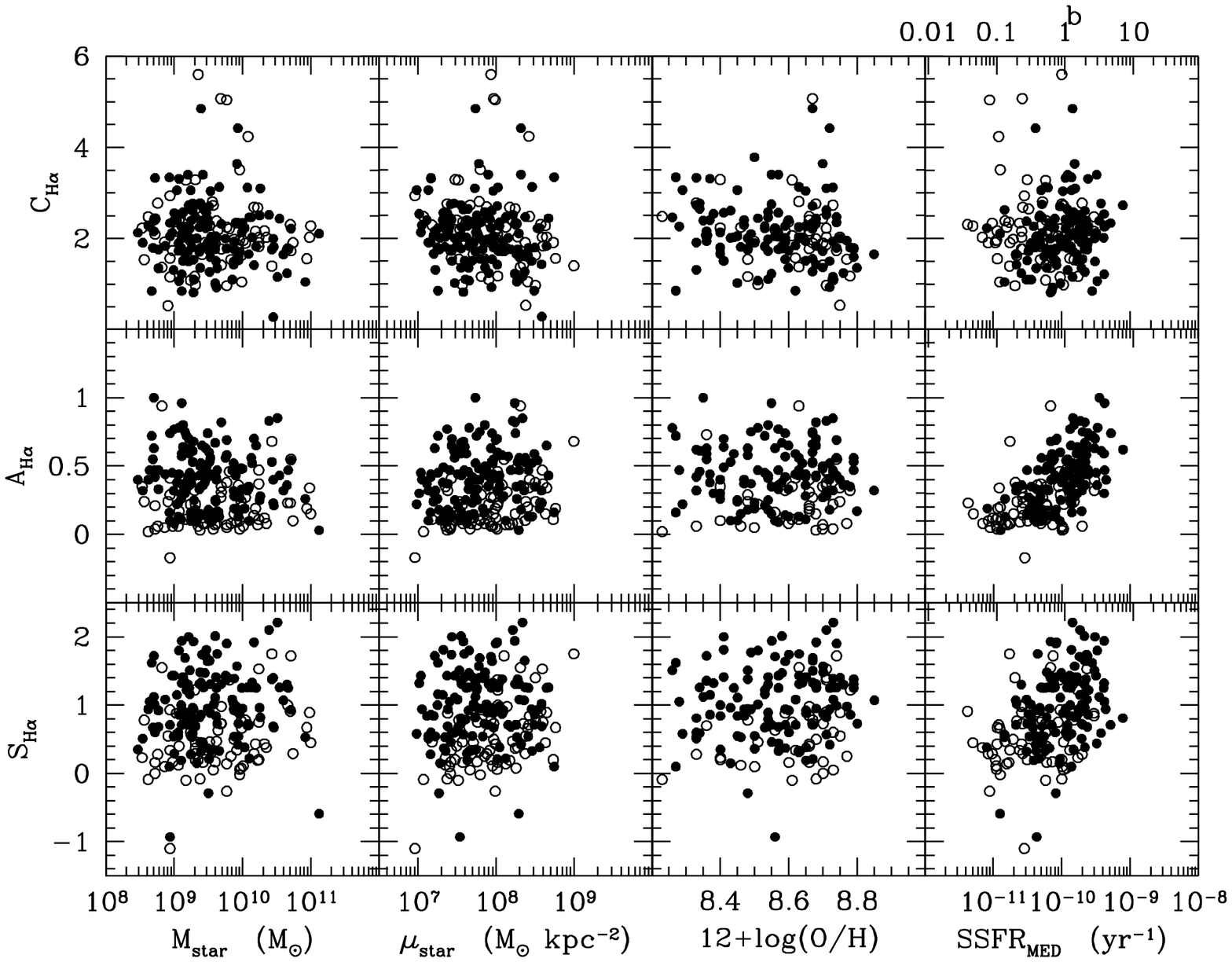}
\caption{Relationship between the CAS parameters determined in the H$\alpha$+[NII] narrow band filter 
and the total stellar mass (left), the stellar mass surface density (centre left), 
the metallicity (centre right), and the specific star formation rate (right) for HI-normal ($HI-def$ $\leq$ 0.4; filled dots) and HI-deficient
($HI-def$ $>$ 0.4; empty circles) galaxies. }
\label{casha}
\end{figure*}

We can also study how the morphological properties of the HRS late-type galaxies traced by the CAS parameters
determined both in the broad band $r$ filter and in the narrow band H$\alpha$+[NII] filter 
are related to the scaling parameters used in the previous section (Figs. \ref{casr} and \ref{casha}). Figure \ref{casr}
shows that the asymmetry $A_r$ and clumpiness $S_r$ parameters measured in the $r$-band do not change as a function of stellar mass,
stellar surface density, and metallicity of galaxies. The concentration parameter $C_r$, instead, is fairly constant for galaxies 
of stellar mass $M_{star}$ $\lesssim$ 10$^{9.5}$ M$_{\odot}$, and increases at higher mass. An increase of the concentration index with
stellar mass, generally attributed to the presence of a dominant bulge, has been already noticed in the past (e.g. Gavazzi et al. 1996; Fossati et al. 2013).
A similar increase of $C_r$ with stellar mass surface density and metallicity is also present, and expected given that all these scaling variables
are tightly connected. The three CAS parameters also do not change significantly with the specific star formation rate of galaxies.
We notice, however, a very moderate increase of $A_r$ with $SSFR$.

The same parameters measured in the H$\alpha$+[NII] band show much more dispersed distributions
and do not show any evident trend when plotted against the stellar mass, stellar mass surface density, and metallicity.
Contrary to the $r$-band, the concentration parameter $C_{H\alpha}$ is fairly constant over the whole stellar mass range
covered by the sample, consistent with the fact that the star formation activity of galaxies is mainly localised 
on the disc. Furthermore, the mean values of the three CAS parameters measured in the H$\alpha$+[NII] band
are systematically shifted with respect to those estimated in the $r$-band, as already noticed by Fossati et al. (2013).
A similar increase of the CAS parameters with decreasing wavelength has been also observed by Lauger et al. (2005) and Taylor-Mager et al. (2007)
in $z$ $\sim$ 1 galaxies. The CAS parameters are often used to quantify the merging rate of different samples of galaxies (e.g. Conselice 2014).
None of the 205 galaxies with available CAS parameters in the $r$-band satisfies the condition:

\begin{equation}
{(A>0.35)~ \& ~(A>S)}
\end{equation} 

\noindent
proposed to identify mergers (Conselice 2014). In the H$\alpha$ band this condition is satisfied by 7/202 objects (HRS 12, 27, 31, 189, 237, 256, 265),
but the clumpy nature of the star forming regions makes the condition given in eq. 14 probably not valid in this band.

The asymmetry and clumpiness parameters increase systematically with the specific star formation rate $SSFR$.
Galaxies with the highest specific star formation rate,
thus those objects which are forming most of their stars at the present epoch, have a strongly asymmetric distribution of the star forming
regions over their disc. They have also a clumpy structure in H$\alpha$+[NII] indicating that their star formation activity is 
mostly due to giant HII regions sporadically distributed over their discs (distance and blending problems should not introduce strong systematic effects since 
galaxies are all at $\sim$ the same distance, Thilker et al. 2000, Liu et al. 2013). This result can be explained if the HII region luminosity function and the contribution 
of the diffuse emission change 
in galaxies of different stellar mass or morphological type, as first suggested by Kennicutt et al. (1989), Caldwell et al. (1991), and Youngblood \& Hunter (1999). 
These studies have shown that in early-type massive spirals the bulk of the star formation occurs in several small HII
regions, whilst late-type, low-mass systems form most of their massive stars in a few giant HII regions. These results, however,
still need to be confirmed on large homogeneous samples with high-quality imaging data (Thilker et al. 2002; Helmboldt et al. 2005;
Schombert et al. 2013; Liu et al. 2013). Nevertheless, since the number of HII regions decreases with the decreasing size of galaxies, and that their distribution
over the disc is random, we expect an increase of the asymmetry parameter in dwarf systems.
Variations of the clumpiness index $S$ as a function of the H$\alpha$ equivalent width, which is a proxy for the $SSFR$, have been already reported by Conselice (2003),
while that of the asymmetry parameter $A$ with the $B-V$ colour index by Conselice et al. (2000).

\begin{figure*}
\centering
\includegraphics[width=17cm]{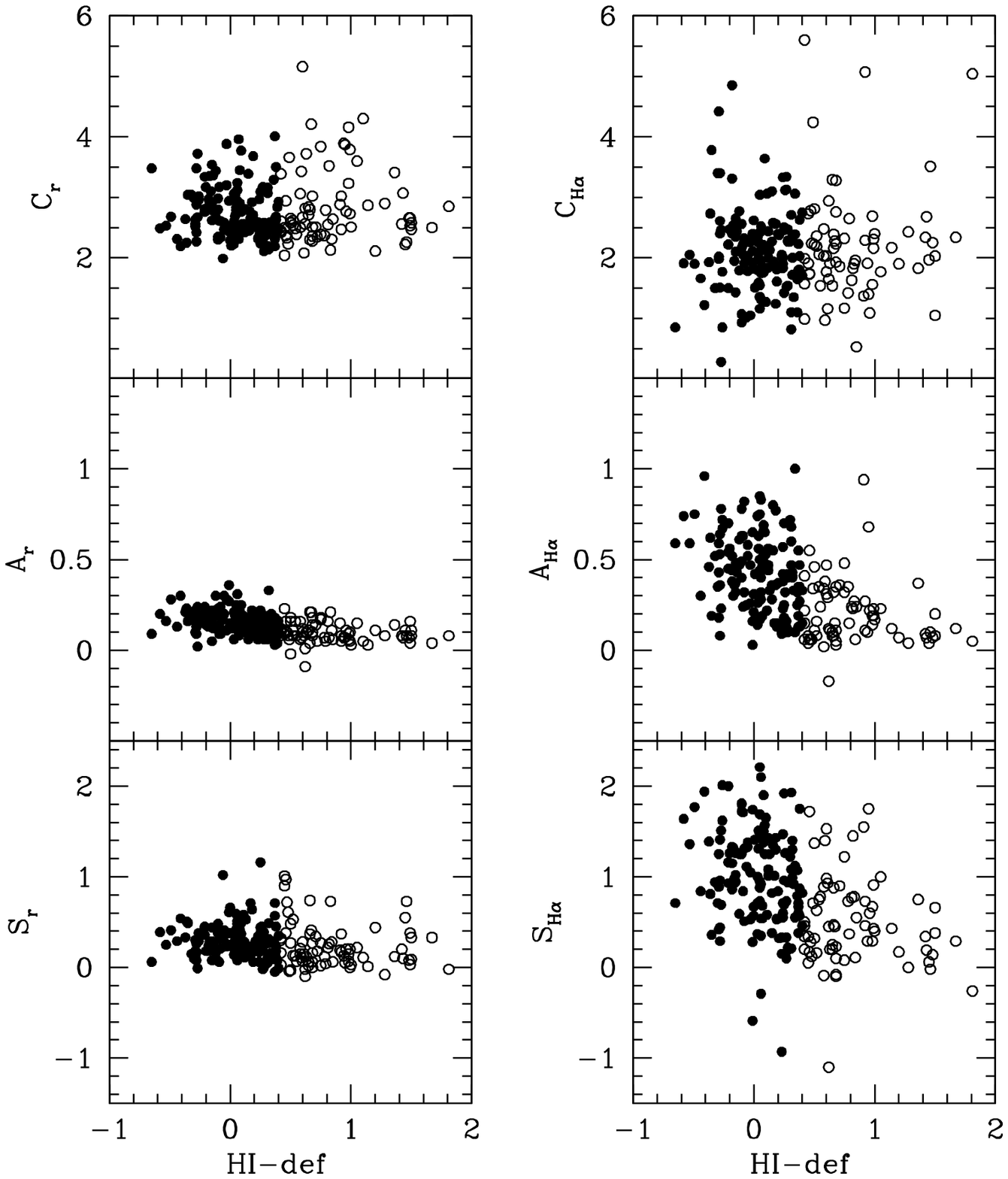}
\caption{Relationship between the CAS parameters determined in the $r$ (left) and H$\alpha$+[NII] narrow band filter (right)
and the HI-deficiency parameter for HI-normal ($HI-def$ $\leq$ 0.4; filled dots) and HI-deficient
($HI-def$ $>$ 0.4; empty circles) galaxies. }
\label{casdef}
\end{figure*}

The CAS parameters can also be plotted against the HI-deficiency parameter, often taken as a direct tracer of the 
perturbation induced by the hostile environment in cluster galaxies (e.g. Boselli \& Gavazzi 2006).
Figure \ref{casdef} shows that while the CAS parameters measured in the $r$-band are fairly constant, $A_{H\alpha}$ and $S_{H\alpha}$
systematically decreases with increasing HI-deficiency. Given that HI-deficient galaxies have, on average, lower $SSFR$s than unperturbed systems, 
this observed decrease might results from the $A_{H\alpha}$ and $S_{H\alpha}$ increase with $SSFR$ observed in Fig. \ref{casha}.
%An increase of $A_{H\alpha}$ with $HI-def$ would be expected if the origin of the perturbation
%is the gravitational interaction with nearby companions during a flyby encounter (galaxy harassment), as
%observed in hydrodynamic simulations. A symmetric distribution of the star forming regions over the disc of galaxies
%is instead consistent with a ram-pressure stripping scenario, where the interstellar medium is removed during the 
%dynamical friction exerted by the hot and dense intracluster medium on galaxies moving at high velocity within the
%potential well of the cluster. In this scenario, the apparent asymmetric and clumpy distribution of the HII regions
%in unperturbed systems is mainly related to the stochastic formation of HII regions over their disc 
%rather than to environmental effects. It is thus not surprising than, on average, HI-deficient objects have lower
%values of $A_{H\alpha}$ and $C_{H\alpha}$ than normal unperturbed objects.

\section{Conclusions}

We present new H$\alpha$+[NII] narrow band imaging observations of late-type galaxies in the \textit{Herschel} Reference Survey
done with the 2.1 m San Pedro Martir telescope. Combined with those already available in the literature, H$\alpha$+[NII] 
data are now available for 281/323 galaxies of the sample, and for 254/260 of the late-type systems. These data are used to extract fluxes and equivalent widths, 
as well as the CAS morphological parameters, with the aim of studying the star formation properties of a volume-limited, K-band-selected complete sample of nearby 
late-type (Sa-Im-BCD) galaxies.
The H$\alpha$+[NII] imaging data are first corrected for [NII] contamination and dust attenuation using the integrated spectroscopy and the 24 
$\mu$m flux densities available for this sample. We then compare the star formation rate determined using either independent 
monochromatic star formation tracers (H$\alpha$, FUV \textit{GALEX}, radio continuum luminosities) or the output of the 
CIGALE SED fitting code done under different assumptions. The comparison of the different tracers shows that:\\
1) All tracers are strongly correlated with each other (e.g. Calzetti et al. 2007, 2010; Kennicutt et al. 2009). 
The typical dispersion in the derived star formation rates is of the order of 24\%. 
This value, however, should be considered as a lower limit for the uncertainty of the $SFR$ determination since it has been determined on non fully independent variables
and making similar assumptions on the IMF and on the star formation history of the target galaxies.\\
2) The H$\alpha$ luminosities corrected for dust attenuation using the Balmer decrement and the 24 $\mu$m emission give consistent results only whenever
the Balmer decrement is determined with an accuracy $\sigma[C(H\beta)]$ $\leq$ 0.1. The most distant objects from the $L(H\alpha)_{BD}$ vs. $L(H\alpha)_{24\mu m}$ relation
all have physical properties typical of quiescent galaxies.
This observational evidence suggests that the 24 $\mu$m based attenuation correction derived for star forming galaxies could be non universal and, as pointed out by Kennicutt et al.
(2009), should be used with extreme caution in massive galaxies characterised by a low activity of star formation, where the heating of the dust is done also by the evolved stellar populations.\\
3) The comparison of the star formation rates determined using monochromatic tracers and UV-to-far infrared SED fitting codes gives consistent results in unperturbed 
late-type systems (e.g. Wuyts et al. 2009; Pforr et al. 2012; Buat et al. 2014), 
while small but systematic differences in the two tracers are present in HI-deficient cluster galaxies. These differences cannot be explained by a 
non optimised parametrisation of the star formation history of cluster galaxies that is unable to reproduce an abrupt truncation of their activity once entered into the cluster.
They might be a further indication that the prescription used to correct the H$\alpha$ for dust attenuation based on the 24 $\mu$m emission are not valid
in red and quiescent spirals, where the dust heating comes primarily from the evolved stellar population.\\

The new set of data is used to trace the star formation rate distribution of a K-band-selected, volume-limited sample of nearby galaxies and to derive the typical scaling
relations between the specific star formation rate and the morphological type, the stellar mass and stellar surface density, the metallicity, and the CAS parameters
for the late-type systems. The specific star formation rate is anticorrelated with the stellar mass (e.g. Boselli et al. 2001), 
the stellar mass surface density, and the metallicity of galaxies
in the range sampled by these parameters, while it is fairly constant with the morphological type (e.g. Kennicutt et al. 1994). 
All relations show a systematic difference between cluster and isolated galaxies confirming the quenching of the star formation activity in high
density environments (e.g. Boselli \& Gavazzi 2006, 2014). They also show a clear relation between the asymmetry and clumpiness parameters 
and the specific star formation rate. This relation can be explained 
if the HII luminosity function of low-mass actively star forming galaxies is dominated by a few giant HII regions while that of massive and quiescent systems
by sevreal HII region of intermediate-to-low luminosity, a result that still needs to be confirmed observationally.

\begin{acknowledgements}
We are grateful to the referee for constructive comments and suggestions, and to Luca Cortese for his long term collaboration on the HRS project.
This research has made use of data from the HRS project. HRS is a Herschel Key Programme utilising 
guaranteed time from the SPIRE instrument team, ESAC scientists and a mission scientist. 
The HRS data was accessed through the Herschel Database in Marseille (HeDaM - http://hedam.lam.fr) 
operated by CeSAM and hosted by the Laboratoire d'Astrophysique de Marseille. 
We acknowledge financial support from Programme National de Cosmologie and Galaxies (PNCG) of CNRS/INSU, France.
MF acknowledges the support of the Deutsche Forschungsgemeinschaft via Project ID 387/1-1.
The H$\alpha$+[NII] images of the Virgo cluster galaxies belonging to the HRS have been taken 
from the GOLDMine database (Gavazzi et al. 2003).
\end{acknowledgements}

\begin{appendix}
\section{{\sc GANDALF} spectroscopy}

\begin{figure}
\centering
\includegraphics[width=15cm]{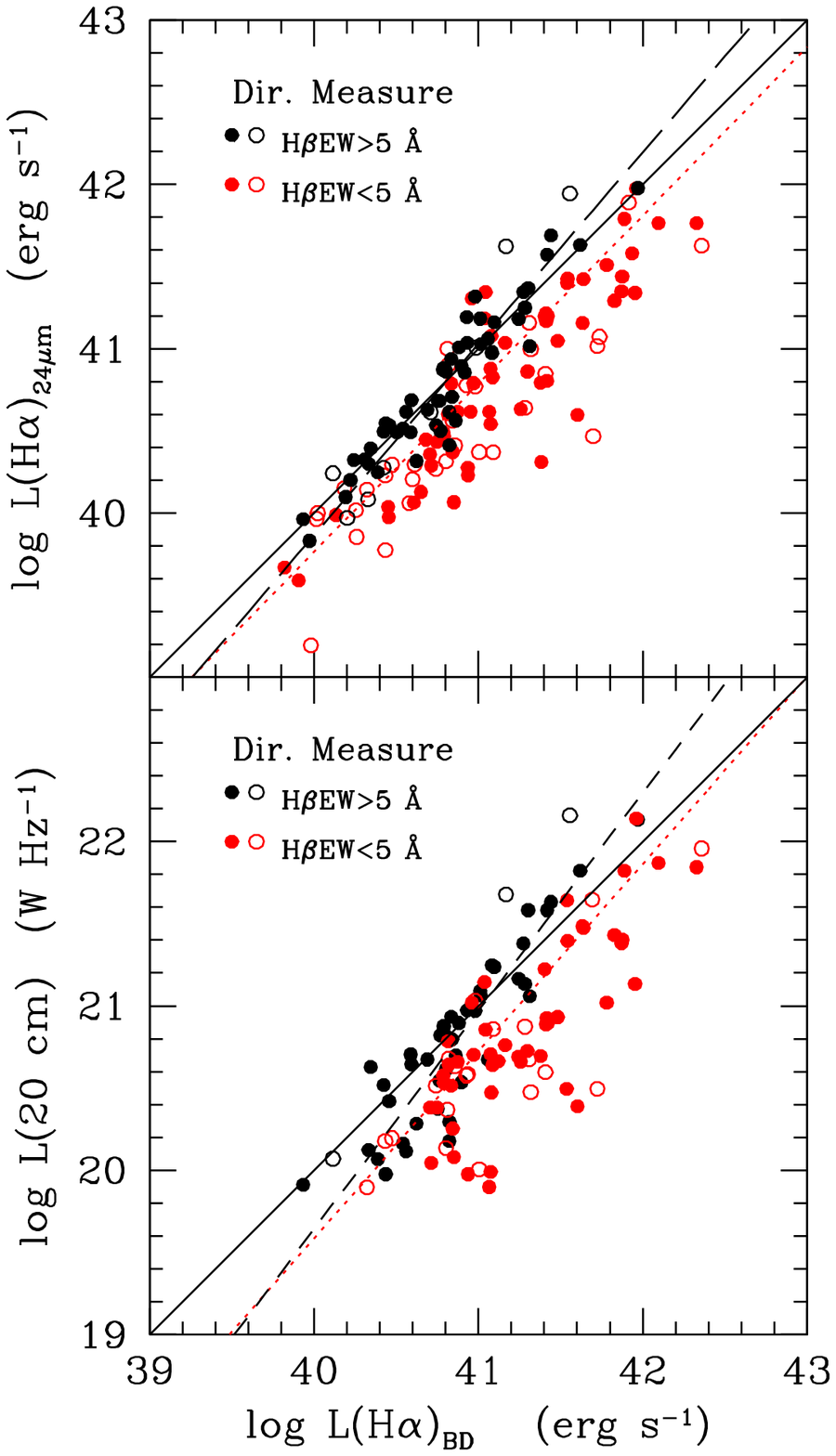}
\caption{Relationship between the H$\alpha$ luminosity corrected for dust attenuation using the 24 $\mu$m emission (upper panel), the 20 cm radio continuum luminosity (lower panel)
and the H$\alpha$ luminosity corrected for Balmer decrement using the spectroscopic set of data published in Boselli et al. (2013), where the line emission is directly measured
on the data. Filled dots are for  HI-normal ($HI-def$ $\leq$ 0.4) galaxies, empty symbols for HI-deficient
($HI-def$ $>$ 0.4) objects. Black symbols indicate galaxies with an H$\beta$E.W. $>$ 5 \AA, red symbols those with H$\beta$E.W. $<$ 5 \AA. The
black solid line shows the 1:1 relation, the black long dashed line the bisector fit to the high-quality data (black symbols), the red dotted line the best fit to the whole dataset.}
\label{haradcal}
\end{figure}

The comparison of the H$\alpha$ luminosity corrected for dust attenuation using the Balmer decrement measured in Boselli et al. (2013) 
and the H$\alpha$ luminosity corrected using the 24 $\mu$m emission, or the 20 cm radio continuum luminosity
(Fig. \ref{haradcal}) shows a systematic shift of $\sim$ 0.5 dex in the relations between galaxies with strong (H$\beta$E.W. $>$ 5 \AA) 
and weak (H$\beta$E.W. $<$ 5 \AA) H$\beta$ emission. Given that the three variables are roughly equivalent, the observed shift probably results
from a systematic bias in the spectroscopic data used to correct $L(H\alpha)_{BD}$. We recall that in Boselli et al. (2013) the intensity of 
the H$\beta$ line has been measured by fitting the spectra with a double Gaussian, one for the emission line and the other for the underlying stellar absorption.
The observed shift can thus be explained if the Balmer decrement is overestimated, or the H$\beta$ emission is underestimated by $\sim$ 60 \%.
To test whether the observed shift in the relations shown in Fig. \ref{haradcal} results from a systematic bias introduced by the extraction procedure 
adopted in Boselli et al. (2013) we have extracted the emission line fluxes of all the HRS galaxies with available spectra (Gavazzi et al. 2004; Boselli et al. 2013; 264/323 objects)
using the {\sc GANDALF} fitting code (Sarzi et al. 2006; Falcon-Barroso et al. 2006).
{\sc GANDALF} is a simultaneous emission and absorption lines fitting algorithm designed to 
separate the relative contribution of the stellar continuum and of nebular emission in the spectra of galaxies. This code implements the pPXF method (Cappellari \& Emsellem 2004)
which combines and adjusts the
observed spectra of several stars of all spectral type to the stellar continuum to first quantify and remove the underlying absorption
contaminating the emission of the most important emission lines, including H$\beta$ and H$\alpha$. With this procedure, the code is expected to correctly 
account for the underlying stellar absorption that contaminates several emission lines, in particular the Balmer lines which are crucial for the 
determination of the dust attenuation of the H$\alpha$ line. Figures \ref{hacal} and \ref{radiocal} show the same relationships between $L(H\alpha)_{24 \mu m}$, $L(20cm)$,
and $L(H\alpha)_{BD}$ as those plotted in Fig. \ref{haradcal} but using {\sc GANDALF} data for the Balmer decrement and [NII] contamination corrections. Most of the systematic differences 
observed between galaxies with high and low H$\beta$E.W., or high and low signal-to-noise, are removed using this new set of data. We thus use in the analysis presented in this work 
the spectroscopic dataset extracted using {\sc GANDALF}. The fluxes of the different emission lines are given 
in Table \ref{GANDALF}, arranged as follows:

\begin{itemize}
\item {Column 1:\textit{Herschel} Reference Sample (HRS) name.}
\item {Columns 2-8: observed line intensities normalised to H$\alpha$. The {\sc GANDALF} fitting code assumes that [OIII]$\lambda$4958 = 0.3$\times$[OIII]$\lambda$5007
and that [NII]$\lambda$6548 = 0.3$\times$[NII]$\lambda$6584.}
\item {Column 9-10: signal-to-noise SN measured empirically from the spectrum as the amplitude of the H$\alpha$ and H$\beta$ lines divided by the noise in the nearby continuum.}
\item {Column 11: Balmer decrement $C(H\beta)$. The contribution of the Milky Way is subtracted using the Galactic extinction map of 
Schlegel et al. (1998) combined with the Fitzpatrick \& Massa (2007) Galactic extinction law.}
\item {Column 12: uncertainty on the Balmer decrement $\sigma[C(H\beta)]$ obtained from standard error propagation of the uncertainties on the line
fluxes.}
\end{itemize}

\begin{figure*}
\centering
\includegraphics[width=10cm]{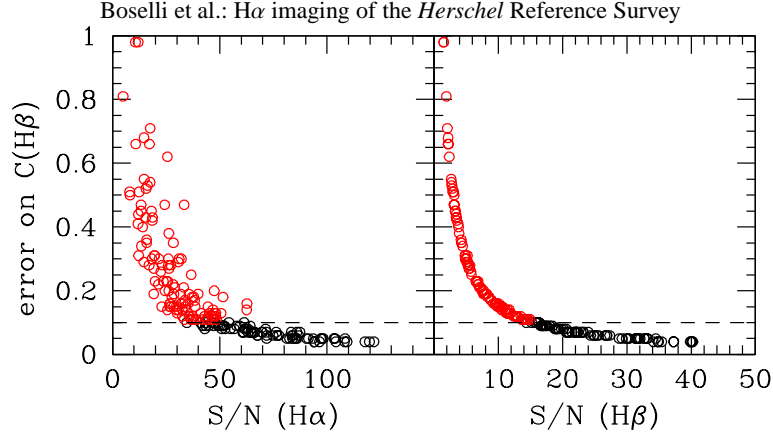}
\caption{Relationship between the uncertainty on the Balmer decrement $\sigma[C(H\beta)$ and the H$\alpha$ and H$\beta$ signal to noise. The horizontal dashed line shows the 
$\sigma[C(H\beta)$ = 0.1 limit adopted in this work to identify good quality objects (black symobls) from uncertain measurements (red symbols).}
\label{sigmaCHb}
\end{figure*}

The uncertainty on the Balmer decrement $\sigma[C(H\beta)]$ is tightly correlated with the signal-to-noise measured at the H$\alpha$ and H$\beta$ lines, as depicted in Fig. \ref{sigmaCHb}.
For the typical spectral resolution of $R$ $\sim$ 1000 of the spectroscopic data used in this work (Boselli et al. 2013), the required limiting uncertainty of $\sigma[C(H\beta)]$ = 0.1 
is reached for a signal-to-noise of $\simeq$ 50 for H$\alpha$ and $\simeq$ 15 for H$\beta$. A spectral resolution of $R$ $\gtrsim$ 1000 is required for an accurate 
deblending of the H$\alpha$ line from the nearby [NII] lines and for an accurate determination of the underlying Balmer absorption.

\section{Radio continuum data}

We have collected radio continuum data at 1.49 GHz (20 cm) obtained at the VLA for 191 HRS galaxies. These data
have been taken, in order of preference, from the pointed survey of \textit{IRAS} bright galaxies of Condon et al. (1990),
from the survey of optically selected galaxies of Condon (1987), or from the survey of galaxies with H-band magnitudes
of Condon et al. (1987). The data presented in these papers have been obtained under similar VLA configurations (generally D and C/D)
and thus have a comparable angular resolution ($\sim$ 45 arcsec). The typical rms noise level of these observations is 0.1-0.2 mJy per beam.
Flux densities of the remaining galaxies have been taken from the NVSS survey of Condon et al. (1998)
and published in Condon et al. (2002) whenever available, or extracted by ourselves from the NVSS survey catalogue assuming a 
search radius of 15 arcsec. These data have been homogeneously taken in VLA-D configuration, and have a typical resolution of 45 arcsec and a sensitivity of 0.45 mJy rms. 
For three galaxies, we adopt the integrated flux densities obtained by the FIRST survey (Becker et al. 1995). The FIRST survey has been completed using the VLA in B configuration, and thus have 
an angular resolution of $\lesssim$ 5 arcsec. Because of the adopted configuration, this survey is sensitive to compact sources and might miss some of the low surface brightness extended emission.
Its typical sensitivity is 0.15 mJy rms. 

Table \ref{Tabradio} gives the radio data for all the HRS galaxies. Table \ref{Tabradio} is arranged as follows:

\begin{itemize}
\item {Column 1:\textit{Herschel} Reference Sample (HRS) name.}
\item {Column 2: sign indicating detections (1) and undetections (0)}
\item {Column 3: flux density $S(20cm)$, or upper limit, in mJy. For the undetected objects we estimate an upper limit to the 
flux density $S_{ul}(20cm)$ = 4 $\times$ $rms$, and assuming for all galaxies the same typical rms noise of the NVSS 
survey of 0.45 mJy.}
\item {Column 4: logarithm of the 20 cm radio luminosity, or upper limit, in W Hz$^{-1}$.}
\item {Column 5: flag to the data, as indicated in the original papers. 1 stands for galaxies with good quality observations, 2 for uncertain values.}
\item {Column 6: references to the data: 1: Condon et al. (1990); 2: Condon (1987); 3: Condon et al. (1987); 4: Condon et al. (2002); 
5: extracted as described in the text from the NVSS survey (Condon et al. 1998); 6: extracted from the FIRST survey (Becker et al. 1995).}
\end{itemize}

\section{Adopted calibrations of the different star formation tracers}

The H$\alpha$ and FUV star formation rates used in this work have been determined using the relation:

\begin{equation}
{SFR_{\lambda} ~~~[\rm M_{\odot} yr^{-1}] = K_{\lambda} L(\lambda)_{cor} ~~~[\rm erg~s^{-1}] }
\end{equation}

\noindent
where the different values of $K_{\lambda}$ used in this work are given in Table \ref{calconst} (adapted from Hao et al. 2011).

\begin{table}
\caption{Calibrations of the different star formation tracers}
\label{calconst}
{%\scriptsize
\[
\begin{tabular}{cc}
\hline
\hline
\noalign{\smallskip}
%\hline
$\lambda$	&  $K_{\lambda}$ \\
\hline
H$\alpha$	& 8.79 $\times$ 10$^{-42}$ \\
FUV		& 6.21 $\times$ 10$^{-44}$ \\
%NUV		& 9.48 $\times$ 10$^{-44}$ \\
\noalign{\smallskip}
\hline
\end{tabular}
\]
Note: for a Salpeter IMF in the stellar mass range 0.1 $<$ $m$ $<$ 100 M$_{\odot}$, from Hao et al. (2011). }
\end{table}

\noindent
The attenuation corrected H$\alpha$ luminosity have been determined using either the Balmer decrement or
the following relation (from Calzetti et al. 2010):

\begin{equation}
{log(L(H\alpha)_{cor} = logL(H\alpha)_{obs} + 0.020 \times logL(24 \mu m) ~~\rm{for}~~ L(24 \mu m) < 10^{42} ~~[\rm{erg s^{-1}]}}
\end{equation}

and

\begin{equation}
{log(L(H\alpha)_{cor} = logL(H\alpha)_{obs} + 0.031 \times logL(24 \mu m) ~~\rm{for}~~ L(24 \mu m) \geq 10^{42} ~~[\rm{erg s^{-1}]}}
\end{equation}

\noindent
where the observed ($L(H\alpha)_{obs}$)\footnote{The observed H$\alpha$ luminosity must be also corrected
for Galactic extinction and [NII] contamination.} and corrected ($L(H\alpha)_{cor}$) H$\alpha$ luminosities and 
the 24 $\mu$m luminosity ($L(24 \mu m)$) are expressed in erg s$^{-1}$.

The corrected UV luminosities are determined using the recipes of Hao et al. (2011):

\begin{equation}
{L(FUV)_{24\mu m} = L(FUV)_{obs} + 3.89 \times L(24\mu m) }
\end{equation}

%\noindent
%and

%\begin{equation}
%{L(NUV)_{24\mu m} = L(NUV)_{obs} + 2.26 \times L(24\mu m) }
%\end{equation}

Star formation rates from radio continuum luminosities at 1.49 GHz (20 cm) have been determined using the calibration of Bell (2003):

\begin{equation}
{SFR_{radio} ~~[\rm M_{\odot} yr^{-1}] = 5.52 \times 10^{-22} L(20cm) ~~\rm{for}~~ L(20cm)>L_c}
\end{equation} 
\begin{equation}
{SFR_{radio} ~~[\rm M_{\odot} yr^{-1}] = \frac{5.52 \times 10^{-22}}{0.1+0.9(L(20cm)/L_c)^0.3} L(20cm) ~~\rm{for}~~ L(20cm)\leq L_c}
\end{equation} 

\noindent
where the radio luminosity is expressed in W Hz$^{-1}$, and $L_c$ = 6.4 $\times$ 10$^{21}$ W Hz$^{-1}$.

The different estimates of the star formation rate are listed in Table \ref{SFR}, arranged as follows:

\begin{itemize}
\item {Column 1:\textit{Herschel} Reference Sample (HRS) name.}
\item {Column 2: $SFR_{H\alpha +BD}$ determined by correcting the H$\alpha$ luminosity using the Balmer decrement derived using the {\sc GANDALF} code for galaxies with SN $>$ 50.}
\item {Column 3: $SFR_{H\alpha +24\mu m}$ determined by correcting the H$\alpha$ luminosity using the 24 $\mu$m emission and eq. C.2 and C.3.}
\item {Column 4: $SFR_{FUV +24\mu m}$ determined by correcting the FUV luminosity using the 24 $\mu$m emission and eq. C4.}
\item {Column 5: $SFR_{radio}$ determined using the relation C.5 and C.6. This value has been determined only in galaxies with high-quality radio data (flag = 1 in Table \ref{Tabradio}).}
\item {Column 6: $SFR_{MED}$, mean value of the different monochromatic estimates given in columns 2-5.}
\item {Column 7: $SFR_{1exp}$ derived by fitting the SED with the CIGALE code using one exponentially declining star formation history. }
\item {Column 8: $SFR_{2exp}$ derived by fitting the SED with the CIGALE code using two exponentially declining star formation histories.}
\item {Column 9: $SFR_{del}$ derived by fitting the SED with the CIGALE code using a delayed exponentially declining star formation history.}
\end{itemize}

The star formation rates determined using the CIGALE fitting code have been measured only whenever FUV data are available. All galaxies are assumed coeval, with an age of 13 Gyr.
We recall that these $SFR$ have been measured assuming consistently a Salpeter IMF
in the stellar mass range 0.1 $<$ $m_{star}$ $<$ 100 M$_{\odot}$. These values of $SFR$ can be converted into $SFR$ measured using a 
Chabrier (2003) or Kroupa (2001) IMF by dividing them by a factor of $\sim$ 1.58. This value slightly changes with the assumed 
population synthesis model, and might vary from reference to reference (Brinchmann et al. 2004; Salim et al. 2007; Bigiel et al. 2008; Argence \& Lamareille 2009; 
Schiminovich et al. 2010; Peng et al. 2010; Chomiuk \& Povich 2011).

\end{appendix}

\onecolumn
\begin{landscape}
\begin{center} \tiny
\begin{longtable}{c c c c c c c c c c c c c c c c }
\caption{\textit{Herschel} Reference Survey.\label{TabHRS}}\\
\hline \hline 
\multicolumn{1}{c}{HRS} &
\multicolumn{1}{c}{CGCG}&      
\multicolumn{1}{c}{VCC} &    
\multicolumn{1}{c}{UGC} &     
\multicolumn{1}{c}{NGC} &    
\multicolumn{1}{c}{IC}  & 
\multicolumn{1}{c}{RA(2000)} &   
\multicolumn{1}{c}{dec}      & 
\multicolumn{1}{c}{type}     &
\multicolumn{1}{c}{$D$}    & 
\multicolumn{1}{c}{$M_{star}$}& 
\multicolumn{1}{c}{$D_{24.5}(g)$}       &
\multicolumn{1}{c}{incl}  &    
\multicolumn{1}{c}{vel}           &   
\multicolumn{1}{c}{memb}  & 
\multicolumn{1}{c}{H$\alpha$+[NII]}    \\
\multicolumn{1}{c}{} &
\multicolumn{1}{c}{} &      
\multicolumn{1}{c}{} &    
\multicolumn{1}{c}{} &     
\multicolumn{1}{c}{} &    
\multicolumn{1}{c}{}  & 
\multicolumn{1}{c}{h m s} &   
\multicolumn{1}{c}{$^{\circ}$ ' "}      & 
\multicolumn{1}{c}{}       &
\multicolumn{1}{c}{Mpc}    & 
\multicolumn{1}{c}{M$_{\odot}$}    & 
\multicolumn{1}{c}{'}       &  
\multicolumn{1}{c}{deg}	&  
\multicolumn{1}{c}{km s$^{-1}$}           &   
\multicolumn{1}{c}{}    & 
\multicolumn{1}{c}{}    \\
\multicolumn{1}{c}{(1)} &
\multicolumn{1}{c}{(2)}&      
\multicolumn{1}{c}{(3)} &    
\multicolumn{1}{c}{(4)} &     
\multicolumn{1}{c}{(5)} &    
\multicolumn{1}{c}{(6)}  & 
\multicolumn{1}{c}{(7)} &   
\multicolumn{1}{c}{(8)}      & 
\multicolumn{1}{c}{(9)}       &
\multicolumn{1}{c}{(10)}    & 
\multicolumn{1}{c}{(11)}    & 
\multicolumn{1}{c}{(12)}       &    
\multicolumn{1}{c}{(13)}           &   
\multicolumn{1}{c}{(14)}    & 
\multicolumn{1}{c}{(15)}    & 
\multicolumn{1}{c}{(16)}    \\
\hline
  \\
\endhead 
%This is the footer for all pages except the last page of the table...
 \\ 
 \hline
  \multicolumn{16}{c}{{Continued on next page\ldots}} \\
\endfoot

%This is the footer for the last page of the table...
  \\ 
\hline \hline
\endlastfoot
       1&   123035&        - &       -&        -&        -&101739.66& 224835.9&Pec  		   &   16.79 &    8.72 &   0.92&   67.09 & 1175&Leo Cl. 	     &   1  \\
       2&   124004&        - &    5588&        -&        -&102057.13& 252153.4&S?  		   &   18.44 &    8.77 &   0.73&   44.82 & 1291&Leo Cl. 	     &   1  \\
       3&    94026&        - &    5617&     3226&        -&102327.01& 195354.7&E2:pec;LINER;Sy3	   &   16.70 &   10.21 &   3.15&   34.97 & 1169&Leo Cl. 	     &   1  \\
       4&    94028&        - &    5620&     3227&        -&102330.58& 195154.2&SAB(s)pec;Sy1.5     &   16.40 &   10.26 &   4.80&   61.37 & 1148&Leo Cl. 	     &   1  \\
       5&    94052&        - &       -&        -&      610&102628.37& 201341.5&Sc   		   &   16.71 &    9.62 &   1.96&   77.41 & 1170&Leo Cl. 	     &   1  \\
       6&   154016&        - &    5662&    3245A&        -&102701.16& 283821.9&SB(s)b  		   &   18.89 &    8.94 &   2.21&   90.00 & 1322&Leo Cl. 	     &   1  \\
       7&   154017&        - &    5663&     3245&        -&102718.39& 283026.6&SA(r)0:?;HII;LINER  &   18.77 &   10.29 &   3.13&   55.29 & 1314&Leo Cl. 	     &   0  \\
       8&   154020&        - &    5685&     3254&        -&102919.92& 292929.2&SA(s)bc;Sy2   	   &   19.37 &    9.88 &   3.92&   70.43 & 1356&Leo Cl. 	     &   1  \\
\end{longtable}
\end{center}
\end{landscape}

%\twocolumn
%\clearpage

%\onecolumn
%\input{obs_short.tex}
%\begin{landscape}
\begin{center} \tiny \begin{longtable}{c c c c c c c c c c c c}
\caption{H$\alpha$ observational parameters of the 138 target galaxies.\label{TabHaobs}}\\
\hline \hline 
\multicolumn{1}{c}{HRS} &
\multicolumn{1}{c}{Filter}&      
\multicolumn{1}{c}{Tel} &    
\multicolumn{1}{c}{CCD} &     
\multicolumn{1}{c}{Pixel} &    
\multicolumn{1}{c}{Year}  & 
\multicolumn{1}{c}{N. exp} &   
\multicolumn{1}{c}{T. exp} &
\multicolumn{1}{c}{AM} &    
\multicolumn{1}{c}{Phot}  & 
\multicolumn{1}{c}{Z. point} &   
\multicolumn{1}{c}{Norm}     \\
\multicolumn{1}{c}{} &
\multicolumn{1}{c}{\AA} &      
\multicolumn{1}{c}{} &
\multicolumn{1}{c}{} &
\multicolumn{1}{c}{"} &
\multicolumn{1}{c}{} &
\multicolumn{1}{c}{} &
\multicolumn{1}{c}{s} &
\multicolumn{1}{c}{} &
\multicolumn{1}{c}{} &
\multicolumn{1}{c}{erg cm$^{-2}$ s$^{-1}$} &     
\multicolumn{1}{c}{}      \\
\multicolumn{1}{c}{(1)} &
\multicolumn{1}{c}{(2)} &      
\multicolumn{1}{c}{(3)} &    
\multicolumn{1}{c}{(4)} &     
\multicolumn{1}{c}{(5)} &    
\multicolumn{1}{c}{(6)} & 
\multicolumn{1}{c}{(7)} &   
\multicolumn{1}{c}{(8)} &
\multicolumn{1}{c}{(9)} &    
\multicolumn{1}{c}{(10)} & 
\multicolumn{1}{c}{(11)} &   
\multicolumn{1}{c}{(12)} 
     \\     \\
\hline
  \\
\endhead 
%This is the footer for all pages except the last page of the table...
 \\ 
 \hline
  \multicolumn{12}{c}{{Continued on next page\ldots}} \\
\endfoot

%This is the footer for the last page of the table...
  \\ 
\hline \hline
\endlastfoot
% HRS      on        tel     ccd     pix   date Noexp tinton amon  phot     Zp        n
  1     & 6603 &     2.1 &   E2V  & 0.35 & 2011 & 3 &    300.0  & 1.02   & 1  & -15.36   &  0.190    \\ 
  2     & 6603 &     2.1 &   E2V  & 0.35 & 2011 & 3 &    300.0  & 1.02   & 1  & -15.36   &  0.130    \\ 
  3     & 6603 &     2.1 &   E2V  & 0.35 & 2012 & 4 &    300.0  & 1.06   & 0  & -        &  0.438    \\ 
  4     & 6603 &     2.1 &   E2V  & 0.35 & 2012 & 4 &    300.0  & 1.06   & 0  & -        &  0.438    \\ 
  5     & 6603 &     2.1 &   sit3 & 0.31 & 2008 & 3 &    420.0  & 1.04   & 1  & -15.45   &  0.153    \\ 
  6     & 6607 &     1.5 &   e2vm & 0.28 & 2010 & 2 &    720.0  & 1.00   & 0  & -        &  0.166    \\ 
  8     & 6603 &     2.1 &   sit3 & 0.31 & 2006 & 1 &    900.0  & 1.00   & 1  & -15.52   &  0.460    \\ 
  9     & 6603 &     2.1 &   sit3 & 0.31 & 2008 & 1 &    900.0  & 1.05   & 1  & -15.45   &  0.446    \\
\end{longtable}
\end{center}
%\end{landscape}

%\twocolumn
%\clearpage

%\onecolumn
%\input{flussi_short.tex}
%\begin{landscape}
\begin{center} \tiny \begin{longtable}{c c c c c c c c }
\caption{H$\alpha$+[NII] fluxes and equivalent widths.\label{TabHa}}\\
\hline \hline 
\multicolumn{1}{c}{HRS} &
\multicolumn{1}{c}{H$\alpha$+[NII]E.W.}&      
\multicolumn{1}{c}{err} &    
\multicolumn{1}{c}{log F(H$\alpha$+[NII])} &     
\multicolumn{1}{c}{err} &    
\multicolumn{1}{c}{Ref}  & 
\multicolumn{1}{c}{Alt. Ref.} &   
\multicolumn{1}{c}{Notes}       \\
\multicolumn{1}{c}{} &
\multicolumn{1}{c}{\AA} &      
\multicolumn{1}{c}{\AA} &    
\multicolumn{1}{c}{erg cm$^{-2}$ s$^{-1}$} &     
\multicolumn{1}{c}{erg cm$^{-2}$ s$^{-1}$} &    
\multicolumn{1}{c}{}  & 
\multicolumn{1}{c}{} &   
\multicolumn{1}{c}{}      \\
\multicolumn{1}{c}{(1)} &
\multicolumn{1}{c}{(2)} &      
\multicolumn{1}{c}{(3)} &    
\multicolumn{1}{c}{(4)} &     
\multicolumn{1}{c}{(5)} &    
\multicolumn{1}{c}{(6)} & 
\multicolumn{1}{c}{(7)} &   
\multicolumn{1}{c}{(8)}      \\
\hline
  \\
\endhead 
%This is the footer for all pages except the last page of the table...
 \\ 
 \hline
  \multicolumn{8}{c}{{Continued on next page\ldots}} \\
\endfoot

%This is the footer for the last page of the table...
  \\ 
\hline \hline
\endlastfoot
% HRS     HALFA  ERRHA  FLUXHA       err REFHA altref note    
  1     &   41  &   4  &    -12.80 & 0.04  & TW      &               &   \\    %37  04  12.82 56				   
  2     &   41  &   3  &    -12.40 & 0.03  & TW      &               &   \\  				    
  3     &   8   &   4  &    -12.82 & 0.21  & TW      & 14,17         & c \\     %flusso da 61 = 13.32 (EW 45 da ref 48), EW noi da 57		c: CALIBRATAA PARTIRE DALLA 3 CHE E" SULLA STESSA IMMAGINE		   
  4     &   34  &   5  &    -11.90 & -     & TW,7    & 22            &   \\     %flusso da 46 (EW 78), EW noi 57				     
  5     &   3   &   3  &    -13.27 & 0.40  & TW      &               &   \\  				    
  6     &   0   &   36 &    -      &       & TW      &               &   \\  				     
  7     &   -   &   -  &    -      & -     & -       &               &   \\  				    
  8     &   11  &   4  &    -12.16 & 0.14  & TW      & 22            &   \\
\end{longtable}
\end{center}
%\end{landscape}

%TW: ref 44, 45, 50, 51, 52, 53, 54, 55, 56, 57
%1: Boselli & Gavazzi 2002 (ref 15, 20, 24, 30)
%2: Boselli et al 2002 NOT/INT (ref 26, 31)
%3: Gavazzi et al 2002 (ref 25, 27, 33, 37)
%4: Gavazzi et al 2006 (ref 39, 41, 43)
%5: Koopman et al 2001 (ref 32)
%6: Young et al 1996 (ref 10)
%7: Kennicutt et al 1987 (ref 46)
%8: Macchetto et al 1996 (ref 21)
%9: James et al 2004 (ref 40)
%10: Hameed et al 2005 (ref 47)
%11: Koopmann & Kenney 2006 (ref 49)
%12: Usui et al 1998 (ref 23)
%13: Domingue et al 2003 (ref 48)
%14: Trinchieri & Di Serego 1991 (ref 13)
%15: Finkelman et al 2010 (ref 59)
%16: Kim 1989 (ref 60)
%17: Martel et al 2004 (ref 61)
%18: Shields 1991 (ref 64)
%19: Singh et al 1995 (ref 63)
%20: Kennicutt & Kent 1983 (ref 1)
%21: Romanishin 1990 (ref 7)
%22: Sanchez-Gallego et al 2012 (ref 65) aggiunta a mano dopo

%\twocolumn
%\clearpage

%\onecolumn
%Last Update MF 9 Nov 2014
\newpage
\tiny
\begin{longtable}{lccccccccc}
\caption{CAS parameters for HRS galaxies}\\
\hline \hline
HRS & $r_e(r)$ & $r_e(\rm{H\alpha})$ & $r_e(\rm{EW_{H\alpha}})$ & $C_r$  &${A}_r$  & ${S}_r$ &${C}_{\rm{H\alpha}}$  &${A}_{\rm{H\alpha}}$  & ${S}_{\rm{H\alpha}}$\\ 
    &  arcsec   &  arcsec & arcsec &      &   &    &   &  & \\
(1) &   (2)       &  (3)   & (4)  & (5) & (6) & (7)& (8)& (9)             &  (10)      \\
\hline                                                                                                                              
\endfirsthead
\multicolumn{10}{l}{\textbf{Table \ref{Tabstruc}.} CAS parameters for HRS galaxies} \\ \\
\hline \hline
HRS & $r_e(r)$ & $r_e(\rm{H\alpha})$ & $r_e(\rm{EW_{H\alpha}})$ & $C_r$  &${A}_r$  & ${S}_r$ &${C}_{\rm{H\alpha}}$  &${A}_{\rm{H\alpha}}$  & ${S}_{\rm{H\alpha}}$ \\ 
    &  arcsec   &  arcsec & arcsec &      &   &    &   &  & \\
(1) &   (2)       &  (3)   & (4)  & (5) & (6) & (7)& (8)& (9)             &  (10)     \\
\hline
\endhead

%This is the footer for all pages except the last page of the table...
 \\ 
 \hline
  \multicolumn{10}{c}{{Continued on next page\ldots}} \\
\endfoot

%This is the footer for the last page of the table...
  \\ 
\hline \hline
\endlastfoot
  1   &      8.37   &	   7.80   &	 9.54	  &   2.66  &	  0.21 &   0.08 &   2.10  &	0.21  &   0.16    \\  
  2   &      8.01   &	   9.00   &	 11.09    &   2.31  &	  0.10 &   0.14 &   1.76  &	0.47  &   0.68    \\  
  5   &      25.00  &	   21.27  &	 20.72    &   2.30  &	  0.04 &   0.74 &   2.39  &	0.08  &   0.47    \\  
  6   &      35.88  &	   27.52  &	 31.11    &   2.82  &	 -0.09 &  -0.10 &   2.94  &    -0.17  &  -1.10    \\  
  8   &      46.10  &	   71.12  &	 82.15    &   3.96  &	  0.08 &   0.14 &   1.76  &	0.19  &   0.54    \\  
  9   &      12.89  &	   9.52   &	 24.43    &   3.66  &	  0.06 &  -0.04 &   4.24  &	0.34  &   0.71    \\  
  10  &      9.56   &	   11.48  &	 17.80    &   2.81  &	  0.10 &   0.11 &   2.44  &	0.47  &   1.01    \\  
  11  &      27.87  &	   27.07  &	 30.30    &   2.37  &	  0.19 &   0.17 &   1.83  &	0.38  &   1.47    \\
\end{longtable}

\normalsize

%\input{HRS_CAS_short.tex}
%\twocolumn
%\clearpage

%\onecolumn
%\input{HRSradio_short.tex}
%\begin{landscape}
\begin{center} \tiny \begin{longtable}{c c c c c c}
\caption{Radio continuum data.\label{Tabradio}}\\
\hline \hline 
\multicolumn{1}{c}{HRS} &
\multicolumn{1}{c}{sgn}&      
\multicolumn{1}{c}{S(20cm)} &
\multicolumn{1}{c}{log L(20cm)} &    
\multicolumn{1}{c}{flag} &     
\multicolumn{1}{c}{ref} \\
\multicolumn{1}{c}{} &
\multicolumn{1}{c}{} &      
\multicolumn{1}{c}{mJy} &
\multicolumn{1}{c}{W Hz$^{-1}$} &   
\multicolumn{1}{c}{} &     
\multicolumn{1}{c}{} \\
\multicolumn{1}{c}{(1)} &
\multicolumn{1}{c}{(2)} &      
\multicolumn{1}{c}{(3)} &    
\multicolumn{1}{c}{(4)} &     
\multicolumn{1}{c}{(5)} &
\multicolumn{1}{c}{(6)} \\
\hline
  \\
\endhead 
%This is the footer for all pages except the last page of the table...
 \\ 
 \hline
  \multicolumn{6}{c}{{Continued on next page\ldots}} \\
\endfoot

%This is the footer for the last page of the table...
  \\ 
\hline \hline
\endlastfoot
           1  &        0 &       1.8  &      19.76  & -   &	  5 \\ 
           2  &        1 &       5.0  &      20.28  & 1   &	  4 \\ 
           3  &        1 &       6.4  &      20.30  & 1   &	  1 \\ 
           4  &        1 &     126.0  &      21.58  & 1   &	  1 \\ 
           5  &        1 &       5.0  &      20.20  & -   &	  5 \\ 
           6  &        0 &       1.8  &      19.86  & -   &	  5 \\ 
           7  &        1 &       6.7  &      20.43  & 1   &	  4 \\ 
           8  &        0 &       1.8  &      19.88  & -   &	  5 \\
\end{longtable}
\end{center}
%\end{landscape}

%\twocolumn
%\clearpage

%\onecolumn
%\input{HRS_SFR_short.tex}
%\begin{landscape}
\begin{center} \tiny \begin{longtable}{c c c c c c c c c}
\caption{Star formation rates.\label{SFR}}\\
\hline \hline 
\multicolumn{1}{c}{HRS} &
\multicolumn{1}{c}{$SFR_{H\alpha + BD}$}&      
\multicolumn{1}{c}{$SFR_{H\alpha + 24\mu m}$} &    
\multicolumn{1}{c}{$SFR_{FUV + 24\mu m}$} &     
\multicolumn{1}{c}{$SFR_{radio}$}  &
\multicolumn{1}{c}{$SFR_{MED}$} & 
\multicolumn{1}{c}{$SFR_{1exp}$} &   
\multicolumn{1}{c}{$SFR_{2exp}$}      & 
\multicolumn{1}{c}{$SFR_{del}$}     \\
\multicolumn{1}{c}{} &
\multicolumn{1}{c}{M$_{\odot}$ yr$^{-1}$} &      
\multicolumn{1}{c}{M$_{\odot}$ yr$^{-1}$} &    
\multicolumn{1}{c}{M$_{\odot}$ yr$^{-1}$} &     
\multicolumn{1}{c}{M$_{\odot}$ yr$^{-1}$} &    
\multicolumn{1}{c}{M$_{\odot}$ yr$^{-1}$} & 
\multicolumn{1}{c}{M$_{\odot}$ yr$^{-1}$} & 
\multicolumn{1}{c}{M$_{\odot}$ yr$^{-1}$} &   
\multicolumn{1}{c}{M$_{\odot}$ yr$^{-1}$}        \\
\multicolumn{1}{c}{(1)} &
\multicolumn{1}{c}{(2)}&      
\multicolumn{1}{c}{(3)} &    
\multicolumn{1}{c}{(4)} &     
\multicolumn{1}{c}{(5)} &    
\multicolumn{1}{c}{(6)}  & 
\multicolumn{1}{c}{(7)} &   
\multicolumn{1}{c}{(8)} &
\multicolumn{1}{c}{(9)}     \\
\hline
  \\
\endhead 
%This is the footer for all pages except the last page of the table...
 \\ 
 \hline
  \multicolumn{9}{c}{{Continued on next page\ldots}} \\
\endfoot

%This is the footer for the last page of the table...
  \\ 
\hline \hline
\endlastfoot
           1 &    -      &    0.0501 &    -       &   -       &   0.0501  &   -       &   -       &   -      \\ 
           2 &    0.2801 &    0.1808 &    0.1661  &   0.2560  &   0.2208  &   0.2169  &   0.2425  &   0.2378 \\ 
           4 &    -      &    -      &    -       &   2.4262  &   2.4262  &   -       &   -       &   -      \\ 
           5 &    -      &    0.0987 &    -       &   -       &   0.0987  &   -       &   -       &   -      \\ 
           6 &    -      &    -      &    0.0393  &   -       &   0.0393  &   0.0357  &   0.0384  &   0.0491 \\ 
           8 &    -      &    -      &    0.4053  &   -       &   0.4053  &   0.4002  &   0.4167  &   0.4939 \\ 
           9 &    -      &    -      &    0.1838  &   0.2765  &   0.2301  &   0.1798  &   0.1208  &   0.1811 \\ 
          10 &    0.1566 &    0.1542 &    -       &   0.1737  &   0.1615  &   -       &   -       &   -      \\
\end{longtable}
\end{center}
%\end{landscape}

%\twocolumn
%\clearpage

%\onecolumn
%\begin{landscape}
\begin{center} \tiny \begin{longtable}{c c c c c c c c c c c c}
\caption{Fluxes of the emission lines normalised to H$\alpha$ determined using GANDALF.\label{GANDALF}}\\
\hline \hline 
\multicolumn{1}{c}{HRS} &
\multicolumn{1}{c}{[OII]} &    
\multicolumn{1}{c}{H$\beta$} &    
\multicolumn{1}{c}{[OIII]}  & 
\multicolumn{1}{c}{H$\alpha$} &   
\multicolumn{1}{c}{[NII]}      & 
\multicolumn{1}{c}{[SII]}     &
\multicolumn{1}{c}{[SII]}   &
\multicolumn{1}{c}{S/N(H$\alpha$)}&
\multicolumn{1}{c}{S/N(H$\beta$)} &
\multicolumn{1}{c}{$C(H\beta)$}&      
\multicolumn{1}{c}{$\sigma[C(H\beta)]$} \\      
\multicolumn{1}{c}{}       &
\multicolumn{1}{c}{$\lambda$3727} &    
\multicolumn{1}{c}{$\lambda$4861} &    
\multicolumn{1}{c}{$\lambda$5007} & 
\multicolumn{1}{c}{$\lambda$6563} &   
\multicolumn{1}{c}{$\lambda$6584} & 
\multicolumn{1}{c}{$\lambda$6716} &   
\multicolumn{1}{c}{$\lambda$6731} & 
\multicolumn{1}{c}{}       &
\multicolumn{1}{c}{}       &
\multicolumn{1}{c}{}       &
\multicolumn{1}{c}{}       \\
\multicolumn{1}{c}{(1)} &
\multicolumn{1}{c}{(2)} &      
\multicolumn{1}{c}{(3)} &    
\multicolumn{1}{c}{(4)} &     
\multicolumn{1}{c}{(5)} &    
\multicolumn{1}{c}{(6)} & 
\multicolumn{1}{c}{(7)} &   
\multicolumn{1}{c}{(8)} & 
\multicolumn{1}{c}{(9)} &  
\multicolumn{1}{c}{(10)} & 
\multicolumn{1}{c}{(11)} &
\multicolumn{1}{c}{(12)}     \\
\hline
  \\
\endhead 
%This is the footer for all pages except the last page of the table...
 \\ 
 \hline
  \multicolumn{9}{c}{{Continued on next page\ldots}} \\
\endfoot

%This is the footer for the last page of the table...
  \\ 
\hline \hline
\endlastfoot
% HRS   [OII]   Hb   &  [OIII]5007   Ha   &  NII   & SII_100  SII_1a0  AoN_Ha	AoN_Hb        CHb     SCHb    
  1  &  0.41 &  0.17 &  0.07   &     1.0  &  0.37  & 0.25 &   0.17  &  50.64  & 11.91   &    1.00  & 	0.13  \\
  2  &  0.48 &  0.22 &  0.09   &     1.0  &  0.32  & 0.20 &   0.14  &  108.46 & 31.77   &    0.64  & 	0.05  \\
  4  &  -    &  -    &  -      &     -    &  -     & -    &   -     &  -      & -       &    -     & 	-     \\
  5  &  0.65 &  0.00 &  0.00   &     1.0  &  0.89  & 0.62 &   0.52  &  12.78  & -       &    -     & 	-     \\
  10 &  0.84 &  0.28 &  0.24   &     1.0  &  0.24  & 0.21 &   0.14  &  91.33  & 34.49   &    0.28  & 	0.05  \\
  11 &  0.40 &  0.22 &  0.12   &     1.0  &  0.27  & 0.25 &   0.18  &  84.19  & 25.35   &    0.64  & 	0.06  \\
  12 &  0.86 &  0.25 &  0.32   &     1.0  &  0.18  & 0.24 &   0.18  &  97.14  & 33.10   &    0.46  & 	0.05  \\
  13 &  0.67 &  0.20 &  0.04   &     1.0  &  0.39  & 0.21 &   0.16  &  48.13  & 12.70   &    0.79  & 	0.12  \\
\end{longtable}
\end{center}
%\end{landscape}

%\input{HRS_GANDALF_short.tex}
%\twocolumn
\clearpage

\begin{table*}
\caption{Comparison with the literature}
\label{comparisontab}
{%\scriptsize
\[
\begin{tabular}{ccccc}
\hline
\hline
\noalign{\smallskip}
%\hline
Sample			&  H$\alpha$+[NII]E.W.$_{T.W.}$/H$\alpha$+[NII]E.W.$_{Lit.}$ 	& N. obj.& $F(H\alpha+[NII]_{T.W.})$/$F(H\alpha+[NII]_{Lit.})$ & N. obj. \\
\hline
All			& 1.16$\pm$0.90							& 116	& 1.07$\pm$0.82						& 191	\\
This work		& 1.22$\pm$0.63							& 8	& 1.07$\pm$0.10						& 4	\\
Our previous works 	& 1.68$\pm$1.02							& 3	& 1.21$\pm$0.53						& 3	\\
Kennicutt \& Kent (1983)$^a$& 1.17$\pm$0.81						& 29	& 1.02$\pm$0.39						& 20	\\
Romanishin (1990)	& 1.05$\pm$0.33							& 17	& 0.96$\pm$0.25						& 15	\\
Sanchez-Gallego et al. (2012) & 1.05$\pm$0.97						& 29	& 1.12$\pm$0.87						& 30	\\
\noalign{\smallskip}
\hline
\end{tabular}
\] }
Note: $a$ the Kennicutt \& Kent (1983) fluxes and equivalent widths have been multiplied by a factor of 1.16 to take into account a possible
contamination of telluric absorption features in the red continuum as suggested by Kennicutt et al. (1994).
\end{table*}

\begin{table*}
\caption{Bivariate fit of the luminosity-luminosity relations }
\label{calibrazione}
{%\scriptsize
\[
\begin{tabular}{cccccccc}
\hline
\hline
\noalign{\smallskip}
%\hline
%\multicolumn{3}{c}{}				&\multicolumn{1}{c}{}		&{c}				&\multicolumn{1}{c}{}\\
x					&	y			& sample			&$a$	& $b$	& $\rho$& $\sigma$	& N  \\
\hline
%log$L(H\alpha)_{BD}(BOS13)$			&log$L(H\alpha)_{24\mu m}(BOS13)$& all				&1.02	&-1.19	& 0.84 	& 0.21		& 151 \\ 
%						&				& H$\beta$E.W.$>$5\AA		&1.17	&-6.77	& 0.94	& 0.11		& 61 \\  
%log$L(H\alpha)_{BD}(BOS13)$			& log$L(20 cm)$			& all				&1.14	&-26.00	& 0.73 	& 0.23		& 90 \\ %
%						&				& H$\beta$E.W.$>$5\AA		&1.33	&-33.45	& 0.88	& 0.14		& 43 \\ %
%log$L(H\alpha)_{24\mu m}(BOS13)$		& 				& all				&1.18	&-27.31	& 0.90 	& 0.14		& 87 \\ %
%						&				& H$\beta$E.W.$>$5\AA		&1.19	&-28.03	& 0.88	& 0.13		& 43 \\ %
\hline
log$L(H\alpha)_{BD}$			&log$L(H\alpha)_{24\mu m}$      & all   			&1.02	&-1.08	& 0.90 	& 0.17		& 152 \\    %
					&				& $\sigma[C(H\beta)]\leq$ 0.1 	&1.16	&-6.45	& 0.95	& 0.10		& 69 \\     %
log$L(H\alpha)_{BD}$			& log$L(20 cm)$			& all				&1.18	&-27.58	& 0.80 	& 0.20		& 93 \\ %   %
					&				& $\sigma[C(H\beta)]\leq$ 0.1	&1.36	&-34.80	& 0.84	& 0.14		& 46 \\ %   %
log$L(H\alpha)_{24\mu m}$		& 				& all				&1.17	&-27.09	& 0.89 	& 0.15		& 95 \\ %   %
					&				& $\sigma[C(H\beta)]\leq$ 0.1	&1.20	&-28.27	& 0.91	& 0.11		& 43 \\ %   %
\noalign{\smallskip}
\hline
\end{tabular}
\]
Note: all fits are done using only galaxies with $HI-def$ $\leq$ 0.4 (Figs. \ref{hacal} and \ref{radiocal}). }
\end{table*}

\begin{table*}
\caption{Bivariate fit of the relations between the different star formation tracers  }
\label{sfrsfr}
{%\scriptsize
\[
\begin{tabular}{ccccccc}
\hline
\hline
\noalign{\smallskip}
%\hline
%\multicolumn{3}{c}{}	&\multicolumn{1}{c}{}		&\multicolumn{1}{c}{}\\
x			&	y			&  $a$	& $b$	& $\rho$& $\sigma$	& N  \\
\hline
log$SFR_{H\alpha +BD}^a$  & log$SFR_{H\alpha +24\mu m}$	& 1.14	& 0.00	& 0.94 	& 0.09		& 54 \\ %
			& log$SFR_{radio}$		& 1.05	&-0.03  & 0.84	& 0.13		& 46 \\ %
			& log$SFR_{FUV +24\mu m}$	& 1.10  &-0.08  & 0.87	& 0.11		& 49 \\ %
%			& log$SFR_{NUV +24\mu m}$	& 1.00  &-0.05  & 0.85	& 0.14		& 46 \\
log$SFR_{FUV +24\mu m}$	& log$SFR_{H\alpha +24\mu m}$	& 1.10	& 0.08	& 0.97	& 0.07		& 109\\ %
			& log$SFR_{radio}$		& 1.00  & 0.03	& 0.88  & 0.13		& 90 \\	%
%			& log$SFR_{H\alpha +BD}^a$	& 0.95  & 0.10  & 0.85  & 0.13		& 45 \\
%			& log$SFR_{NUV +24\mu m}$	& 0.96	& 0.04  & 0.99  & 0.03		& 114\\		
\noalign{\smallskip}
\hline
\end{tabular}
\]
Note: all fits are done using only galaxies with $HI-def$ $\leq$ 0.4. $a$ determined only on galaxies with {\sc GANDALF} reduced spectra and $\sigma[C(H\beta)]\leq$ 0.1 (Fig. \ref{sfrha}). }
\end{table*}

\begin{table*}
\caption{Bivariate fit of the relations between the different H$\alpha$ and FUV attenuations }
\label{ahacortab}
{%\scriptsize
\[
\begin{tabular}{ccccccccc}
\hline
\hline
\noalign{\smallskip}
%\hline
%\multicolumn{3}{c}{}	&\multicolumn{1}{c}{}		&{c} 			& correction		&\multicolumn{1}{c}{}\\
x			&	y			& sample		& 			& $a$	& $b$	& $\rho$& $\sigma$	& N  \\
\hline
$A(H\alpha)_{BD}$       & $A(H\alpha)_{24\mu m}$        & all			& Calzetti et al. 2010	& 0.67	& 0.08	& 0.61 	& 0.36		& 119 \\ %
			&				&$\sigma[C(H\beta)]\leq$0.1&			& 1.07	&-0.15	& 0.55	& 0.24		& 54  \\ %
			&				& all			& Calzetti et al. 2007	& 0.60	& 0.25	& 0.67	& 0.31		& 119 \\ %
			&				&$\sigma[C(H\beta)]\leq$0.1&			& 0.97	& 0.03	& 0.65	& 0.20		& 54  \\ %
			&				& all			& Kennicutt et al. 2009 & 0.51	& 0.12	& 0.67	& 0.27		& 119 \\ %
			&				&$\sigma[C(H\beta)]\leq$0.1&			& 0.79	&-0.05	& 0.65	& 0.18		& 54  \\ %
$A(FUV)_{24\mu m}$	& $A(H\alpha)_{24\mu m}$        &all			& Calzetti et al. 2010	& 0.65	& 0.09	& 0.86	& 0.21		& 109 \\ %
			&				&$\sigma[C(H\beta)]\leq$0.1&			& 0.66	& 0.05	& 0.79	& 0.21		& 49  \\ %
			&				&all			& Calzetti et al. 2007	& 0.59	& 0.24	& 0.86	& 0.19		& 109 \\ %			
			&				&$\sigma[C(H\beta)]\leq$0.1&   			& 0.59	& 0.21	& 0.83	& 0.17		& 49  \\ %
			&				&all			& Kennicutt et al. 2009 & 0.49	& 0.12	& 0.86	& 0.17		& 109 \\ %
			&				&$\sigma[C(H\beta)]\leq$0.1&   			& 0.47	& 0.11	& 0.83	& 0.14		& 49  \\ %
\noalign{\smallskip}
\hline
\end{tabular}
\]
Note: all fits are done using only galaxies with $HI-def$ $\leq$ 0.4 (Fig. \ref{AHacor}). }
\end{table*}

%\clearpage
\begin{table*}
\caption{The star formation rate properties as a function of the morphological type} 
\label{type}
{%\tiny
\[
\begin{tabular}{ccccccccccccc}
\hline
\noalign{\smallskip}
\hline
Gas	               	& Sample		& Sa 		 	& Sab 		 	& Sb 			 & Sbc 			 & Sc 			 & Scd 		 	& Sd		   & Sdm 	     & Sm 	 	& Im              & Pec \\
\hline
log$SSFR_{H\alpha}$    	&  $HI-def$ $\leq$0.4	&-10.22$\pm$0.75(8)	& -10.64$\pm$0.55(4)	& -9.94$\pm$0.41(20)	 & -10.25$\pm$0.35(19)	 & -10.08$\pm$0.29(32)	 & -9.95$\pm$0.38(22)   &-9.75$\pm$0.21(11)&-9.74$\pm$0.15(6) &-9.73$\pm$0.13(4) &-9.90$\pm$0.39(7)&-9.62$\pm$0.25(4)\\
\hline
log$SSFR_{H\alpha}$    	&  $HI-def$ $>$0.4	&-10.70$\pm$0.57(14)	& -10.71$\pm$0.46(14)	&-10.51$\pm$0.47(26)	 & -10.34$\pm$0.36(11)	 & -10.41$\pm$0.48(7)	 &-10.16$\pm$0.28(6)    &-10.74$\pm$0.03(3)&- 		      &-		 &-		   &-10.87$\pm$0.61(4) \\
\noalign{\smallskip}
\hline
\end{tabular}
\]
%\scriptsize
Note: mean values and standard deviations correspond to the big blue filled dots ($HI-def$ $\leq$0.4) and big red empty circles ($HI-def$ $>$0.4) in Fig. \ref{sfrtype}. For each morphological class, the number in parenthesis gives the number of objects used to determine the mean values. }
\end{table*}

\clearpage

\begin{table*}
\caption{Coefficients of the scaling relations: y = ax + b }
\label{fit}
{%\scriptsize
\[
\begin{tabular}{ccccccccc}
\hline
\noalign{\smallskip}
\hline
y		&x			& fit		&	sample         	& a     & b     & $\rho$& $\sigma$ &	N.obj\\
\hline	
$SFR_{MED}$	& $M_{star}$		& linear	& $HI-def$ $\leq$0.4    &0.59   &-6.10  & 0.64	& 0.30	& 138	\\ % a
		&			& bisector	&			&0.88	&-8.84	& 0.64	& 0.29	& 138	\\
		& $M_{star}$		& linear	& $HI-def$ $>$0.4    	&0.58   &-6.44  & 0.66	& 0.36	& 89	\\ 
		&			& bisector	&			&0.90	&-9.55	& 0.66	& 0.34	& 89	\\
$SSFR_{MED}$	& $M_{star}$		& bisector	& $HI-def$ $\leq$0.4    &-0.79	&-2.46	& -0.53	& 0.32  & 138	\\ %
		& $\mu_{star}$		& bisector	& $HI-def$ $\leq$0.4    &-0.96  &-2.49  & -0.29 & 0.35  & 138	\\
		& 12+log(O/H)		& bisector	& $HI-def$ $\leq$0.4    &-2.00  & 7.20  & -0.44 & 0.15  & 138	\\					
\noalign{\smallskip}
\hline
\end{tabular}
\]
Notes: 
$\rho$ is the Spearman correlation coefficient, $\sigma$ the dispersion in the relations (solid lines in Figs. \ref{main} and \ref{scaling}).\\
}
\end{table*}

%\clearpage
\begin{table*}
\caption{Average scaling relations }
\label{Tabscaling}
{%\scriptsize
\[
\begin{tabular}{cccccc}
\hline
\hline
\noalign{\smallskip}
%\hline
\multicolumn{3}{c}{}&\multicolumn{1}{c}{}&\multicolumn{1}{c}{}&\multicolumn{1}{c}{}\\
y		&	x	&  sample	&$<x>$ 		&	 $<y>$    &  N  \\
\hline
log$SFR_{MED}$	& log$M_{star}$	&$HI-def \leq$0.4&10.72$\pm$0.21	&  0.33$\pm$0.30 & 8    \\ %
		&		&		 &10.23$\pm$0.15	& -0.12$\pm$0.51 & 18	\\ %
		&		&		 & 9.76$\pm$0.16	& -0.37$\pm$0.28 & 34	\\ %
		&		&		 & 9.25$\pm$0.14	& -0.63$\pm$0.36 & 59	\\ %
		&		&		 & 8.79$\pm$0.14	& -0.94$\pm$0.30 & 18	\\ %
\hline
log$SFR_{MED}$	& log$M_{star}$	&$HI-def >$0.4   &10.81$\pm$0.13	& -0.19$\pm$0.44 & 11   \\
\hline
\end{tabular}
\]
Note: mean values and standard deviations for the scaling relations (big symbols in Fig. \ref{main} and \ref{scaling}). }
\end{table*}


\begin{thebibliography}{}

\bibitem[Argence \& Lamareille(2009)]{2009A&A...495..759A} Argence, B., \& Lamareille, F.\ 2009, \aap, 495, 759 
\bibitem[Bakes \& Tielens(1994)]{1994ApJ...427..822B} Bakes, E.~L.~O., \& Tielens, A.~G.~G.~M.\ 1994, \apj, 427, 822 
\bibitem[Bauer et al.(2005)]{2005ApJ...621L..89B} Bauer, A.~E., Drory, N., Hill, G.~J., \& Feulner, G.\ 2005, \apjl, 621, L89 
\bibitem[Becker et al.(1995)]{1995ApJ...450..559B} Becker, R.~H., White, R.~L., \& Helfand, D.~J.\ 1995, \apj, 450, 559 
\bibitem[Bell(2003)]{2003ApJ...586..794B} Bell, E.~F.\ 2003, \apj, 586, 794 
\bibitem[Bell et al.(2005)]{2005ApJ...625...23B} Bell, E.~F., Papovich, C., Wolf, C., et al.\ 2005, \apj, 625, 23 
\bibitem[Bendo et al.(2010)]{2010A&A...518L..65B} Bendo, G.~J., Wilson, C.~D., Pohlen, M., et al.\ 2010, \aap, 518, LL65 
\bibitem[Bendo et al.(2012)]{2012MNRAS.419.1833B} Bendo, G.~J., Boselli, A., Dariush, A., et al.\ 2012, \mnras, 419, 1833 
\bibitem[Bigiel et al.(2008)]{2008AJ....136.2846B} Bigiel, F., Leroy, A., Walter, F., et al.\ 2008, \aj, 136, 2846 
\bibitem[Binggeli et al.(1985)]{1985AJ.....90.1681B} Binggeli, B., Sandage, A., \& Tammann, G.~A.\ 1985, \aj, 90, 1681 
\bibitem[Boissier(2013)]{2013pss6.book..141B} Boissier, S.\ 2013, Planets, Stars and Stellar Systems.~Volume 6: Extragalactic Astronomy and Cosmology, 141 
\bibitem[Boselli(2011)]{2011pvg..book.....B} Boselli, A.\ 2011, A Panchromatic View of Galaxies, by Alessandro Boselli.~- Practical Approach Book - ISBN-10: 3-527-40991-2.~ISBN-13: 978-3-527-40991-4 - Wiley-VCH, Berlin 2011.~XVI, 324pp, Hardcover,  
\bibitem[Boselli \& Gavazzi(2002)]{2002A&A...386..124B} Boselli, A., \& Gavazzi, G.\ 2002, \aap, 386, 124 
\bibitem[Boselli \& Gavazzi(2006)]{2006PASP..118..517B} Boselli, A., \& Gavazzi, G.\ 2006, \pasp, 118, 517 
\bibitem[Boselli \& Gavazzi(2014)]{2014A&ARv..22...74B} Boselli, A., \& Gavazzi, G.\ 2014, \aapr, 22, 74 
\bibitem[Boselli et al.(2001)]{2001AJ....121..753B} Boselli, A., Gavazzi, G., Donas, J., \& Scodeggio, M.\ 2001, \aj, 121, 753 
\bibitem[Boselli et al.(2002a)]{2002A&A...386..134B} Boselli, A., Iglesias-P{\'a}ramo, J., V{\'{\i}}lchez, J.~M., \& Gavazzi, G.\ 2002a, \aap, 386, 134 
\bibitem[Boselli et al.(2002b)]{2002A&A...384...33B} Boselli, A., Lequeux, J., \& Gavazzi, G.\ 2002b, \aap, 384, 33 
\bibitem[Boselli et al.(2005)]{2005ApJ...629L..29B} Boselli, A., Cortese, L., Deharveng, J.~M., et al.\ 2005, \apjl, 629, L29 
\bibitem[Boselli et al.(2006)]{2006ApJ...651..811B} Boselli, A., Boissier, S., Cortese, L., et al.\ 2006, \apj, 651, 811 
\bibitem[Boselli et al.(2008a)]{2008ApJ...674..742B} Boselli, A., Boissier, S., Cortese, L., \& Gavazzi, G.\ 2008, \apj, 674, 742 
\bibitem[Boselli et al.(2008b)]{2008A&A...489.1015B} Boselli, A., Boissier, S., Cortese, L., \& Gavazzi, G.\ 2008, \aap, 489, 1015 
\bibitem[Boselli et al.(2009)]{2009ApJ...706.1527B} Boselli, A., Boissier, S., Cortese, L., et al.\ 2009, \apj, 706, 1527 
\bibitem[Boselli et al.(2010)]{2010PASP..122..261B} Boselli, A., Eales, S., Cortese, L., et al.\ 2010, \pasp, 122, 261 
\bibitem[Boselli et al.(2011)]{2011A&A...528A.107B} Boselli, A., Boissier, S., Heinis, S., et al.\ 2011, \aap, 528, AA107 
\bibitem[Boselli et al.(2012)]{2012A&A...540A..54B} Boselli, A., Ciesla, L., Cortese, L., et al.\ 2012, \aap, 540, AA54 
\bibitem[Boselli et al.(2013)]{2013A&A...550A.114B} Boselli, A., Hughes, T.~M., Cortese, L., Gavazzi, G., \& Buat, V.\ 2013, \aap, 550, A114 
\bibitem[Boselli et al.(2014)]{2014A&A...564A..65B} Boselli, A., Cortese, L., \& Boquien, M.\ 2014a, \aap, 564, AA65 
\bibitem[Boselli et al.(2014b)]{2014A&A...564A..66B} Boselli, A., Cortese, L., Boquien, M., et al.\ 2014b, \aap, 564, AA66 
\bibitem[Boselli et al.(2014c)]{2014A&A...564A..67B} Boselli, A., Cortese, L., Boquien, M., et al.\ 2014c, \aap, 564, AA67 
\bibitem[Boselli et al.(2014d)]{2014A&A...570A..69B} Boselli, A., Voyer, E., Boissier, S., et al.\ 2014d, \aap, 570, AA69 
\bibitem[Boquien et al.(2011)]{2011AJ....142..111B} Boquien, M., Calzetti, D., Combes, F., et al.\ 2011, \aj, 142, 111 
\bibitem[Boquien et al.(2014)]{2014A&A...571A..72B} Boquien, M., Buat, V., \& Perret, V.\ 2014, \aap, 571, AA72 
\bibitem[Brinchmann \& Ellis(2000)]{2000ApJ...536L..77B} Brinchmann, J., \& Ellis, R.~S.\ 2000, \apjl, 536, L77 
\bibitem[Brinchmann et al.(2004)]{2004MNRAS.351.1151B} Brinchmann, J., Charlot, S., White, S.~D.~M., et al.\ 2004, \mnras, 351, 1151 
\bibitem[Bruzual \& Charlot(2003)]{2003MNRAS.344.1000B} Bruzual, G., \& Charlot, S.\ 2003, \mnras, 344, 1000 
\bibitem[Buat et al.(2002)]{2002A&A...383..801B} Buat, V., Boselli, A., Gavazzi, G., \& Bonfanti, C.\ 2002, \aap, 383, 801 
\bibitem[Buat et al.(2014)]{2014A&A...561A..39B} Buat, V., Heinis, S., Boquien, M., et al.\ 2014, \aap, 561, AA39 
\bibitem[Caldwell et al.(1991)]{1991ApJ...370..526C} Caldwell, N., Kennicutt, R., Phillips, A.~C., \& Schommer, R.~A.\ 1991, \apj, 370, 526 
\bibitem[Calzetti(2001)]{2001PASP..113.1449C} Calzetti, D.\ 2001, \pasp, 113, 1449 
\bibitem[Calzetti et al.(2007)]{2007ApJ...666..870C} Calzetti, D., Kennicutt, R.~C., Engelbracht, C.~W., et al.\ 2007, \apj, 666, 870 
\bibitem[Calzetti et al.(2010)]{2010ApJ...714.1256C} Calzetti, D., Wu, S.-Y., Hong, S., et al.\ 2010, \apj, 714, 1256 
\bibitem[Cappellari \& Emsellem(2004)]{2004PASP..116..138C} Cappellari, M., \& Emsellem, E.\ 2004, \pasp, 116, 138 
\bibitem[Chabrier(2003)]{2003PASP..115..763C} Chabrier, G.\ 2003, \pasp, 115, 763 
\bibitem[Chomiuk \& Povich(2011)]{2011AJ....142..197C} Chomiuk, L., \& Povich, M.~S.\ 2011, \aj, 142, 197 
\bibitem[Ciesla et al.(2012)]{2012A&A...543A.161C} Ciesla, L., Boselli, A., Smith, M.~W.~L., et al.\ 2012, \aap, 543, AA161 
\bibitem[Ciesla et al.(2014)]{2014A&A...565A.128C} Ciesla, L., Boquien, M., Boselli, A., et al.\ 2014, \aap, 565, AA128 
\bibitem[Ciesla et al.(2015)]{2015arXiv150103672C} Ciesla, L., Charmandaris, V., Georgakakis, A., et al.\ 2015, arXiv:1501.03672 
\bibitem[Condon(1987)]{1987ApJS...65..485C} Condon, J.~J.\ 1987, \apjs, 65, 485 
\bibitem[Condon(1992)]{1992ARA&A..30..575C} Condon, J.~J.\ 1992, \araa, 30, 575 
\bibitem[Condon et al.(1987)]{1987ApJS...65..543C} Condon, J.~J., Yin, Q.~F., \& Burstein, D.\ 1987, \apjs, 65, 543 
\bibitem[Condon et al.(1990)]{1990ApJS...73..359C} Condon, J.~J., Helou, G., Sanders, D.~B., \& Soifer, B.~T.\ 1990, \apjs, 73, 359 
\bibitem[Condon et al.(1991)]{1991ApJ...376...95C} Condon, J.~J., Anderson, M.~L., \& Helou, G.\ 1991, \apj, 376, 95 
\bibitem[Condon et al.(1998)]{1998AJ....115.1693C} Condon, J.~J., Cotton, W.~D., Greisen, E.~W., et al.\ 1998, \aj, 115, 1693 
\bibitem[Condon et al.(2002)]{2002AJ....124..675C} Condon, J.~J., Cotton, W.~D., \& Broderick, J.~J.\ 2002, \aj, 124, 675 
\bibitem[Conselice(2003)]{2003ApJS..147....1C} Conselice, C.~J.\ 2003, \apjs, 147, 1 
\bibitem[Conselice(2014)]{2014ARA&A..52..291C} Conselice, C.~J.\ 2014, \araa, 52, 291 
\bibitem[Conselice et al.(2000)]{2000A&A...354L..21C} Conselice, C.~J., Bershady, M.~A., \& Gallagher, J.~S., III 2000, \aap, 354, L21 
\bibitem[Cortese \& Hughes(2009)]{2009MNRAS.400.1225C} Cortese, L., \& Hughes, T.~M.\ 2009, \mnras, 400, 1225 
\bibitem[Cortese et al.(2008)]{2008MNRAS.386.1157C} Cortese, L., Boselli, A., Franzetti, P., et al.\ 2008, \mnras, 386, 1157 
\bibitem[Cortese et al.(2011)]{2011MNRAS.415.1797C} Cortese, L., Catinella, B., Boissier, S., Boselli, A., \& Heinis, S.\ 2011, \mnras, 415, 1797 
\bibitem[Cortese et al.(2012a)]{2012A&A...544A.101C} Cortese, L., Boissier, S., Boselli, A., et al.\ 2012a, \aap, 544, A101 
\bibitem[Cortese et al.(2012b)]{2012A&A...540A..52C} Cortese, L., Ciesla, L., Boselli, A., et al.\ 2012b, \aap, 540, AA52 
\bibitem[Cortese et al.(2014)]{2014MNRAS.440..942C} Cortese, L., Fritz, J., Bianchi, S., et al.\ 2014, \mnras, 440, 942 
\bibitem[da Cunha et al.(2008)]{2008MNRAS.388.1595D} da Cunha, E., Charlot, S., \& Elbaz, D.\ 2008, \mnras, 388, 1595 
\bibitem[Daddi et al.(2007)]{2007ApJ...670..156D} Daddi, E., Dickinson, M., Morrison, G., et al.\ 2007, \apj, 670, 156 
\bibitem[da Silva et al.(2014)]{2014MNRAS.444.3275D} da Silva, R.~L., Fumagalli, M., \& Krumholz, M.~R.\ 2014, \mnras, 444, 3275 
\bibitem[de Jong et al.(1985)]{1985A&A...147L...6D} de Jong, T., Klein, U., Wielebinski, R., \& Wunderlich, E.\ 1985, \aap, 147, L6 
\bibitem[Domingue et al.(2003)]{2003AJ....125..555D} Domingue, D.~L., Sulentic, J.~W., Xu, C., et al.\ 2003, \aj, 125, 555 
\bibitem[Draine \& Li(2007)]{2007ApJ...657..810D} Draine, B.~T., \& Li, A.\ 2007, \apj, 657, 810 
\bibitem[Draine et al.(2007)]{2007ApJ...663..866D} Draine, B.~T., Dale, D.~A., Bendo, G., et al.\ 2007, \apj, 663, 866 
\bibitem[Dreyer(1895)]{1895MmRAS..51..185D} Dreyer, J.~L.~E.\ 1895, \memras, 51, 185 
\bibitem[Dreyer(1888)]{1888MmRAS..49....1D} Dreyer, J.~L.~E.\ 1888, \memras, 49, 1 
\bibitem[Elbaz et al.(2007)]{2007A&A...468...33E} Elbaz, D., Daddi, E., Le Borgne, D., et al.\ 2007, \aap, 468, 33 
\bibitem[Falc{\'o}n-Barroso et al.(2006)]{2006MNRAS.369..529F} Falc{\'o}n-Barroso, J., Bacon, R., Bureau, M., et al.\ 2006, \mnras, 369, 529 
\bibitem[Finkelman et al.(2010)]{2010MNRAS.407.2475F} Finkelman, I., Brosch, N., Funes, J.~G., Kniazev, A.~Y., Vaumlis{\"a}nen, P.\ 2010, \mnras, 407, 2475 
\bibitem[Fitzpatrick \& Massa(2007)]{2007ApJ...663..320F} Fitzpatrick, E.~L., \& Massa, D.\ 2007, \apj, 663, 320 
\bibitem[Fossati et al.(2013)]{2013A&A...553A..91F} Fossati, M., Gavazzi, G., Savorgnan, G., et al.\ 2013, \aap, 553, A91 
\bibitem[Fumagalli et al.(2011)]{2011ApJ...741L..26F} Fumagalli, M., da Silva, R.~L., \& Krumholz, M.~R.\ 2011, \apjl, 741, LL26 
\bibitem[Gallego et al.(1995)]{1995ApJ...455L...1G} Gallego, J., Zamorano, J., Aragon-Salamanca, A., \& Rego, M.\ 1995, \apjl, 455, L1 
\bibitem[Gavazzi \& Boselli(1999a)]{1999A&A...343...93G} Gavazzi, G., \& Boselli, A.\ 1999, \aap, 343, 93 
\bibitem[Gavazzi \& Boselli(1999b)]{1999A&A...343...86G} Gavazzi, G., \& Boselli, A.\ 1999, \aap, 343, 86 
\bibitem[Gavazzi et al.(1991)]{1991AJ....101.1207G} Gavazzi, G., Boselli, A., \& Kennicutt, R.\ 1991, \aj, 101, 1207 
\bibitem[Gavazzi et al.(1998)]{1998AJ....115.1745G} Gavazzi, G., Catinella, B., Carrasco, L., Boselli, A., \& Contursi, A.\ 1998, \aj, 115, 1745 
\bibitem[Gavazzi et al.(1999)]{1999MNRAS.304..595G} Gavazzi, G., Boselli, A., Scodeggio, M., Pierini, D., \& Belsole, E.\ 1999, \mnras, 304, 595 
%\bibitem[Gavazzi et al.(2002)]{2002A&A...396..449G} Gavazzi, G., Boselli, A., Pedotti, P., Gallazzi, A., \& Carrasco, L.\ 2002, \aap, 396, 449 
\bibitem[Gavazzi et al.(2002)]{2002A&A...386..114G} Gavazzi, G., Boselli, A., Pedotti, P., Gallazzi, A., \& Carrasco, L.\ 2002, \aap, 386, 114 
\bibitem[Gavazzi et al.(2003)]{2003A&A...400..451G} Gavazzi, G., Boselli, A., Donati, A., Franzetti, P., \& Scodeggio, M.\ 2003, \aap, 400, 451 
\bibitem[Gavazzi et al.(2006)]{2006A&A...446..839G} Gavazzi, G., Boselli, A., Cortese, L., et al.\ 2006, \aap, 446, 839 
\bibitem[Gavazzi et al.(2012)]{2012A&A...545A..16G} Gavazzi, G., Fumagalli, M., Galardo, V., et al.\ 2012, \aap, 545, A16 
\bibitem[Gavazzi et al.(2013a)]{2013A&A...553A..89G} Gavazzi, G., Fumagalli, M., Fossati, M., et al.\ 2013a, \aap, 553, AA89 
\bibitem[Gavazzi et al.(2013b)]{2013A&A...553A..90G} Gavazzi, G., Savorgnan, G., Fossati, M., et al.\ 2013b, \aap, 553, AA90 
\bibitem[Gavazzi et al.(2015a)]{} Gavazzi, G., Consolandi, G., Dotti M., et al. 2015a, A\&A, submitted
\bibitem[Gavazzi et al.(2015b)]{} Gavazzi, G., Consolandi, G., Viscardi E., et al. 2015b, A\&A, in press
\bibitem[Groves et al.(2012)]{2012MNRAS.419.1402G} Groves, B., Brinchmann, J., \& Walcher, C.~J.\ 2012, \mnras, 419, 1402 
\bibitem[Gunawardhana et al.(2013)]{2013MNRAS.433.2764G} Gunawardhana, M.~L.~P., Hopkins, A.~M., Bland-Hawthorn, J., et al.\ 2013, \mnras, 433, 2764 
\bibitem[Guzm{\'a}n et al.(1997)]{1997ApJ...489..559G} Guzm{\'a}n, R., Gallego, J., Koo, D.~C., et al.\ 1997, \apj, 489, 559 
\bibitem[James et al.(2004)]{2004A&A...414...23J} James, P.~A., Shane, N.~S., Beckman, J.~E., et al.\ 2004, \aap, 414, 23 
\bibitem[Hameed \& Devereux(2005)]{2005AJ....129.2597H} Hameed, S., \& Devereux, N.\ 2005, \aj, 129, 2597 
\bibitem[Hao et al.(2011)]{2011ApJ...741..124H} Hao, C.-N., Kennicutt, R.~C., Johnson, B.~D., et al.\ 2011, \apj, 741, 124 
\bibitem[Haynes \& Giovanelli(1984)]{1984AJ.....89..758H} Haynes, M.~P., \& Giovanelli, R.\ 1984, \aj, 89, 758 
\bibitem[Helmboldt et al.(2005)]{2005ApJ...630..824H} Helmboldt, J.~F., Walterbos, R.~A.~M., Bothun, G.~D., \& O'Neil, K.\ 2005, \apj, 630, 824 
\bibitem[Hollenbach \& Salpeter(1971)]{1971ApJ...163..155H} Hollenbach, D., \& Salpeter, E.~E.\ 1971, \apj, 163, 155 
\bibitem[Hollenbach \& Tielens(1997)]{1997ARA&A..35..179H} Hollenbach, D.~J., \& Tielens, A.~G.~G.~M.\ 1997, \araa, 35, 179 
\bibitem[Hughes \& Cortese(2009)]{2009MNRAS.396L..41H} Hughes, T.~M., \& Cortese, L.\ 2009, \mnras, 396, L41 
\bibitem[Hughes et al.(2013)]{2013A&A...550A.115H} Hughes, T.~M., Cortese, L., Boselli, A., Gavazzi, G., \& Davies, J.~I.\ 2013, \aap, 550, AA115 
\bibitem[Karim et al.(2011)]{2011ApJ...730...61K} Karim, A., Schinnerer, E., Mart{\'{\i}}nez-Sansigre, A., et al.\ 2011, \apj, 730, 61 
\bibitem[Kennicutt(1998)]{1998ARA&A..36..189K} Kennicutt, R.~C., Jr.\ 1998, \araa, 36, 189 
\bibitem[Kennicutt \& Kent(1983)]{1983AJ.....88.1094K} Kennicutt, R.~C., Jr., \& Kent, S.~M.\ 1983, \aj, 88, 1094 
\bibitem[Kennicutt \& Evans(2012)]{2012ARA&A..50..531K} Kennicutt, R.~C., \& Evans, N.~J.\ 2012, \araa, 50, 531 
\bibitem[Kennicutt et al.(1987)]{1987AJ.....93.1011K} Kennicutt, R.~C., Jr., Roettiger, K.~A., Keel, W.~C., van der Hulst, J.~M., \& Hummel, E.\ 1987, \aj, 93, 1011 
\bibitem[Kennicutt et al.(1989)]{1989ApJ...337..761K} Kennicutt, R.~C., Jr., Edgar, B.~K., \& Hodge, P.~W.\ 1989, \apj, 337, 761 
\bibitem[Kennicutt et al.(1994)]{1994ApJ...435...22K} Kennicutt, R.~C., Jr., Tamblyn, P., \& Congdon, C.~E.\ 1994, \apj, 435, 22 
\bibitem[Kennicutt et al.(2007)]{2007ApJ...671..333K} Kennicutt, R.~C., Jr., Calzetti, D., Walter, F., et al.\ 2007, \apj, 671, 333 
\bibitem[Kennicutt et al.(2008)]{2008ApJS..178..247K} Kennicutt, R.~C., Jr., Lee, J.~C., Funes, S.~J., Jos{\'e} G., Sakai, S., \& Akiyama, S.\ 2008, \apjs, 178, 247 
\bibitem[Kennicutt et al.(2009)]{2009ApJ...703.1672K} Kennicutt, R.~C., Jr., Hao, C.-N., Calzetti, D., et al.\ 2009, \apj, 703, 1672 
\bibitem[Kim(1989)]{1989ApJ...346..653K} Kim, D.-W.\ 1989, \apj, 346, 653 
\bibitem[Koopmann \& Kenney(2006)]{2006ApJS..162...97K} Koopmann, R.~A., \& Kenney, J.~D.~P.\ 2006, \apjs, 162, 97 
\bibitem[Koopmann et al.(2001)]{2001ApJS..135..125K} Koopmann, R.~A., Kenney, J.~D.~P., \& Young, J.\ 2001, \apjs, 135, 125 
\bibitem[Kroupa(2001)]{2001MNRAS.322..231K} Kroupa, P.\ 2001, \mnras, 322, 231 
\bibitem[Iglesias-P{\'a}ramo et al.(2002)]{2002A&A...384..383I} Iglesias-P{\'a}ramo, J., Boselli, A., Cortese, L., V{\'{\i}}lchez, J.~M., \& Gavazzi, G.\ 2002, \aap, 384, 383 
\bibitem[Isobe et al.(1990)]{1990ApJ...364..104I} Isobe, T., Feigelson, E.~D., Akritas, M.~G., \& Babu, G.~J.\ 1990, \apj, 364, 104 
\bibitem[Lauger et al.(2005)]{2005A&A...434...77L} Lauger, S., Burgarella, D., \& Buat, V.\ 2005, \aap, 434, 77 
\bibitem[Lee et al.(2009)]{2009ApJ...706..599L} Lee, J.~C., Gil de Paz, A., Tremonti, C., et al.\ 2009, \apj, 706, 599 
\bibitem[Lequeux(1971)]{1971A&A....15...42L} Lequeux, J.\ 1971, \aap, 15, 42 
\bibitem[Lequeux et al.(1981)]{1981A&A...103..305L} Lequeux, J., Maucherat-Joubert, M., Deharveng, J.~M., \& Kunth, D.\ 1981, \aap, 103, 305 
\bibitem[Liu et al.(2013)]{2013ApJ...772...27L} Liu, G., Calzetti, D., Kennicutt, R.~C., Jr., et al.\ 2013, \apj, 772, 27 
\bibitem[Macchetto et al.(1996)]{1996A&AS..120..463M} Macchetto, F., Pastoriza, M., Caon, N., et al.\ 1996, \aaps, 120, 463 
\bibitem[Martel et al.(2004)]{2004AJ....128.2758M} Martel, A.~R., Ford, H.~C., Bradley, L.~D., et al.\ 2004, \aj, 128, 2758 
\bibitem[Massey et al.(1988)]{1988ApJ...328..315M} Massey, P., Strobel, K., Barnes, J.~V., \& Anderson, E.\ 1988, \apj, 328, 315 
\bibitem[Momcheva et al.(2013)]{2013AJ....145...47M} Momcheva, I.~G., Lee, J.~C., Ly, C., et al.\ 2013, \aj, 145, 47 
\bibitem[Nilson(1973)]{1973ugcg.book.....N} Nilson, P.\ 1973, Acta Universitatis Upsaliensis.~Nova Acta Regiae Societatis Scientiarum Upsaliensis - Uppsala Astronomiska Observatoriums Annaler, Uppsala: Astronomiska Observatorium, 1973,  
\bibitem[Noeske et al.(2007)]{2007ApJ...660L..43N} Noeske, K.~G., Weiner, B.~J., Faber, S.~M., et al.\ 2007, \apjl, 660, L43 
\bibitem[Noll et al.(2009)]{2009A&A...507.1793N} Noll, S., Burgarella, D., Giovannoli, E., et al.\ 2009, \aap, 507, 1793 
\bibitem[Nolthenius(1993)]{1993ApJS...85....1N} Nolthenius, R.\ 1993, \apjs, 85, 1 
\bibitem[O'Connell(1999)]{1999ARA&A..37..603O} O'Connell, R.~W.\ 1999, \araa, 37, 603 
\bibitem[Osterbrock \& Ferland(2006)]{2006agna.book.....O} Osterbrock, D.~E., \& Ferland, G.~J.\ 2006, Astrophysics of gaseous nebulae and active galactic nuclei, 2nd.~ed.~by D.E.~Osterbrock and G.J.~Ferland.~Sausalito, CA: University Science Books, 2006,  
\bibitem[Pannella et al.(2009)]{2009ApJ...698L.116P} Pannella, M., Carilli, C.~L., Daddi, E., et al.\ 2009, \apjl, 698, L116 
\bibitem[Papovich et al.(2006)]{2006ApJ...640...92P} Papovich, C., Moustakas, L.~A., Dickinson, M., et al.\ 2006, \apj, 640, 92 
\bibitem[Peng et al.(2010)]{2010ApJ...721..193P} Peng, Y.-j., Lilly, S.~J., Kova{\v c}, K., et al.\ 2010, \apj, 721, 193 
\bibitem[P{\'e}rez-Gonz{\'a}lez et al.(2003)]{2003ApJ...591..827P} P{\'e}rez-Gonz{\'a}lez, P.~G., Zamorano, J., Gallego, J., Arag{\'o}n-Salamanca, A., \& Gil de Paz, A.\ 2003, \apj, 591, 827 
\bibitem[Pettini \& Pagel(2004)]{2004MNRAS.348L..59P} Pettini, M., \& Pagel, B.~E.~J.\ 2004, \mnras, 348, L59 
\bibitem[Pforr et al.(2012)]{2012MNRAS.422.3285P} Pforr, J., Maraston, C., \& Tonini, C.\ 2012, \mnras, 422, 3285 
\bibitem[Reddy et al.(2006)]{2006ApJ...644..792R} Reddy, N.~A., Steidel, C.~C., Fadda, D., et al.\ 2006, \apj, 644, 792 
\bibitem[Rodighiero et al.(2010)]{2010A&A...518L..25R} Rodighiero, G., Cimatti, A., Gruppioni, C., et al.\ 2010, \aap, 518, LL25 
\bibitem[Rodighiero et al.(2011)]{2011ApJ...739L..40R} Rodighiero, G., Daddi, E., Baronchelli, I., et al.\ 2011, \apjl, 739, LL40 
\bibitem[Rodighiero et al.(2014)]{2014MNRAS.443...19R} Rodighiero, G., Renzini, A., Daddi, E., et al.\ 2014, \mnras, 443, 19 
\bibitem[Romanishin(1990)]{1990AJ....100..373R} Romanishin, W.\ 1990, \aj, 100, 373 
\bibitem[Salim \& Lee(2012)]{2012ApJ...758..134S} Salim, S., \& Lee, J.~C.\ 2012, \apj, 758, 134 
\bibitem[Salim et al.(2007)]{2007ApJS..173..267S} Salim, S., Rich, R.~M., Charlot, S., et al.\ 2007, \apjs, 173, 267 
\bibitem[Salim et al.(2009)]{2009ApJ...700..161S} Salim, S., Dickinson, M., Michael Rich, R., et al.\ 2009, \apj, 700, 161 
\bibitem[S{\'a}nchez-Gallego et al.(2012)]{2012MNRAS.422.3208S} S{\'a}nchez-Gallego, J.~R., Knapen, J.~H., Wilson, C.~D., et al.\ 2012, \mnras, 422, 3208
\bibitem[Sarzi et al.(2006)]{2006MNRAS.366.1151S} Sarzi, M., Falc{\'o}n-Barroso, J., Davies, R.~L., et al.\ 2006, \mnras, 366, 1151 
\bibitem[Schiminovich et al.(2010)]{2010MNRAS.408..919S} Schiminovich, D., Catinella, B., Kauffmann, G., et al.\ 2010, \mnras, 408, 919 
\bibitem[Schlegel et al.(1998)]{1998ApJ...500..525S} Schlegel, D.~J., Finkbeiner, D.~P., \& Davis, M.\ 1998, \apj, 500, 525 
\bibitem[Schombert et al.(2013)]{2013AJ....146...41S} Schombert, J., McGaugh, S., \& Maciel, T.\ 2013, \aj, 146, 41 
\bibitem[Shields(1991)]{1991AJ....102.1314S} Shields, J.~C.\ 1991, \aj, 102, 1314 
\bibitem[Silva et al.(1998)]{1998ApJ...509..103S} Silva, L., Granato, G.~L., Bressan, A., \& Danese, L.\ 1998, \apj, 509, 103 
\bibitem[Singh et al.(1995)]{1995A&A...302..658S} Singh, K.~P., Bhat, P.~N., Prabhu, T.~P., \& Kembhavi, A.~K.\ 1995, \aap, 302, 658 
\bibitem[Speagle et al.(2014)]{2014ApJS..214...15S} Speagle, J.~S., Steinhardt, C.~L., Capak, P.~L., \& Silverman, J.~D.\ 2014, \apjs, 214, 15 
\bibitem[Spector et al.(2012)]{2012MNRAS.419.2156S} Spector, O., Finkelman, I., \& Brosch, N.\ 2012, \mnras, 419, 2156 
\bibitem[Tacconi et al.(2013)]{2013ApJ...768...74T} Tacconi, L.~J., Neri, R., Genzel, R., et al.\ 2013, \apj, 768, 74 
\bibitem[Taylor-Mager et al.(2007)]{2007ApJ...659..162T} Taylor-Mager, V.~A., Conselice, C.~J., Windhorst, R.~A., \& Jansen, R.~A.\ 2007, \apj, 659, 162 
\bibitem[Thilker et al.(2000)]{2000AJ....120.3070T} Thilker, D.~A., Braun, R., \& Walterbos, R.~A.~M.\ 2000, \aj, 120, 3070 
\bibitem[Thilker et al.(2002)]{2002AJ....124.3118T} Thilker, D.~A., Walterbos, R.~A.~M., Braun, R., \& Hoopes, C.~G.\ 2002, \aj, 124, 3118
\bibitem[Tremonti et al.(2004)]{2004ApJ...613..898T} Tremonti, C.~A., Heckman, T.~M., Kauffmann, G., et al.\ 2004, \apj, 613, 898 
\bibitem[Tresse \& Maddox(1998)]{1998ApJ...495..691T} Tresse, L., \& Maddox, S.~J.\ 1998, \apj, 495, 691 
\bibitem[Trinchieri \& di Serego Alighieri(1991)]{1991AJ....101.1647T} Trinchieri, G., \& di Serego Alighieri, S.\ 1991, \aj, 101, 1647 
\bibitem[Tully(1987)]{1987ApJ...321..280T} Tully, R.~B.\ 1987, \apj, 321, 280 
\bibitem[Usui et al.(1998)]{1998AJ....116.2166U} Usui, T., Saito, M., \& Tomita, A.\ 1998, \aj, 116, 2166 
\bibitem[Valiante et al.(2009)]{2009MNRAS.397.1661V} Valiante, R., Schneider, R., Bianchi, S., \& Andersen, A.~C.\ 2009, \mnras, 397, 1661 
\bibitem[Young et al.(1996)]{1996AJ....112.1903Y} Young, J.~S., Allen, L., Kenney, J.~D.~P., Lesser, A., \& Rownd, B.\ 1996, \aj, 112, 1903 
\bibitem[Youngblood \& Hunter(1999)]{1999ApJ...519...55Y} Youngblood, A.~J., \& Hunter, D.~A.\ 1999, \apj, 519, 55 
\bibitem[Yun et al.(2001)]{2001ApJ...554..803Y} Yun, M.~S., Reddy, N.~A., \& Condon, J.~J.\ 2001, \apj, 554, 803 
\bibitem[Weisz et al.(2012)]{2012ApJ...744...44W} Weisz, D.~R., Johnson, B.~D., Johnson, L.~C., et al.\ 2012, \apj, 744, 44 
\bibitem[Whitaker et al.(2012)]{2012ApJ...754L..29W} Whitaker, K.~E., van Dokkum, P.~G., Brammer, G., \& Franx, M.\ 2012, \apjl, 754, LL29 
\bibitem[Whitaker et al.(2014)]{2014ApJ...795..104W} Whitaker, K.~E., Franx, M., Leja, J., et al.\ 2014, \apj, 795, 104 
\bibitem[Waller(1990)]{1990PASP..102.1217W} Waller, W.~H.\ 1990, \pasp, 102, 1217 
\bibitem[Wolfire et al.(1995)]{1995ApJ...443..152W} Wolfire, M.~G., Hollenbach, D., McKee, C.~F., Tielens, A.~G.~G.~M., \& Bakes, E.~L.~O.\ 1995, \apj, 443, 152 
\bibitem[Wolfire et al.(2008)]{2008ApJ...680..384W} Wolfire, M.~G., Tielens, A.~G.~G.~M., Hollenbach, D., \& Kaufman, M.~J.\ 2008, \apj, 680, 384 
\bibitem[Wuyts et al.(2009)]{2009ApJ...696..348W} Wuyts, S., Franx, M., Cox, T.~J., et al.\ 2009, \apj, 696, 348 
\bibitem[Zhu et al.(2008)]{2008ApJ...686..155Z} Zhu, Y.-N., Wu, H., Cao, C., \& Li, H.-N.\ 2008, \apj, 686, 155 
\bibitem[Zibetti et al.(2009)]{2009MNRAS.400.1181Z} Zibetti, S., Charlot, S., \& Rix, H.-W.\ 2009, \mnras, 400, 1181 
\bibitem[Zwicky et al.(1968)]{1968cgcg.book.....Z} Zwicky, F., Herzog, E., \& Wild, P.\ 1968, Pasadena: California Institute of Technology (CIT), 1961-1968,  

\end{thebibliography}
\end{document}